\documentclass[aps,prl,reprint,superscriptaddress]{revtex4-2}

\usepackage{graphicx}
\usepackage{amsmath}
\usepackage{amssymb}
\usepackage{bm}
\usepackage{xcolor}
\usepackage{hyperref}
\hypersetup{
  colorlinks=true,
  breaklinks=true,
  linkcolor=blue,
  urlcolor=blue,
  citecolor=blue,
}
\usepackage{changes}
\usepackage{comment}
\def\<#1>{\mathinner{\langle#1\rangle}}
\mathchardef\up="0222
\mathchardef\dn="0223

\begin{document}
\preprint{arXiv}

\title{Rise and Fall of the Pseudogap in the Emery model: Insights for Cuprates}

\author{M.~O.~Malcolms}
\email{m.deoliveira@fkf.mpg.de}
\affiliation{Max-Planck-Institut für Festkörperforschung, Heisenbergstraße 1, 70569 Stuttgart, Germany}

\author{Henri Menke}
\affiliation{Max-Planck-Institut für Festkörperforschung, Heisenbergstraße 1, 70569 Stuttgart, Germany}
\affiliation{Department of Physics, Friedrich-Alexander-Universit\"at Erlangen/N\"urnberg, 91058 Erlangen, Germany}

\author{Yi-Ting Tseng}
\affiliation{Department of Physics, Friedrich-Alexander-Universit\"at Erlangen/N\"urnberg, 91058 Erlangen, Germany}

\author{\\Eric Jacob}
\affiliation{Institute of Solid State Physics, TU Wien, 1040 Vienna, Austria}

\author{Karsten Held}
\affiliation{Institute of Solid State Physics, TU Wien, 1040 Vienna, Austria}

\author{Philipp Hansmann}
\affiliation{Department of Physics, Friedrich-Alexander-Universit\"at Erlangen/N\"urnberg, 91058 Erlangen, Germany}

\author{Thomas~Sch{\"a}fer}
\email{t.schaefer@fkf.mpg.de}
\affiliation{Max-Planck-Institut für Festkörperforschung, Heisenbergstraße 1, 70569 Stuttgart, Germany}

\begin{abstract}

The pseudogap in high-temperature superconducting cuprates is an exotic state of matter, displaying emerging Fermi arcs and a momentum-selective suppression of states upon cooling. We show how these phenomena are originating in the three-band Emery model by performing cutting-edge dynamical vertex approximation calculations for its normal state. For the hole-doped parent compound our results demonstrate the formation of a pseudogap due to short-ranged commensurate antiferromagnetic fluctuations. At larger doping values, progressively, incommensurate correlations and a metallic regime appear. Our results are in qualitative agreement with the normal state of cuprates, and, hence, represent a crucial step towards the uniform description of their phase diagrams within a single theoretical framework.
\end{abstract}

\maketitle

\textit{Introduction.} The intriguing properties of layered cuprate
superconductors have not lost their fascination since their discovery in 1986 by Bednorz and M{\"uller} \cite{Bednorz1986}. The reason for the unbroken interest is at least three-fold: First, their temperature/doping phase diagram is unusually rich \cite{Keimer2015}, exhibiting quantum magnetism, unconventional superconductivity, strange metallicity and the famous pseudogap regime, as a function of temperature and (hole-)doping. Second, cuprates hold the potential for immensely impactful technological applications of high-temperature superconductivity. Third, the physical mechanisms behind this rich phenomenology are --after almost 40 years of intense community effort-- still  highly debated. The reason for the latter is deeply rooted in the fact that cuprates are strongly interacting quantum many-body systems, whose properties cannot be explained by a simple single-particle picture: their electrons are strongly correlated in space and in time.

Adding to the complexity of this material class, it has been realized early on that, due to 
the close vicinity of the oxygen $p$-orbitals to the Fermi level, the oxygen $p_{x/y}$-orbitals of the CuO$_2$ two-dimensional layers are relevant besides the
copper $d_{x^2-y^2}$-orbitals; and that the undoped parent compounds are charge-transfer, rather than (single-orbital) Mott-Hubbard insulators \cite{Zaanen1985}. 
In order to explicitly include charge-transfer processes the minimal model thus has to treat the oxygen $p$- on top of the copper $d$-orbitals, enabling, among other properties, (Zhang-Rice) singlet-formation between these orbitals \cite{Zhang1988}. These considerations led to the famous three-band model proposed by Emery \cite{Emery1987}, Varma, Schmitt-Rinks, and Abrahams \cite{Varma1987} and later refined by Andersen and coworkers \cite{Andersen1995}. 
In this respect, the situation in cuprates is largely differing from the case of the recently discovered infinite-layer nickelate superconductors \cite{Li2019}. For the latter case first-principle-based calculations indicate that indeed a single-band Hubbard model could suffice for the low-energy description of superconductivity and a putative pseudogap  \cite{Kitatani2020,Klett2022}; for recent reviews on the single-band Hubbard model see  \cite{Qin2022,Arovas2022}.

\begin{figure}[b!]
\centering
\includegraphics[width=1.\columnwidth]{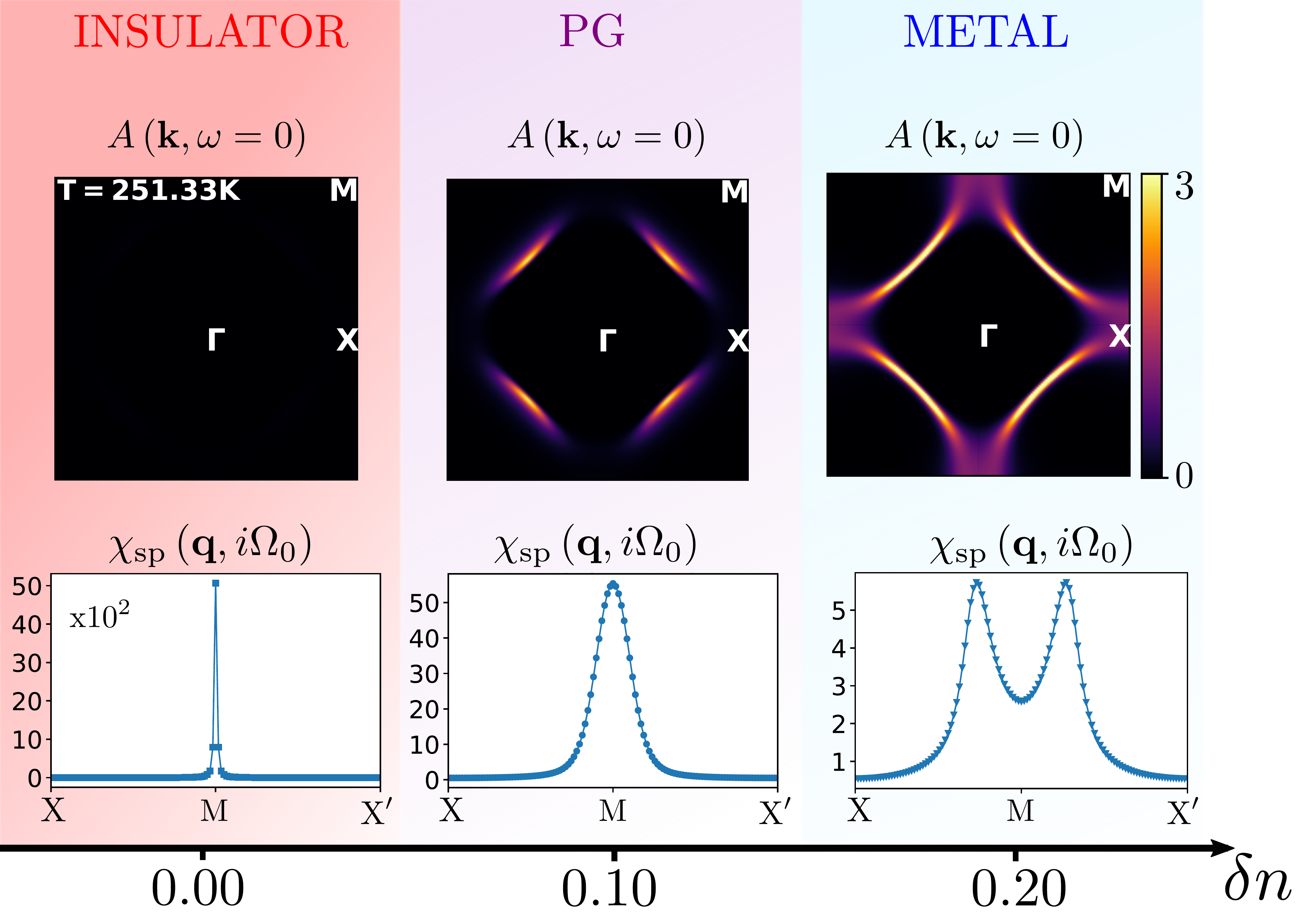}
\caption{Sketch of the phase diagram of the normal phase regime of the three-orbital Emery model for cuprates, for a fixed temperature $T=251.33$ K as a function of hole doping, as actually calculated by the dynamical vertex approximation. After an insulating regime the pseudogap (PG) first rises and then falls off again with the system eventually becoming a conventional metal. In the PG regime, the Fermi surface is reduced to Fermi arcs (upper row). Concomitantly, commensurate antiferromagnetic and incommensurate magnetic fluctuations are present, demonstrated by the momentum-dependent spin susceptibility (lower row). The shadings serve as a guide to the eye throughout the manuscript.}
\label{fig:phase_diagram}
\end{figure}

\begin{figure*}[t!]
\includegraphics[width=0.99\textwidth]{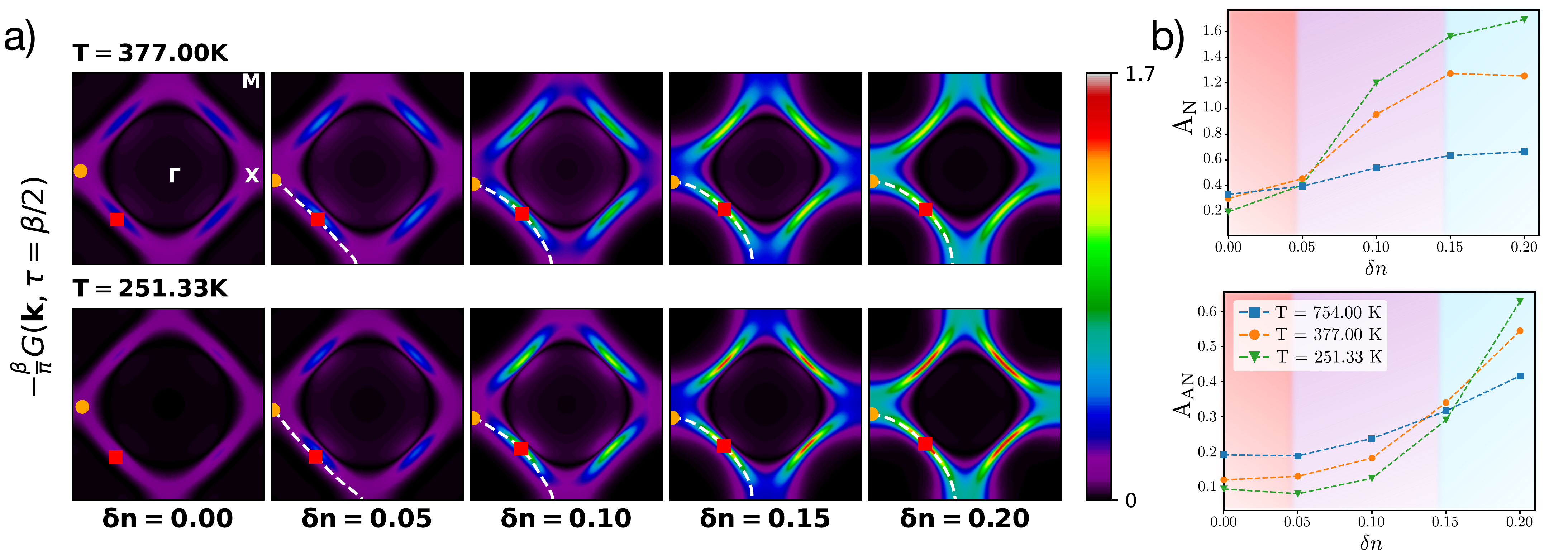}
\caption{(a) Hole-doping evolution of D$\Gamma$A $\mathbf{k}$-resolved spectral intensities at the Fermi energy for $T=t_{pp}/20=377$ K (upper panels) and $T=t_{pp}/30=251.33 $ K (lower panels) with displayed interacting Fermi plus Luttinger surface (dashed white line) and the position of the nodal (red square) and anti-nodal (orange circle) points. (b) Hole-doping and temperature dependence of the spectral function at the interacting Fermi surface at the nodal (upper panel) and anti-nodal (lower panel) points. The shaded background highlights the different crossovers revealed by D$\Gamma$A, indicated in Fig.~\ref{fig:phase_diagram} by the same color.} 
\label{fig:spectral_functions}
\end{figure*}

One of the hallmark features of the normal state of the cuprates is the so-called pseudogap, a suppression of the spectral function near the antinode $(\pi,0)$ at low hole-doping levels. This intriguing regime, which shows no signs of a thermodynamic phase transition, can be detected experimentally, for example, by nuclear magnetic resonance (NMR) Knight shift \cite{Alloul1989}, or, even more directly, by angle-resolved photoemission spectroscopy (ARPES) \cite{Damascelli2003,Kanigel2006,Yoshida2006,Shen2005,Taillefer2010,Boschini2020,Boschini2024}. 

A reliable modeling of cuprates, not only has to take into account their multi-orbital nature, but should also be able to reveal the crossover into the pseudogap regime upon hole doping.
While the pseudogap in the Hubbard model has been firmly established by numerous approaches \cite{Kyung2004,Schaefer2021,Krien2022,Vilk2024,Simkovic2024}, the presence and nature of the pseudogap in the hole-doped Emery model has been, to the best of our knowledge, not thoroughly investigated yet: Despite efforts with a variety of techniques \cite{Avella2013,Fratino2016b,Dash2019,Mao2024,Gauvin2024,Sordi2024,Bacq2024} and calculations for the superconducting properties \cite{Kowalski2021,Mai2021,Mai2021b}, a detailed analysis of the evolution of the Fermi surface with temperature and doping, and its connection to the precise nature of spin fluctuations, essential hallmarks of the pseudogap regime, have not been reported before.

In this Letter we now make significant progress in the universal description of the cuprate phase diagram by extending the state-of-the-art $\lambda$-corrected dynamical vertex approximation (D$\Gamma$A) to the three-band Emery model \cite{Supplemental,Galler2017}. With our extension to three bands and arbitrarily accurate resolution of momenta and correlation lengths we establish the normal phase regime of the Emery model in two dimensions with realistic cuprate parameters as a function of temperature and hole doping. We document the fate of the pseudogap from its rise  out of the insulating parent compound to its eventual fall in the metallic regime (see Fig.~\ref{fig:phase_diagram}). We further discuss the relevance of our calculations for spectroscopic experiments, a crucial step on the way to a comprehensive theoretical description of the cuprates.

\textit{Model and Method.} We consider the three-band Emery model that accounts for the copper-oxygen hybridized character of the single band that crosses the Fermi surface (see Supplemental Materials \cite{Supplemental}) and whose Hamiltonian reads 

\begin{equation}
H = \sum_{\mathbf{k},\sigma} \psi^{\dagger}_{\mathbf{k},\sigma}\Bar{h}_{0}(\mathbf{k})\psi_{\mathbf{k},\sigma} + U\sum_{i}n^{d}_{i,\uparrow}n^{d}_{i,\downarrow}, 
\label{eq:hamiltoniam}
\end{equation}
where $\psi^{\dagger}_{\mathbf{k},\sigma}=(d^{\dagger}_{\mathbf{k},\sigma}, p^{\dagger}_{x,\mathbf{k},\sigma}, p^{\dagger}_{y,\mathbf{k},\sigma})$ contains the electronic creation operators on the copper $d_{x^{2}-y^{2}}$- and the oxygen $p_{x}$- and $p_{y}$-orbitals with spin $\sigma$ and momentum $\mathbf{k}$. $U$ quantifies the local part of the Coulomb repulsion restricted to the copper $d_{x^{2}-y^{2}}$-orbital and $n^{d}_{i,\sigma}=d^{\dagger}_{i,\sigma}d_{i,\sigma}$  is the number operator per spin for the $d$-orbital. On the square lattice, setting the distance between unit cells to unity, the non-interacting Hamiltonian $\Bar{h}_{0}(\mathbf{k})$ in Eq.~(\ref{eq:hamiltoniam}) reads
\begin{equation}
\bar{h}_{0}\left(\mathbf{k}\right)=\left(\begin{array}{ccc}
\epsilon_{d} & t_{pd}s_{k_{x}} & t_{pd}s_{k_{y}}\\
t_{pd}s_{k_{x}}^{*} & \epsilon_{p}+t'_{pp}\varepsilon_{k_{x}} & t_{pp}s_{k_{x}}^{*}s_{k_{y}}\\
t_{pd}s_{k_{y}}^{*} & t_{pp}s_{k_{y}}^{*}s_{k_{x}} & \epsilon_{p}+t'_{pp}\varepsilon_{k_{y}}
\end{array}\right),
\label{eq:nonint_hamiltoniam}
\end{equation}
with $\varepsilon_{k} = 2 \cos k$ and $s_{k} = 2i e^{i\frac{k}{2}}\sin \frac{k}{2}$. The on-site energies on the Cu and O orbitals are denoted $\epsilon_{d}$ and $\epsilon_{p}$, respectively. The nearest-neighbor Cu-O hopping is denoted by $t_{pd}$, while $t'_{pp}$ and $t_{pp}$  denote the direct nearest-neighbor O-O hopping and the indirect one through the Cu site, respectively.

In our calculations, following Refs. \cite{Weber2012,Kowalski2021},  we adopt $t_{pp} \sim 650\,$meV, $\epsilon_{p} = 2.3t_{pp} = 1.5\,$eV, $t_{pd} = 2.1t_{pp} = 1.37\,$eV and $t'_{pp} = 0.2t_{pp} = 130\,$meV for the non-interacting system in Eq.~(\ref{eq:nonint_hamiltoniam});
$\epsilon_{p}$ has been obtained by subtracting the same double-counting contribution as in Ref. \cite{Weber2012}. Motivated by X-ray photoemission spectroscopy (XPES) experiments \cite{XAS_1, XAS_2}, in this work, we fix the interaction value to $U=10t_{pp}=6.5$ eV, while the total electronic density $n=n_{\text{d}_{x^{2}-y^{2}}}+n_{\text{p}_{x}}+n_{\text{p}_{y}}$ and the temperature $T$ are varied.

We analyze the normal state of the model by means of ladder D$\Gamma$A \cite{Toschi2007}, a diagrammatic extension \cite{Rohringer2018} of the dynamical mean-field theory \cite{Georges1996} which is particularly suited for treating spatial and temporal correlations. Specific properties of this method are its arbitrarily fine momentum resolution and applicability in the strong-coupling regime, both of which being (i) highly desired for the description of the pseudogap regime and (ii) of advantage over cluster (see, e.g., \cite{Maier2005,Fratino2016b,Dash2019,Kowalski2021,Mai2021,Mai2021b,Mai2024,Sordi2024}) or other diagrammatic (see, e.g., \cite{Gauvin2024}) techniques, respectively. As such, D$\Gamma$A has already been very successful for single-band systems over the past years \cite{Rohringer2018,Kitatani2020,Schaefer2021}. We now extend this method to the Emery model \cite{Supplemental}.

\begin{figure}[t!]
\includegraphics[width=0.85\columnwidth]{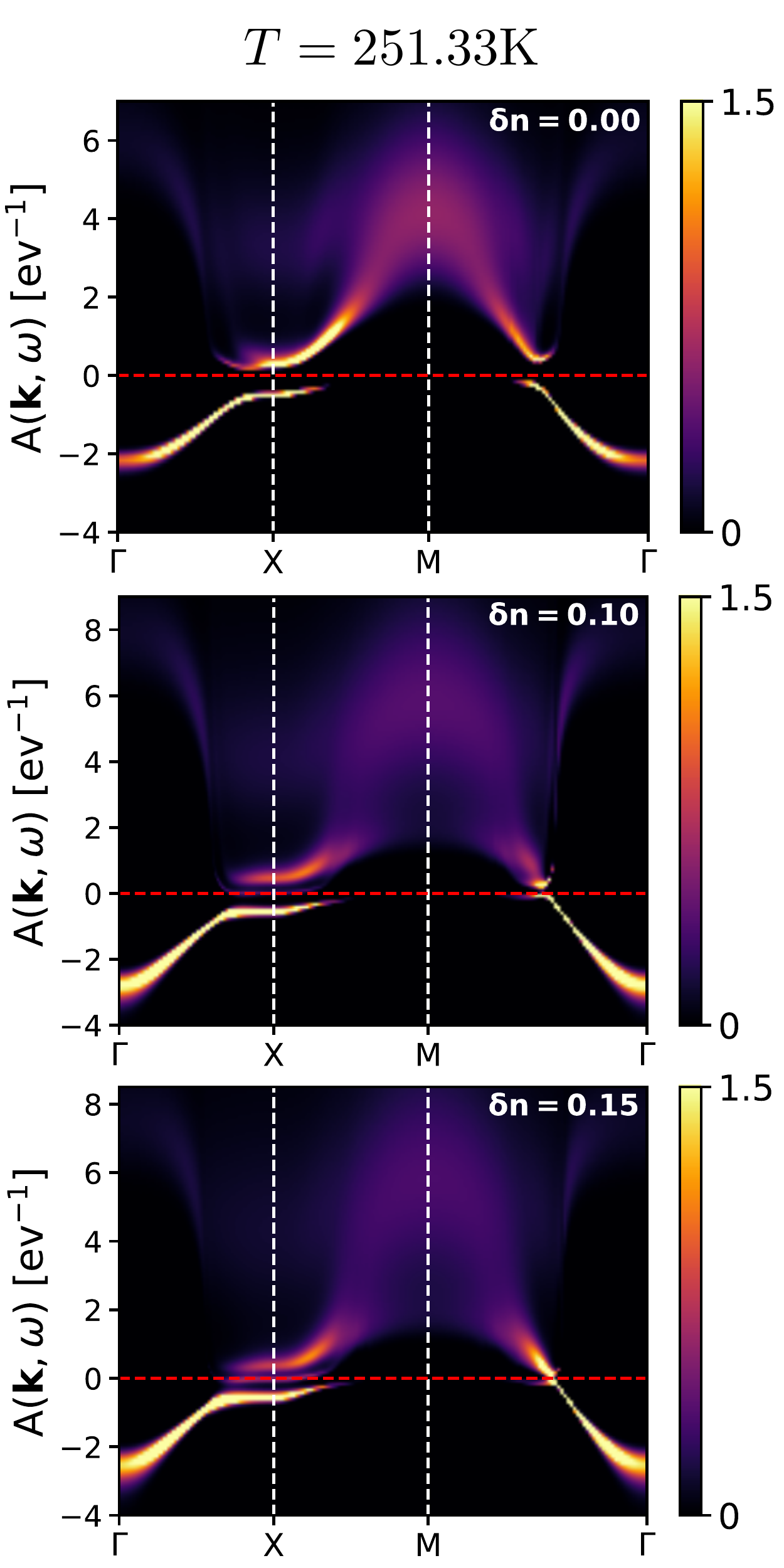}
\caption{Hole-doping evolution of ladder D$\Gamma$A spectral intensities along the high-symmetry path $\Gamma$XM$\Gamma$ of the Brillouin zone for $T=377$ K obtained by MaxEnt analytical continuation \cite{Reymbaut2017}. The dashed red line denotes the Fermi level of the interacting system. The intensity grows from purple to yellow.}
\label{fig:ImG_AN_N}
\end{figure}

 \begin{figure*}[ht!]
\includegraphics[width=0.99\textwidth]{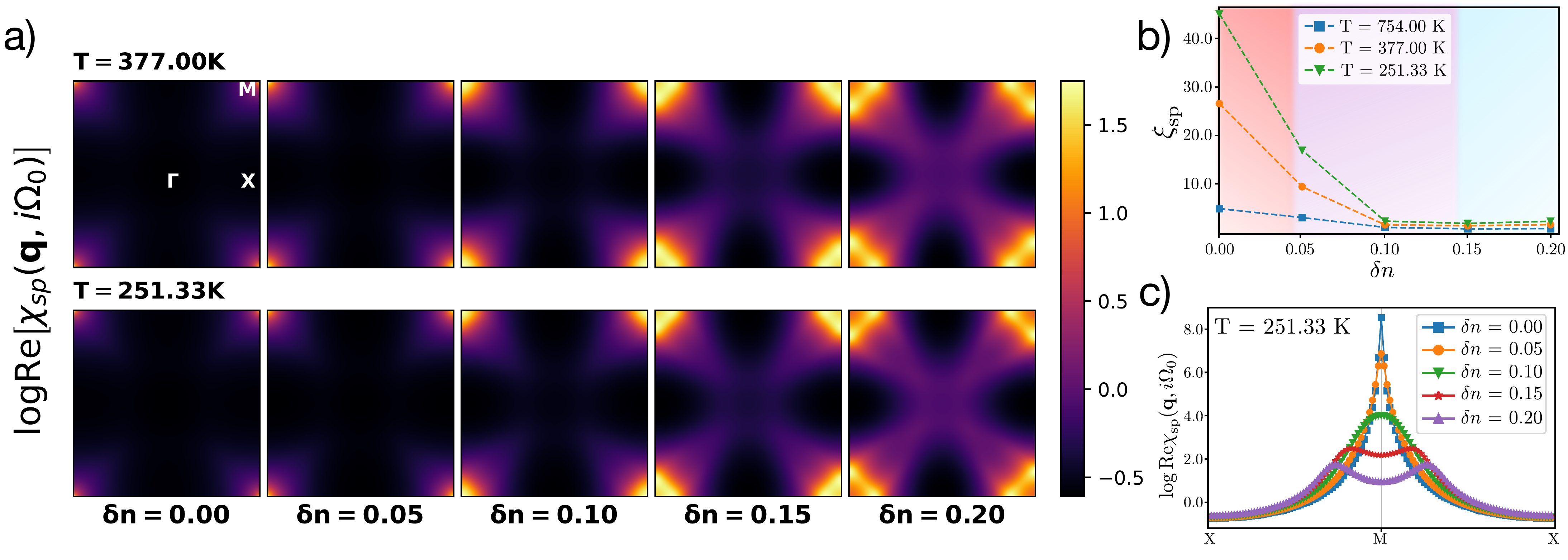}
\caption{(a) Hole-doping evolution of the logarithm of the $\mathbf{q}$-resolved ladder D$\Gamma$A static spin susceptibility for $T=377$ K (upper panels) and $T=251.33$ K (lower panels). (b) Hole-doping and temperature dependence of ladder D$\Gamma$A spin correlation length. (c) Hole-doping evolution of the logarithm of the ladder D$\Gamma$A static spin susceptibility along the high symmetry path XMX of the Brillouin zone for $T=251.33$ K.} 
\label{fig:2p_qnts}
\end{figure*}

\textit{Fermiology: Fermi surfaces, Fermi arcs, and spectral functions.}
We first investigate the hole-doping and temperature evolution of the electronic spectrum. Fig.~\ref{fig:spectral_functions}(a) displays the momentum dependence of the Green function in imaginary time $\tau\!=\!\beta/2$ $\left(\beta\!=\!1/T\right)$ for different temperatures: $T=377$ K, $251.33$ K and various values of hole-doping: $\delta n=0.00, 0.05, 0.10, 0.15, 0.20$ with $\delta n=5-n$.
It is directly related to the spectrum in an energy interval $\sim\!T$ around the Fermi energy, $-\frac{\beta}{\pi}G\left(\mathbf{k},\tau=\beta/2\right)=\frac{\beta}{2\pi}\int\text{d}\omega\frac{A\left(\mathbf{k},\omega\right)}{\cosh\left(\beta\omega/2\right)}$, and avoids the perils of the ill-conditioned analytical continuation.

Fig.~\ref{fig:spectral_functions}(a) also displays the correlated Fermi plus Luttinger surface, indicated as dashed white lines and determined as $\text{Re }G\left(\mathbf{k}_{\text{FS}},i\omega_{n}\rightarrow0\right)=0$, where $i\omega_{n}\rightarrow0$ is obtained as a linear extrapolation of the two first fermionic Matsubara frequencies. The minima (maxima) of $-\frac{\beta}{\pi}G\left(\mathbf{k}_{\text{FS}},\tau=\beta/2\right)$ on the correlated Fermi surface defines the anti-nodal (nodal) point indicated by the orange (red) dot (square) in Fig.~\ref{fig:spectral_functions}(a). For both  temperatures, $T=377$ K and $251.33$ K, the electronic spectrum evolves upon doping from an insulating parent compound (undoped system, $\delta n$=0.00) characterized by a very low spectral intensity over the Fermi surface into a metallic regime ($\delta n$=0.20). In between, a pseudogap phase, marked by the appearance of a Fermi arc structure at the Fermi level, emerges ($\delta n=0.10$), which is a hallmark feature of cuprates \cite{Yoshida2006}. Upon cooling, the spectrum intensity over the Fermi surface reduces for the undoped system while it increases for the highest doping analyzed here $\left(\delta n=0.20\right)$, which documents the evolution of the system from an insulating into a metallic state upon hole-doping. In addition, from $\delta n=0.05$ to $\delta n=0.10$ in Fig.~\ref{fig:spectral_functions}(a), for both temperatures $T=377$ K and $251.33$ K, the reduction in the scattering processes between the quasiparticles leads to a topological reconstruction of the Fermi surface from a electron-like to an hole-like shape (Lifshitz transition) \cite{Wu2018,Meixner2024}. For $\delta n=0.00$ (insulating state) no Fermi surface can be defined, since the system is completely gapped.

In Fig.~\ref{fig:spectral_functions}(a) we further observe that even deep inside the metallic state $\left(\delta n = 0.20\right)$, non-local correlation effects alter the Fermi surface (see \cite{Supplemental} for a comparison with DMFT]). These changes of the Fermi surface arise from the momentum dependence of the real part of the electronic self-energy (Re $\Sigma_{\mathbf{k},i\omega_{n}}$) and are      minor compared to the much more dramatic opening of the pseudogap at the antinode. This pseudogap opening instead  
is due to the momentum dependence of Im$\Sigma_{\mathbf{k},i\omega_{n}}$ and can severely affect the superconducting instabilities as previously shown  \cite{Kitatani2023}.

Fig.~\ref{fig:spectral_functions}(b) shows the doping evolution of the nodal (upper panel) and anti-nodal (lower pannel) spectral weight for different temperatures. For the undoped system we observe that upon cooling the spectral weight at both node and anti-node decreases, corroborating the notion of an insulating regime. However, at $\delta n=0.05$, while the spectral function at the antinodal point still decreases upon cooling,  it saturates at the nodal one. Such behavior indicates a crossover into the pseudogap phase since at a higher doping $\delta n=0.10$ the spectral function at the nodal point shows the inverse behavior of the antinodal one. That is, it increases now upon decreasing temperature indicating the metallicity of the nodal point. With further doping to $\delta n=0.15$, a saturation at the antinodal point is also observed. This point introduces a second crossover into a metallic phase for all momenta that can be observed at $\delta n=0.20$, where both spectral weights at node and antinode increase when cooling the system, the characteristic feature of a metal. 

In order to investigate further the crossovers observed from the Green function on the Matsubara axis, we present in Fig.~\ref{fig:ImG_AN_N} the energy-momentum dispersion of the spectrum around the Fermi level. Here we plot the spectral function at real (instead of Matsubara) frequencies, obtained by MaxEnt analytical continuation \cite{Reymbaut2017} at $T=377$ K for relevant dopings along the $\Gamma$XM$\Gamma$ high-symmetry path of the Brillouin zone. This high-symmetry path includes both the nodal and antinodal points. The undoped system ($\delta n=0.00$) clearly shows an insulating behavior marked by the presence of a complete gap (upper pannel) at the Fermi level. For $\delta n=0.10$, at the insulator-to-pseudogap crossover (middle pannel), electronic states emerge along the M$\Gamma$-direction while a gap still persists along the $\Gamma$X-direction. Deep inside the pseudogap phase at $\delta n=0.15$ (lower pannel), there remains a gap-splitting   along the $\Gamma$X-direction
while there is a band crossing the Fermi level along the M$\Gamma$-direction.

\textit{Magnetic susceptibilities and correlation lengths.}
 An interesting question is whether the pseudogap phase is driven by spin fluctuations (see, for instance, \cite{Ye2023,Ye2023b}), as demonstrated for the single-band Hubbard model by fluctuation diagnostics techniques \cite{Gunnarsson2015,Wu2016,Schaefer2021b}. Fig.~\ref{fig:2p_qnts}(b) shows the doping dependence of the magnetic correlation length $\xi$ obtained by an Ornstein-Zernike fit of the static magnetic lattice susceptibility  for different temperatures. Note that, as  we consider an ideal two-dimensional system, neither the magnetic correlation length nor the static magnetic lattice susceptibility diverge at finite temperatures \cite{Mermin1966, Supplemental}. For all temperatures in Fig.~\ref{fig:2p_qnts}(b), the magnetic correlation length rapidly decreases upon hole-doping, reaching always its maxima in the undoped system ($\delta n=0.00$). Differently from the undoped system where the insulating phase seems to be driven by long-ranged spin fluctuations characterized by a large correlation length ($\sim 45$ sites at $T=251.33$ K), deep inside the pseudogap phase ($\delta n=0.10$) the magnetic correlation is rather small ($\lesssim 5$ lattice sites at $T=251.33$ K) and only slightly varying with temperature and doping. This indicates that the pseudogap regime in the Emery model is driven by short-ranged dynamical spin fluctuations included in the ladder D$\Gamma$A construction of the electronic self-energy \cite{Supplemental}.

To extend our analysis, Fig.~\ref{fig:2p_qnts}(a) displays the doping and temperature dependence of the static magnetic lattice susceptibility over the Brillouin zone. For all the temperatures, at low hole-doping ($\delta n<0.15$) where the system is either insulating or in the pseudogap regime, the antiferromagnetic correlations are commensurately peaked at $\mathbf{q}=\mathbf{M}=(\pi,\pi)$, while inside the metallic phase $(\delta n\geq0.15)$ they become incommensurate and peak at $\mathbf{q}=(\pi,\pi\pm\delta)$, $(\pi\pm\delta,\pi)$. The same feature is observed in Fig.~\ref{fig:2p_qnts}(c) where the static magnetic lattice susceptibility is shown along the XMX high-symmetry path of the Brillouin zone as a function of doping at $T=251.33$ K. 

\textit{Discussion and relation to experiments.}
We now compare our results with experimental studies for one of the most prominent members of the cuprate family: La$_{2-x}$Sr$_x$CuO$_4$ (LSCO). The model parameters used for our calculations fit reasonably well with the parameters determined for LSCO previously by ab-initio methods \cite{Weber2012}, except for $\varepsilon_p$, which is slightly smaller in our study. For LSCO a large number of experimental studies that connect to our Emery model calculation in the paramagnetic phase exist: the evolution of the Fermi surface, the nature of the magnetic fluctuations, and the magnetic correlation length have been measured as a function of Sr-doping, corresponding to hole-doping in our calculations. In the following we assume a clean system, see \cite{Alloul2024} for a recent review of the impact of disorder on the cuprates' phase diagram.

For the lowest temperatures that we analyzed, we find the pseudogap regime in a doping range of $\delta n \in [0.05, 0.15]$. This is in good agreement with ARPES \cite{Yoshida2009} and Nernst effect \cite{Cyr2018} measurements performed at $T\approx 100$ K. The underlying Fermi surface determined by ARPES \cite{Yoshida2006} evolves similarly to our calculations: at small dopings the spectral weight is low and concentrated around the node, while the Fermi arcs enlarge at intermediate doping, before the Fermi surface eventually closes at $\delta n \approx 0.2$, indicating the rise and fall of the pseudogap.

Regarding magnetism, a crossover from commensurate to incommensurate magnetic fluctuations has been observed in inelastic neutron scattering experiments for LSCO \cite{Birgeneau1989,Yamada1998}. There, the onset of incommensurability has been determined to start already at very small doping levels, with a direct proportionality of incommensurability and hole doping \cite{Yamada1998}. Our computational results have a non-zero doping threshold for incommensurate fluctuations instead, which we  attribute to the lower  temperatures $T\lesssim 40$ K, at which the experiments have been performed. Let us note that incommensurability has also been measured in YBa$_2$Cu$_3$O$_{6+x}$ \cite{Haug2010}, however it seems to be absent in HgBa$_2$CuO$_{4+\delta}$ \cite{Anderson2024}.

The (in-plane) magnetic correlation length $\xi$ has been shown to significantly drop from about 15 lattice spacings in LSCO near the parent compound, to about 3-4 lattice spacings for $\delta n \in [0.05, 0.15]$ \cite{Birgeneau1988}, being in very good agreement with our calculations in the pseudogap regime.

\textit{Conclusions and outlook.}
With cutting-edge D$\Gamma$A calculations we have demonstrated that the Emery model captures the rise and fall of the pseudogap through a series of crossovers from the insulating to the pseudogap to the metallic regime. Our results are in good agreement with the signatures of the normal phase observed in spectroscopic experiments for cuprates. Both the insulating and pseudogap phases are accompanied by (short-range) commensurate antiferromagnetic correlations while the metallic state is dominated by incommensurate ones. The short-ranged nature of these fluctuations characterize the emerging pseudogap as being of strong-coupling type. Establishing these key features for the normal phase of the fundamental  model for cuprates, the three-band Emery model, represents a significant step towards the universal theoretical description of cuprates.

\begin{acknowledgements}
\textit{Acknowledgements.}
We acknowledge fruitful discussions with Nils Wentzell and Paul Worm on the theoretical and numerical part of our work, as well as with Eva Benckiser and Matthias Hepting on experiments on cuprates and critical reading of our manuscript. The work has been supported in part by the Austrian Science Funds (FWF) through Grant DOI 10.55776/I5398 and  Grant DOI 10.55776/I5868. The authors gratefully acknowledge the scientific support and HPC resources provided by the Erlangen National High Performance Computing Center (NHR@FAU) of the Friedrich-Alexander-Universität Erlangen-Nürnberg (FAU) [hardware funded by the German Research Foundation - DFG], and the computing service facility of the MPI-FKF, responsible for the Bordeaux cluster.
\end{acknowledgements}

\bibliography{main.bib}

\begin{thebibliography}{88}%
\makeatletter
\providecommand \@ifxundefined [1]{%
 \@ifx{#1\undefined}
}%
\providecommand \@ifnum [1]{%
 \ifnum #1\expandafter \@firstoftwo
 \else \expandafter \@secondoftwo
 \fi
}%
\providecommand \@ifx [1]{%
 \ifx #1\expandafter \@firstoftwo
 \else \expandafter \@secondoftwo
 \fi
}%
\providecommand \natexlab [1]{#1}%
\providecommand \enquote  [1]{``#1''}%
\providecommand \bibnamefont  [1]{#1}%
\providecommand \bibfnamefont [1]{#1}%
\providecommand \citenamefont [1]{#1}%
\providecommand \href@noop [0]{\@secondoftwo}%
\providecommand \href [0]{\begingroup \@sanitize@url \@href}%
\providecommand \@href[1]{\@@startlink{#1}\@@href}%
\providecommand \@@href[1]{\endgroup#1\@@endlink}%
\providecommand \@sanitize@url [0]{\catcode `\\12\catcode `\$12\catcode
  `\&12\catcode `\#12\catcode `\^12\catcode `\_12\catcode `\%12\relax}%
\providecommand \@@startlink[1]{}%
\providecommand \@@endlink[0]{}%
\providecommand \url  [0]{\begingroup\@sanitize@url \@url }%
\providecommand \@url [1]{\endgroup\@href {#1}{\urlprefix }}%
\providecommand \urlprefix  [0]{URL }%
\providecommand \Eprint [0]{\href }%
\providecommand \doibase [0]{https://doi.org/}%
\providecommand \selectlanguage [0]{\@gobble}%
\providecommand \bibinfo  [0]{\@secondoftwo}%
\providecommand \bibfield  [0]{\@secondoftwo}%
\providecommand \translation [1]{[#1]}%
\providecommand \BibitemOpen [0]{}%
\providecommand \bibitemStop [0]{}%
\providecommand \bibitemNoStop [0]{.\EOS\space}%
\providecommand \EOS [0]{\spacefactor3000\relax}%
\providecommand \BibitemShut  [1]{\csname bibitem#1\endcsname}%
\let\auto@bib@innerbib\@empty
\bibitem [{\citenamefont {{Bednorz}}\ and\ \citenamefont
  {{M{\"u}ller}}(1986)}]{Bednorz1986}%
  \BibitemOpen
  \bibfield  {author} {\bibinfo {author} {\bibfnamefont {J.~G.}\ \bibnamefont
  {{Bednorz}}}\ and\ \bibinfo {author} {\bibfnamefont {K.~A.}\ \bibnamefont
  {{M{\"u}ller}}},\ }\bibfield  {title} {\bibinfo {title} {{Possible high
  T$_{c}$ superconductivity in the Ba-La-Cu-O system}},\ }\href
  {https://doi.org/10.1007/BF01303701} {\bibfield  {journal} {\bibinfo
  {journal} {Zeitschrift f{\"u}r Physik B Condensed Matter}\ }\textbf {\bibinfo
  {volume} {64}},\ \bibinfo {pages} {189} (\bibinfo {year} {1986})}\BibitemShut
  {NoStop}%
\bibitem [{\citenamefont {{Keimer}}\ \emph {et~al.}(2015)\citenamefont
  {{Keimer}}, \citenamefont {{Kivelson}}, \citenamefont {{Norman}},
  \citenamefont {{Uchida}},\ and\ \citenamefont {{Zaanen}}}]{Keimer2015}%
  \BibitemOpen
  \bibfield  {author} {\bibinfo {author} {\bibfnamefont {B.}~\bibnamefont
  {{Keimer}}}, \bibinfo {author} {\bibfnamefont {S.~A.}\ \bibnamefont
  {{Kivelson}}}, \bibinfo {author} {\bibfnamefont {M.~R.}\ \bibnamefont
  {{Norman}}}, \bibinfo {author} {\bibfnamefont {S.}~\bibnamefont {{Uchida}}},\
  and\ \bibinfo {author} {\bibfnamefont {J.}~\bibnamefont {{Zaanen}}},\
  }\bibfield  {title} {\bibinfo {title} {{From quantum matter to
  high-temperature superconductivity in copper oxides}},\ }\href
  {https://doi.org/10.1038/nature14165} {\bibfield  {journal} {\bibinfo
  {journal} {Nature}\ }\textbf {\bibinfo {volume} {518}},\ \bibinfo {pages}
  {179} (\bibinfo {year} {2015})}\BibitemShut {NoStop}%
\bibitem [{\citenamefont {Zaanen}\ \emph {et~al.}(1985)\citenamefont {Zaanen},
  \citenamefont {Sawatzky},\ and\ \citenamefont {Allen}}]{Zaanen1985}%
  \BibitemOpen
  \bibfield  {author} {\bibinfo {author} {\bibfnamefont {J.}~\bibnamefont
  {Zaanen}}, \bibinfo {author} {\bibfnamefont {G.~A.}\ \bibnamefont
  {Sawatzky}},\ and\ \bibinfo {author} {\bibfnamefont {J.~W.}\ \bibnamefont
  {Allen}},\ }\bibfield  {title} {\bibinfo {title} {{Band gaps and electronic
  structure of transition-metal compounds}},\ }\href
  {https://doi.org/10.1103/PhysRevLett.55.418} {\bibfield  {journal} {\bibinfo
  {journal} {Phys. Rev. Lett.}\ }\textbf {\bibinfo {volume} {55}},\ \bibinfo
  {pages} {418} (\bibinfo {year} {1985})}\BibitemShut {NoStop}%
\bibitem [{\citenamefont {Zhang}\ and\ \citenamefont {Rice}(1988)}]{Zhang1988}%
  \BibitemOpen
  \bibfield  {author} {\bibinfo {author} {\bibfnamefont {F.~C.}\ \bibnamefont
  {Zhang}}\ and\ \bibinfo {author} {\bibfnamefont {T.~M.}\ \bibnamefont
  {Rice}},\ }\bibfield  {title} {\bibinfo {title} {{Effective Hamiltonian for
  the superconducting Cu oxides}},\ }\href
  {https://doi.org/10.1103/PhysRevB.37.3759} {\bibfield  {journal} {\bibinfo
  {journal} {Phys. Rev. B}\ }\textbf {\bibinfo {volume} {37}},\ \bibinfo
  {pages} {3759} (\bibinfo {year} {1988})}\BibitemShut {NoStop}%
\bibitem [{\citenamefont {Emery}(1987)}]{Emery1987}%
  \BibitemOpen
  \bibfield  {author} {\bibinfo {author} {\bibfnamefont {V.~J.}\ \bibnamefont
  {Emery}},\ }\bibfield  {title} {\bibinfo {title} {{Theory of
  high-${\mathrm{T}}_{\mathrm{c}}$ superconductivity in oxides}},\ }\href
  {https://doi.org/10.1103/PhysRevLett.58.2794} {\bibfield  {journal} {\bibinfo
   {journal} {Phys. Rev. Lett.}\ }\textbf {\bibinfo {volume} {58}},\ \bibinfo
  {pages} {2794} (\bibinfo {year} {1987})}\BibitemShut {NoStop}%
\bibitem [{\citenamefont {Varma}\ \emph {et~al.}(1987)\citenamefont {Varma},
  \citenamefont {Schmitt-Rink},\ and\ \citenamefont {Abrahams}}]{Varma1987}%
  \BibitemOpen
  \bibfield  {author} {\bibinfo {author} {\bibfnamefont {C.}~\bibnamefont
  {Varma}}, \bibinfo {author} {\bibfnamefont {S.}~\bibnamefont
  {Schmitt-Rink}},\ and\ \bibinfo {author} {\bibfnamefont {E.}~\bibnamefont
  {Abrahams}},\ }\bibfield  {title} {\bibinfo {title} {{Charge transfer
  excitations and superconductivity in ``ionic” metals}},\ }\href
  {https://doi.org/https://doi.org/10.1016/0038-1098(87)90407-8} {\bibfield
  {journal} {\bibinfo  {journal} {Solid State Communications}\ }\textbf
  {\bibinfo {volume} {62}},\ \bibinfo {pages} {681} (\bibinfo {year}
  {1987})}\BibitemShut {NoStop}%
\bibitem [{\citenamefont {Andersen}\ \emph {et~al.}(1995)\citenamefont
  {Andersen}, \citenamefont {Liechtenstein}, \citenamefont {Jepsen},\ and\
  \citenamefont {Paulsen}}]{Andersen1995}%
  \BibitemOpen
  \bibfield  {author} {\bibinfo {author} {\bibfnamefont {O.}~\bibnamefont
  {Andersen}}, \bibinfo {author} {\bibfnamefont {A.}~\bibnamefont
  {Liechtenstein}}, \bibinfo {author} {\bibfnamefont {O.}~\bibnamefont
  {Jepsen}},\ and\ \bibinfo {author} {\bibfnamefont {F.}~\bibnamefont
  {Paulsen}},\ }\bibfield  {title} {\bibinfo {title} {{LDA energy bands,
  low-energy Hamiltonians, $t'$, $t''$, $t_\perp (k)$, and $J_\perp$}},\ }\href
  {https://doi.org/https://doi.org/10.1016/0022-3697(95)00269-3} {\bibfield
  {journal} {\bibinfo  {journal} {Journal of Physics and Chemistry of Solids}\
  }\textbf {\bibinfo {volume} {56}},\ \bibinfo {pages} {1573} (\bibinfo {year}
  {1995})},\ \bibinfo {note} {proceedings of the Conference on Spectroscopies
  in Novel Superconductors}\BibitemShut {NoStop}%
\bibitem [{\citenamefont {{Li}}\ \emph {et~al.}(2019)\citenamefont {{Li}},
  \citenamefont {{Lee}}, \citenamefont {{Wang}}, \citenamefont {{Osada}},
  \citenamefont {{Crossley}}, \citenamefont {{Lee}}, \citenamefont {{Cui}},
  \citenamefont {{Hikita}},\ and\ \citenamefont {{Hwang}}}]{Li2019}%
  \BibitemOpen
  \bibfield  {author} {\bibinfo {author} {\bibfnamefont {D.}~\bibnamefont
  {{Li}}}, \bibinfo {author} {\bibfnamefont {K.}~\bibnamefont {{Lee}}},
  \bibinfo {author} {\bibfnamefont {B.~Y.}\ \bibnamefont {{Wang}}}, \bibinfo
  {author} {\bibfnamefont {M.}~\bibnamefont {{Osada}}}, \bibinfo {author}
  {\bibfnamefont {S.}~\bibnamefont {{Crossley}}}, \bibinfo {author}
  {\bibfnamefont {H.~R.}\ \bibnamefont {{Lee}}}, \bibinfo {author}
  {\bibfnamefont {Y.}~\bibnamefont {{Cui}}}, \bibinfo {author} {\bibfnamefont
  {Y.}~\bibnamefont {{Hikita}}},\ and\ \bibinfo {author} {\bibfnamefont
  {H.~Y.}\ \bibnamefont {{Hwang}}},\ }\bibfield  {title} {\bibinfo {title}
  {{Superconductivity in an infinite-layer nickelate}},\ }\href
  {https://doi.org/10.1038/s41586-019-1496-5} {\bibfield  {journal} {\bibinfo
  {journal} {Nature}\ }\textbf {\bibinfo {volume} {572}},\ \bibinfo {pages}
  {624} (\bibinfo {year} {2019})}\BibitemShut {NoStop}%
\bibitem [{\citenamefont {Kitatani}\ \emph {et~al.}(2020)\citenamefont
  {Kitatani}, \citenamefont {Si}, \citenamefont {Janson}, \citenamefont
  {Arita}, \citenamefont {Zhong},\ and\ \citenamefont {Held}}]{Kitatani2020}%
  \BibitemOpen
  \bibfield  {author} {\bibinfo {author} {\bibfnamefont {M.}~\bibnamefont
  {Kitatani}}, \bibinfo {author} {\bibfnamefont {L.}~\bibnamefont {Si}},
  \bibinfo {author} {\bibfnamefont {O.}~\bibnamefont {Janson}}, \bibinfo
  {author} {\bibfnamefont {R.}~\bibnamefont {Arita}}, \bibinfo {author}
  {\bibfnamefont {Z.}~\bibnamefont {Zhong}},\ and\ \bibinfo {author}
  {\bibfnamefont {K.}~\bibnamefont {Held}},\ }\bibfield  {title} {\bibinfo
  {title} {{Nickelate superconductors{\textemdash}a renaissance of the one-band
  Hubbard model}},\ }\bibfield  {journal} {\bibinfo  {journal} {npj Quantum
  Materials}\ }\textbf {\bibinfo {volume} {5}},\ \href
  {https://doi.org/10.1038/s41535-020-00260-y} {10.1038/s41535-020-00260-y}
  (\bibinfo {year} {2020})\BibitemShut {NoStop}%
\bibitem [{\citenamefont {Klett}\ \emph {et~al.}(2022)\citenamefont {Klett},
  \citenamefont {Hansmann},\ and\ \citenamefont {Schäfer}}]{Klett2022}%
  \BibitemOpen
  \bibfield  {author} {\bibinfo {author} {\bibfnamefont {M.}~\bibnamefont
  {Klett}}, \bibinfo {author} {\bibfnamefont {P.}~\bibnamefont {Hansmann}},\
  and\ \bibinfo {author} {\bibfnamefont {T.}~\bibnamefont {Schäfer}},\
  }\bibfield  {title} {\bibinfo {title} {Magnetic properties and pseudogap
  formation in infinite-layer nickelates: Insights from the single-band hubbard
  model},\ }\href {https://doi.org/10.3389/fphy.2022.834682} {\bibfield
  {journal} {\bibinfo  {journal} {Frontiers in Physics}\ }\textbf {\bibinfo
  {volume} {10}},\ \bibinfo {pages} {834682} (\bibinfo {year}
  {2022})}\BibitemShut {NoStop}%
\bibitem [{\citenamefont {Qin}\ \emph {et~al.}(2022)\citenamefont {Qin},
  \citenamefont {Schäfer}, \citenamefont {Andergassen}, \citenamefont
  {Corboz},\ and\ \citenamefont {Gull}}]{Qin2022}%
  \BibitemOpen
  \bibfield  {author} {\bibinfo {author} {\bibfnamefont {M.}~\bibnamefont
  {Qin}}, \bibinfo {author} {\bibfnamefont {T.}~\bibnamefont {Schäfer}},
  \bibinfo {author} {\bibfnamefont {S.}~\bibnamefont {Andergassen}}, \bibinfo
  {author} {\bibfnamefont {P.}~\bibnamefont {Corboz}},\ and\ \bibinfo {author}
  {\bibfnamefont {E.}~\bibnamefont {Gull}},\ }\bibfield  {title} {\bibinfo
  {title} {{The Hubbard Model: A Computational Perspective}},\ }\href
  {https://doi.org/10.1146/annurev-conmatphys-090921-033948} {\bibfield
  {journal} {\bibinfo  {journal} {{Annual Review of Condensed Matter Physics}}\
  }\textbf {\bibinfo {volume} {13}},\ \bibinfo {pages} {275} (\bibinfo {year}
  {2022})},\ \Eprint
  {https://arxiv.org/abs/https://doi.org/10.1146/annurev-conmatphys-090921-033948}
  {https://doi.org/10.1146/annurev-conmatphys-090921-033948} \BibitemShut
  {NoStop}%
\bibitem [{\citenamefont {Arovas}\ \emph {et~al.}(2022)\citenamefont {Arovas},
  \citenamefont {Berg}, \citenamefont {Kivelson},\ and\ \citenamefont
  {Raghu}}]{Arovas2022}%
  \BibitemOpen
  \bibfield  {author} {\bibinfo {author} {\bibfnamefont {D.~P.}\ \bibnamefont
  {Arovas}}, \bibinfo {author} {\bibfnamefont {E.}~\bibnamefont {Berg}},
  \bibinfo {author} {\bibfnamefont {S.~A.}\ \bibnamefont {Kivelson}},\ and\
  \bibinfo {author} {\bibfnamefont {S.}~\bibnamefont {Raghu}},\ }\bibfield
  {title} {\bibinfo {title} {{The Hubbard Model}},\ }\bibfield  {journal}
  {\bibinfo  {journal} {{Annual Review of Condensed Matter Physics}}\ }\textbf
  {\bibinfo {volume} {13}},\ \href
  {https://doi.org/10.1146/annurev-conmatphys-031620-102024}
  {10.1146/annurev-conmatphys-031620-102024} (\bibinfo {year} {2022}),\ \Eprint
  {https://arxiv.org/abs/https://doi.org/10.1146/annurev-conmatphys-031620-102024}
  {https://doi.org/10.1146/annurev-conmatphys-031620-102024} \BibitemShut
  {NoStop}%
\bibitem [{\citenamefont {Alloul}\ \emph {et~al.}(1989)\citenamefont {Alloul},
  \citenamefont {Ohno},\ and\ \citenamefont {Mendels}}]{Alloul1989}%
  \BibitemOpen
  \bibfield  {author} {\bibinfo {author} {\bibfnamefont {H.}~\bibnamefont
  {Alloul}}, \bibinfo {author} {\bibfnamefont {T.}~\bibnamefont {Ohno}},\ and\
  \bibinfo {author} {\bibfnamefont {P.}~\bibnamefont {Mendels}},\ }\bibfield
  {title} {\bibinfo {title} {{$^{89}\mathrm{Y}$ NMR evidence for a fermi-liquid
  behavior in
  ${\mathrm{YBa}}_{2}$${\mathrm{Cu}}_{3}$${\mathrm{O}}_{6+\mathrm{x}}$}},\
  }\href {https://doi.org/10.1103/PhysRevLett.63.1700} {\bibfield  {journal}
  {\bibinfo  {journal} {Phys. Rev. Lett.}\ }\textbf {\bibinfo {volume} {63}},\
  \bibinfo {pages} {1700} (\bibinfo {year} {1989})}\BibitemShut {NoStop}%
\bibitem [{\citenamefont {Damascelli}\ \emph {et~al.}(2003)\citenamefont
  {Damascelli}, \citenamefont {Hussain},\ and\ \citenamefont
  {Shen}}]{Damascelli2003}%
  \BibitemOpen
  \bibfield  {author} {\bibinfo {author} {\bibfnamefont {A.}~\bibnamefont
  {Damascelli}}, \bibinfo {author} {\bibfnamefont {Z.}~\bibnamefont
  {Hussain}},\ and\ \bibinfo {author} {\bibfnamefont {Z.-X.}\ \bibnamefont
  {Shen}},\ }\bibfield  {title} {\bibinfo {title} {Angle-resolved photoemission
  studies of the cuprate superconductors},\ }\href
  {https://doi.org/10.1103/RevModPhys.75.473} {\bibfield  {journal} {\bibinfo
  {journal} {Rev. Mod. Phys.}\ }\textbf {\bibinfo {volume} {75}},\ \bibinfo
  {pages} {473} (\bibinfo {year} {2003})}\BibitemShut {NoStop}%
\bibitem [{\citenamefont {{Kanigel}}\ \emph {et~al.}(2006)\citenamefont
  {{Kanigel}}, \citenamefont {{Norman}}, \citenamefont {{Randeria}},
  \citenamefont {{Chatterjee}}, \citenamefont {{Souma}}, \citenamefont
  {{Kaminski}}, \citenamefont {{Fretwell}}, \citenamefont {{Rosenkranz}},
  \citenamefont {{Shi}}, \citenamefont {{Sato}}, \citenamefont {{Takahashi}},
  \citenamefont {{Li}}, \citenamefont {{Raffy}}, \citenamefont {{Kadowaki}},
  \citenamefont {{Hinks}}, \citenamefont {{Ozyuzer}},\ and\ \citenamefont
  {{Campuzano}}}]{Kanigel2006}%
  \BibitemOpen
  \bibfield  {author} {\bibinfo {author} {\bibfnamefont {A.}~\bibnamefont
  {{Kanigel}}}, \bibinfo {author} {\bibfnamefont {M.~R.}\ \bibnamefont
  {{Norman}}}, \bibinfo {author} {\bibfnamefont {M.}~\bibnamefont
  {{Randeria}}}, \bibinfo {author} {\bibfnamefont {U.}~\bibnamefont
  {{Chatterjee}}}, \bibinfo {author} {\bibfnamefont {S.}~\bibnamefont
  {{Souma}}}, \bibinfo {author} {\bibfnamefont {A.}~\bibnamefont {{Kaminski}}},
  \bibinfo {author} {\bibfnamefont {H.~M.}\ \bibnamefont {{Fretwell}}},
  \bibinfo {author} {\bibfnamefont {S.}~\bibnamefont {{Rosenkranz}}}, \bibinfo
  {author} {\bibfnamefont {M.}~\bibnamefont {{Shi}}}, \bibinfo {author}
  {\bibfnamefont {T.}~\bibnamefont {{Sato}}}, \bibinfo {author} {\bibfnamefont
  {T.}~\bibnamefont {{Takahashi}}}, \bibinfo {author} {\bibfnamefont {Z.~Z.}\
  \bibnamefont {{Li}}}, \bibinfo {author} {\bibfnamefont {H.}~\bibnamefont
  {{Raffy}}}, \bibinfo {author} {\bibfnamefont {K.}~\bibnamefont {{Kadowaki}}},
  \bibinfo {author} {\bibfnamefont {D.}~\bibnamefont {{Hinks}}}, \bibinfo
  {author} {\bibfnamefont {L.}~\bibnamefont {{Ozyuzer}}},\ and\ \bibinfo
  {author} {\bibfnamefont {J.~C.}\ \bibnamefont {{Campuzano}}},\ }\bibfield
  {title} {\bibinfo {title} {{Evolution of the pseudogap from Fermi arcs to the
  nodal liquid}},\ }\href {https://doi.org/10.1038/nphys334} {\bibfield
  {journal} {\bibinfo  {journal} {Nature Physics}\ }\textbf {\bibinfo {volume}
  {2}},\ \bibinfo {pages} {447} (\bibinfo {year} {2006})}\BibitemShut {NoStop}%
\bibitem [{\citenamefont {Yoshida}\ \emph {et~al.}(2006)\citenamefont
  {Yoshida}, \citenamefont {Zhou}, \citenamefont {Tanaka}, \citenamefont
  {Yang}, \citenamefont {Hussain}, \citenamefont {Shen}, \citenamefont
  {Fujimori}, \citenamefont {Sahrakorpi}, \citenamefont {Lindroos},
  \citenamefont {Markiewicz}, \citenamefont {Bansil}, \citenamefont {Komiya},
  \citenamefont {Ando}, \citenamefont {Eisaki}, \citenamefont {Kakeshita},\
  and\ \citenamefont {Uchida}}]{Yoshida2006}%
  \BibitemOpen
  \bibfield  {author} {\bibinfo {author} {\bibfnamefont {T.}~\bibnamefont
  {Yoshida}}, \bibinfo {author} {\bibfnamefont {X.~J.}\ \bibnamefont {Zhou}},
  \bibinfo {author} {\bibfnamefont {K.}~\bibnamefont {Tanaka}}, \bibinfo
  {author} {\bibfnamefont {W.~L.}\ \bibnamefont {Yang}}, \bibinfo {author}
  {\bibfnamefont {Z.}~\bibnamefont {Hussain}}, \bibinfo {author} {\bibfnamefont
  {Z.-X.}\ \bibnamefont {Shen}}, \bibinfo {author} {\bibfnamefont
  {A.}~\bibnamefont {Fujimori}}, \bibinfo {author} {\bibfnamefont
  {S.}~\bibnamefont {Sahrakorpi}}, \bibinfo {author} {\bibfnamefont
  {M.}~\bibnamefont {Lindroos}}, \bibinfo {author} {\bibfnamefont {R.~S.}\
  \bibnamefont {Markiewicz}}, \bibinfo {author} {\bibfnamefont
  {A.}~\bibnamefont {Bansil}}, \bibinfo {author} {\bibfnamefont
  {S.}~\bibnamefont {Komiya}}, \bibinfo {author} {\bibfnamefont
  {Y.}~\bibnamefont {Ando}}, \bibinfo {author} {\bibfnamefont {H.}~\bibnamefont
  {Eisaki}}, \bibinfo {author} {\bibfnamefont {T.}~\bibnamefont {Kakeshita}},\
  and\ \bibinfo {author} {\bibfnamefont {S.}~\bibnamefont {Uchida}},\
  }\bibfield  {title} {\bibinfo {title} {{Systematic doping evolution of the
  underlying Fermi surface of
  ${\mathrm{La}}_{2\ensuremath{-}x}{\mathrm{Sr}}_{x}\mathrm{Cu}{\mathrm{O}}_{4}$}},\
  }\href {https://doi.org/10.1103/PhysRevB.74.224510} {\bibfield  {journal}
  {\bibinfo  {journal} {Phys. Rev. B}\ }\textbf {\bibinfo {volume} {74}},\
  \bibinfo {pages} {224510} (\bibinfo {year} {2006})}\BibitemShut {NoStop}%
\bibitem [{\citenamefont {Shen}\ \emph {et~al.}(2005)\citenamefont {Shen},
  \citenamefont {Ronning}, \citenamefont {Lu}, \citenamefont {Baumberger},
  \citenamefont {Ingle}, \citenamefont {Lee}, \citenamefont {Meevasana},
  \citenamefont {Kohsaka}, \citenamefont {Azuma}, \citenamefont {Takano},
  \citenamefont {Takagi},\ and\ \citenamefont {Shen}}]{Shen2005}%
  \BibitemOpen
  \bibfield  {author} {\bibinfo {author} {\bibfnamefont {K.~M.}\ \bibnamefont
  {Shen}}, \bibinfo {author} {\bibfnamefont {F.}~\bibnamefont {Ronning}},
  \bibinfo {author} {\bibfnamefont {D.~H.}\ \bibnamefont {Lu}}, \bibinfo
  {author} {\bibfnamefont {F.}~\bibnamefont {Baumberger}}, \bibinfo {author}
  {\bibfnamefont {N.~J.~C.}\ \bibnamefont {Ingle}}, \bibinfo {author}
  {\bibfnamefont {W.~S.}\ \bibnamefont {Lee}}, \bibinfo {author} {\bibfnamefont
  {W.}~\bibnamefont {Meevasana}}, \bibinfo {author} {\bibfnamefont
  {Y.}~\bibnamefont {Kohsaka}}, \bibinfo {author} {\bibfnamefont
  {M.}~\bibnamefont {Azuma}}, \bibinfo {author} {\bibfnamefont
  {M.}~\bibnamefont {Takano}}, \bibinfo {author} {\bibfnamefont
  {H.}~\bibnamefont {Takagi}},\ and\ \bibinfo {author} {\bibfnamefont {Z.-X.}\
  \bibnamefont {Shen}},\ }\bibfield  {title} {\bibinfo {title} {{Nodal
  Quasiparticles and Antinodal Charge Ordering in
  Ca$_{2-x}$Na$_x$CuO$_2$Cl$_2$}},\ }\href
  {https://doi.org/10.1126/science.1103627} {\bibfield  {journal} {\bibinfo
  {journal} {Science}\ }\textbf {\bibinfo {volume} {307}},\ \bibinfo {pages}
  {901} (\bibinfo {year} {2005})}\BibitemShut {NoStop}%
\bibitem [{\citenamefont {Taillefer}(2010)}]{Taillefer2010}%
  \BibitemOpen
  \bibfield  {author} {\bibinfo {author} {\bibfnamefont {L.}~\bibnamefont
  {Taillefer}},\ }\bibfield  {title} {\bibinfo {title} {{Scattering and Pairing
  in Cuprate Superconductors}},\ }\href
  {https://doi.org/https://doi.org/10.1146/annurev-conmatphys-070909-104117}
  {\bibfield  {journal} {\bibinfo  {journal} {Annual Review of Condensed Matter
  Physics}\ }\textbf {\bibinfo {volume} {1}},\ \bibinfo {pages} {51} (\bibinfo
  {year} {2010})}\BibitemShut {NoStop}%
\bibitem [{\citenamefont {{Boschini}}\ \emph {et~al.}(2020)\citenamefont
  {{Boschini}}, \citenamefont {{Zonno}}, \citenamefont {{Razzoli}},
  \citenamefont {{Day}}, \citenamefont {{Michiardi}}, \citenamefont
  {{Zwartsenberg}}, \citenamefont {{Nigge}}, \citenamefont {{Schneider}},
  \citenamefont {{da Silva Neto}}, \citenamefont {{Erb}}, \citenamefont
  {{Zhdanovich}}, \citenamefont {{Mills}}, \citenamefont {{Levy}},
  \citenamefont {{Giannetti}}, \citenamefont {{Jones}},\ and\ \citenamefont
  {{Damascelli}}}]{Boschini2020}%
  \BibitemOpen
  \bibfield  {author} {\bibinfo {author} {\bibfnamefont {F.}~\bibnamefont
  {{Boschini}}}, \bibinfo {author} {\bibfnamefont {M.}~\bibnamefont {{Zonno}}},
  \bibinfo {author} {\bibfnamefont {E.}~\bibnamefont {{Razzoli}}}, \bibinfo
  {author} {\bibfnamefont {R.~P.}\ \bibnamefont {{Day}}}, \bibinfo {author}
  {\bibfnamefont {M.}~\bibnamefont {{Michiardi}}}, \bibinfo {author}
  {\bibfnamefont {B.}~\bibnamefont {{Zwartsenberg}}}, \bibinfo {author}
  {\bibfnamefont {P.}~\bibnamefont {{Nigge}}}, \bibinfo {author} {\bibfnamefont
  {M.}~\bibnamefont {{Schneider}}}, \bibinfo {author} {\bibfnamefont {E.~H.}\
  \bibnamefont {{da Silva Neto}}}, \bibinfo {author} {\bibfnamefont
  {A.}~\bibnamefont {{Erb}}}, \bibinfo {author} {\bibfnamefont
  {S.}~\bibnamefont {{Zhdanovich}}}, \bibinfo {author} {\bibfnamefont {A.~K.}\
  \bibnamefont {{Mills}}}, \bibinfo {author} {\bibfnamefont {G.}~\bibnamefont
  {{Levy}}}, \bibinfo {author} {\bibfnamefont {C.}~\bibnamefont {{Giannetti}}},
  \bibinfo {author} {\bibfnamefont {D.~J.}\ \bibnamefont {{Jones}}},\ and\
  \bibinfo {author} {\bibfnamefont {A.}~\bibnamefont {{Damascelli}}},\
  }\bibfield  {title} {\bibinfo {title} {{Emergence of pseudogap from
  short-range spin-correlations in electron-doped cuprates}},\ }\href
  {https://doi.org/10.1038/s41535-020-0208-6} {\bibfield  {journal} {\bibinfo
  {journal} {npj Quantum Materials}\ }\textbf {\bibinfo {volume} {5}},\
  \bibinfo {eid} {6} (\bibinfo {year} {2020})}\BibitemShut {NoStop}%
\bibitem [{\citenamefont {Boschini}\ \emph {et~al.}(2024)\citenamefont
  {Boschini}, \citenamefont {Zonno},\ and\ \citenamefont
  {Damascelli}}]{Boschini2024}%
  \BibitemOpen
  \bibfield  {author} {\bibinfo {author} {\bibfnamefont {F.}~\bibnamefont
  {Boschini}}, \bibinfo {author} {\bibfnamefont {M.}~\bibnamefont {Zonno}},\
  and\ \bibinfo {author} {\bibfnamefont {A.}~\bibnamefont {Damascelli}},\
  }\bibfield  {title} {\bibinfo {title} {{Time-resolved ARPES studies of
  quantum materials}},\ }\href {https://doi.org/10.1103/RevModPhys.96.015003}
  {\bibfield  {journal} {\bibinfo  {journal} {Rev. Mod. Phys.}\ }\textbf
  {\bibinfo {volume} {96}},\ \bibinfo {pages} {015003} (\bibinfo {year}
  {2024})}\BibitemShut {NoStop}%
\bibitem [{\citenamefont {Kyung}\ \emph {et~al.}(2004)\citenamefont {Kyung},
  \citenamefont {Hankevych}, \citenamefont {Dar\'e},\ and\ \citenamefont
  {Tremblay}}]{Kyung2004}%
  \BibitemOpen
  \bibfield  {author} {\bibinfo {author} {\bibfnamefont {B.}~\bibnamefont
  {Kyung}}, \bibinfo {author} {\bibfnamefont {V.}~\bibnamefont {Hankevych}},
  \bibinfo {author} {\bibfnamefont {A.-M.}\ \bibnamefont {Dar\'e}},\ and\
  \bibinfo {author} {\bibfnamefont {A.-M.~S.}\ \bibnamefont {Tremblay}},\
  }\bibfield  {title} {\bibinfo {title} {{Pseudogap and Spin Fluctuations in
  the Normal State of the Electron-Doped Cuprates}},\ }\href
  {https://doi.org/10.1103/PhysRevLett.93.147004} {\bibfield  {journal}
  {\bibinfo  {journal} {Phys. Rev. Lett.}\ }\textbf {\bibinfo {volume} {93}},\
  \bibinfo {pages} {147004} (\bibinfo {year} {2004})}\BibitemShut {NoStop}%
\bibitem [{\citenamefont {Sch\"afer}\ \emph {et~al.}(2021)\citenamefont
  {Sch\"afer}, \citenamefont {Wentzell}, \citenamefont {\ifmmode~\check{S}\else
  \v{S}\fi{}imkovic}, \citenamefont {He}, \citenamefont {Hille}, \citenamefont
  {Klett}, \citenamefont {Eckhardt}, \citenamefont {Arzhang}, \citenamefont
  {Harkov}, \citenamefont {Le~R\'egent}, \citenamefont {Kirsch}, \citenamefont
  {Wang}, \citenamefont {Kim}, \citenamefont {Kozik}, \citenamefont {Stepanov},
  \citenamefont {Kauch}, \citenamefont {Andergassen}, \citenamefont {Hansmann},
  \citenamefont {Rohe}, \citenamefont {Vilk}, \citenamefont {LeBlanc},
  \citenamefont {Zhang}, \citenamefont {Tremblay}, \citenamefont {Ferrero},
  \citenamefont {Parcollet},\ and\ \citenamefont {Georges}}]{Schaefer2021}%
  \BibitemOpen
  \bibfield  {author} {\bibinfo {author} {\bibfnamefont {T.}~\bibnamefont
  {Sch\"afer}}, \bibinfo {author} {\bibfnamefont {N.}~\bibnamefont {Wentzell}},
  \bibinfo {author} {\bibfnamefont {F.}~\bibnamefont {\ifmmode~\check{S}\else
  \v{S}\fi{}imkovic}}, \bibinfo {author} {\bibfnamefont {Y.-Y.}\ \bibnamefont
  {He}}, \bibinfo {author} {\bibfnamefont {C.}~\bibnamefont {Hille}}, \bibinfo
  {author} {\bibfnamefont {M.}~\bibnamefont {Klett}}, \bibinfo {author}
  {\bibfnamefont {C.~J.}\ \bibnamefont {Eckhardt}}, \bibinfo {author}
  {\bibfnamefont {B.}~\bibnamefont {Arzhang}}, \bibinfo {author} {\bibfnamefont
  {V.}~\bibnamefont {Harkov}}, \bibinfo {author} {\bibfnamefont
  {F.}~\bibnamefont {Le~R\'egent}}, \bibinfo {author} {\bibfnamefont
  {A.}~\bibnamefont {Kirsch}}, \bibinfo {author} {\bibfnamefont
  {Y.}~\bibnamefont {Wang}}, \bibinfo {author} {\bibfnamefont {A.~J.}\
  \bibnamefont {Kim}}, \bibinfo {author} {\bibfnamefont {E.}~\bibnamefont
  {Kozik}}, \bibinfo {author} {\bibfnamefont {E.~A.}\ \bibnamefont {Stepanov}},
  \bibinfo {author} {\bibfnamefont {A.}~\bibnamefont {Kauch}}, \bibinfo
  {author} {\bibfnamefont {S.}~\bibnamefont {Andergassen}}, \bibinfo {author}
  {\bibfnamefont {P.}~\bibnamefont {Hansmann}}, \bibinfo {author}
  {\bibfnamefont {D.}~\bibnamefont {Rohe}}, \bibinfo {author} {\bibfnamefont
  {Y.~M.}\ \bibnamefont {Vilk}}, \bibinfo {author} {\bibfnamefont {J.~P.~F.}\
  \bibnamefont {LeBlanc}}, \bibinfo {author} {\bibfnamefont {S.}~\bibnamefont
  {Zhang}}, \bibinfo {author} {\bibfnamefont {A.-M.~S.}\ \bibnamefont
  {Tremblay}}, \bibinfo {author} {\bibfnamefont {M.}~\bibnamefont {Ferrero}},
  \bibinfo {author} {\bibfnamefont {O.}~\bibnamefont {Parcollet}},\ and\
  \bibinfo {author} {\bibfnamefont {A.}~\bibnamefont {Georges}},\ }\bibfield
  {title} {\bibinfo {title} {{Tracking the Footprints of Spin Fluctuations: A
  MultiMethod, MultiMessenger Study of the Two-Dimensional Hubbard Model}},\
  }\href {https://doi.org/10.1103/PhysRevX.11.011058} {\bibfield  {journal}
  {\bibinfo  {journal} {Phys. Rev. X}\ }\textbf {\bibinfo {volume} {11}},\
  \bibinfo {pages} {011058} (\bibinfo {year} {2021})}\BibitemShut {NoStop}%
\bibitem [{\citenamefont {Krien}\ \emph {et~al.}(2022)\citenamefont {Krien},
  \citenamefont {Worm}, \citenamefont {Chalupa-Gantner}, \citenamefont
  {Toschi},\ and\ \citenamefont {Held}}]{Krien2022}%
  \BibitemOpen
  \bibfield  {author} {\bibinfo {author} {\bibfnamefont {F.}~\bibnamefont
  {Krien}}, \bibinfo {author} {\bibfnamefont {P.}~\bibnamefont {Worm}},
  \bibinfo {author} {\bibfnamefont {P.}~\bibnamefont {Chalupa-Gantner}},
  \bibinfo {author} {\bibfnamefont {A.}~\bibnamefont {Toschi}},\ and\ \bibinfo
  {author} {\bibfnamefont {K.}~\bibnamefont {Held}},\ }\bibfield  {title}
  {\bibinfo {title} {{Explaining the pseudogap through damping and antidamping
  on the Fermi surface by imaginary spin scattering}},\ }\href
  {https://doi.org/10.1038/s42005-022-01117-5} {\bibfield  {journal} {\bibinfo
  {journal} {Communications Physics}\ }\textbf {\bibinfo {volume} {5}},\
  \bibinfo {pages} {336} (\bibinfo {year} {2022})}\BibitemShut {NoStop}%
\bibitem [{\citenamefont {Vilk}\ \emph {et~al.}(2024)\citenamefont {Vilk},
  \citenamefont {Lahaie},\ and\ \citenamefont {Tremblay}}]{Vilk2024}%
  \BibitemOpen
  \bibfield  {author} {\bibinfo {author} {\bibfnamefont {Y.~M.}\ \bibnamefont
  {Vilk}}, \bibinfo {author} {\bibfnamefont {C.}~\bibnamefont {Lahaie}},\ and\
  \bibinfo {author} {\bibfnamefont {A.-M.~S.}\ \bibnamefont {Tremblay}},\
  }\bibfield  {title} {\bibinfo {title} {{Antiferromagnetic pseudogap in the
  two-dimensional Hubbard model deep in the renormalized classical regime}},\
  }\bibfield  {journal} {\bibinfo  {journal} {Physical Review B}\ }\textbf
  {\bibinfo {volume} {110}},\ \href
  {https://doi.org/10.1103/physrevb.110.125154} {10.1103/physrevb.110.125154}
  (\bibinfo {year} {2024})\BibitemShut {NoStop}%
\bibitem [{\citenamefont {\v{S}imkovic}\ \emph {et~al.}(2024)\citenamefont
  {\v{S}imkovic}, \citenamefont {Rossi}, \citenamefont {Georges},\ and\
  \citenamefont {Ferrero}}]{Simkovic2024}%
  \BibitemOpen
  \bibfield  {author} {\bibinfo {author} {\bibfnamefont {F.}~\bibnamefont
  {\v{S}imkovic}}, \bibinfo {author} {\bibfnamefont {R.}~\bibnamefont {Rossi}},
  \bibinfo {author} {\bibfnamefont {A.}~\bibnamefont {Georges}},\ and\ \bibinfo
  {author} {\bibfnamefont {M.}~\bibnamefont {Ferrero}},\ }\bibfield  {title}
  {\bibinfo {title} {{Origin and fate of the pseudogap in the doped Hubbard
  model}},\ }\href {https://doi.org/10.1126/science.ade9194} {\bibfield
  {journal} {\bibinfo  {journal} {Science}\ }\textbf {\bibinfo {volume}
  {385}},\ \bibinfo {pages} {eade9194} (\bibinfo {year} {2024})}\BibitemShut
  {NoStop}%
\bibitem [{\citenamefont {Avella}\ \emph {et~al.}(2013)\citenamefont {Avella},
  \citenamefont {Mancini}, \citenamefont {Mancini},\ and\ \citenamefont
  {Plekhanov}}]{Avella2013}%
  \BibitemOpen
  \bibfield  {author} {\bibinfo {author} {\bibfnamefont {A.}~\bibnamefont
  {Avella}}, \bibinfo {author} {\bibfnamefont {F.}~\bibnamefont {Mancini}},
  \bibinfo {author} {\bibfnamefont {F.~P.}\ \bibnamefont {Mancini}},\ and\
  \bibinfo {author} {\bibfnamefont {E.}~\bibnamefont {Plekhanov}},\ }\bibfield
  {title} {\bibinfo {title} {{Emery vs. Hubbard model for cuprate
  superconductors: a composite operator method study}},\ }\bibfield  {journal}
  {\bibinfo  {journal} {The European Physical Journal B}\ }\textbf {\bibinfo
  {volume} {86}},\ \href {https://doi.org/10.1140/epjb/e2013-40115-3}
  {10.1140/epjb/e2013-40115-3} (\bibinfo {year} {2013})\BibitemShut {NoStop}%
\bibitem [{\citenamefont {Fratino}\ \emph {et~al.}(2016)\citenamefont
  {Fratino}, \citenamefont {S\'emon}, \citenamefont {Sordi},\ and\
  \citenamefont {Tremblay}}]{Fratino2016b}%
  \BibitemOpen
  \bibfield  {author} {\bibinfo {author} {\bibfnamefont {L.}~\bibnamefont
  {Fratino}}, \bibinfo {author} {\bibfnamefont {P.}~\bibnamefont {S\'emon}},
  \bibinfo {author} {\bibfnamefont {G.}~\bibnamefont {Sordi}},\ and\ \bibinfo
  {author} {\bibfnamefont {A.-M.~S.}\ \bibnamefont {Tremblay}},\ }\bibfield
  {title} {\bibinfo {title} {Pseudogap and superconductivity in two-dimensional
  doped charge-transfer insulators},\ }\href
  {https://doi.org/10.1103/PhysRevB.93.245147} {\bibfield  {journal} {\bibinfo
  {journal} {Phys. Rev. B}\ }\textbf {\bibinfo {volume} {93}},\ \bibinfo
  {pages} {245147} (\bibinfo {year} {2016})}\BibitemShut {NoStop}%
\bibitem [{\citenamefont {Dash}\ and\ \citenamefont
  {Sénéchal}(2019)}]{Dash2019}%
  \BibitemOpen
  \bibfield  {author} {\bibinfo {author} {\bibfnamefont {S.~S.}\ \bibnamefont
  {Dash}}\ and\ \bibinfo {author} {\bibfnamefont {D.}~\bibnamefont
  {Sénéchal}},\ }\bibfield  {title} {\bibinfo {title} {{Pseudogap transition
  within the superconducting phase in the three-band Hubbard model}},\ }\href
  {http://dx.doi.org/10.1103/PhysRevB.100.214509} {\bibfield  {journal}
  {\bibinfo  {journal} {Physical Review B}\ }\textbf {\bibinfo {volume} {100}}
  (\bibinfo {year} {2019})}\BibitemShut {NoStop}%
\bibitem [{\citenamefont {Mao}\ and\ \citenamefont {Jiang}(2024)}]{Mao2024}%
  \BibitemOpen
  \bibfield  {author} {\bibinfo {author} {\bibfnamefont {T.}~\bibnamefont
  {Mao}}\ and\ \bibinfo {author} {\bibfnamefont {M.}~\bibnamefont {Jiang}},\
  }\href {https://arxiv.org/abs/2403.11218} {\bibinfo {title} {{Non-Fermi
  liquid behavior of scattering rate in three-orbital Emery model}}} (\bibinfo
  {year} {2024}),\ \Eprint {https://arxiv.org/abs/2403.11218} {arXiv:2403.11218
  [cond-mat.str-el]} \BibitemShut {NoStop}%
\bibitem [{\citenamefont {Gauvin-Ndiaye}\ \emph {et~al.}(2024)\citenamefont
  {Gauvin-Ndiaye}, \citenamefont {Leblanc}, \citenamefont {Marin},
  \citenamefont {Martin}, \citenamefont {Lessnich},\ and\ \citenamefont
  {Tremblay}}]{Gauvin2024}%
  \BibitemOpen
  \bibfield  {author} {\bibinfo {author} {\bibfnamefont {C.}~\bibnamefont
  {Gauvin-Ndiaye}}, \bibinfo {author} {\bibfnamefont {J.}~\bibnamefont
  {Leblanc}}, \bibinfo {author} {\bibfnamefont {S.}~\bibnamefont {Marin}},
  \bibinfo {author} {\bibfnamefont {N.}~\bibnamefont {Martin}}, \bibinfo
  {author} {\bibfnamefont {D.}~\bibnamefont {Lessnich}},\ and\ \bibinfo
  {author} {\bibfnamefont {A.-M.~S.}\ \bibnamefont {Tremblay}},\ }\bibfield
  {title} {\bibinfo {title} {{Two-particle self-consistent approach for
  multiorbital models: Application to the Emery model}},\ }\href
  {https://doi.org/10.1103/PhysRevB.109.165111} {\bibfield  {journal} {\bibinfo
   {journal} {Phys. Rev. B}\ }\textbf {\bibinfo {volume} {109}},\ \bibinfo
  {pages} {165111} (\bibinfo {year} {2024})}\BibitemShut {NoStop}%
\bibitem [{\citenamefont {Sordi}\ \emph {et~al.}(2024)\citenamefont {Sordi},
  \citenamefont {Reaney}, \citenamefont {Kowalski}, \citenamefont {Sémon},\
  and\ \citenamefont {Tremblay}}]{Sordi2024}%
  \BibitemOpen
  \bibfield  {author} {\bibinfo {author} {\bibfnamefont {G.}~\bibnamefont
  {Sordi}}, \bibinfo {author} {\bibfnamefont {G.~L.}\ \bibnamefont {Reaney}},
  \bibinfo {author} {\bibfnamefont {N.}~\bibnamefont {Kowalski}}, \bibinfo
  {author} {\bibfnamefont {P.}~\bibnamefont {Sémon}},\ and\ \bibinfo {author}
  {\bibfnamefont {A.~M.~S.}\ \bibnamefont {Tremblay}},\ }\href
  {https://arxiv.org/abs/2407.19545} {\bibinfo {title} {{Ambipolar doping of a
  charge-transfer insulator in the Emery model}}} (\bibinfo {year} {2024}),\
  \Eprint {https://arxiv.org/abs/2407.19545} {arXiv:2407.19545
  [cond-mat.str-el]} \BibitemShut {NoStop}%
\bibitem [{\citenamefont {Bacq-Labreuil}\ \emph {et~al.}(2024)\citenamefont
  {Bacq-Labreuil}, \citenamefont {Lacasse}, \citenamefont {Tremblay},
  \citenamefont {Sénéchal},\ and\ \citenamefont {Haule}}]{Bacq2024}%
  \BibitemOpen
  \bibfield  {author} {\bibinfo {author} {\bibfnamefont {B.}~\bibnamefont
  {Bacq-Labreuil}}, \bibinfo {author} {\bibfnamefont {B.}~\bibnamefont
  {Lacasse}}, \bibinfo {author} {\bibfnamefont {A.-M.~S.}\ \bibnamefont
  {Tremblay}}, \bibinfo {author} {\bibfnamefont {D.}~\bibnamefont
  {Sénéchal}},\ and\ \bibinfo {author} {\bibfnamefont {K.}~\bibnamefont
  {Haule}},\ }\href {https://arxiv.org/abs/2410.10019} {\bibinfo {title}
  {{Towards an ab initio theory of high-temperature superconductors: a study of
  multilayer cuprates}}} (\bibinfo {year} {2024}),\ \Eprint
  {https://arxiv.org/abs/2410.10019} {arXiv:2410.10019 [cond-mat.str-el]}
  \BibitemShut {NoStop}%
\bibitem [{\citenamefont {Kowalski}\ \emph {et~al.}(2021)\citenamefont
  {Kowalski}, \citenamefont {Dash}, \citenamefont {S{\'e}mon}, \citenamefont
  {S{\'e}n{\'e}chal},\ and\ \citenamefont {Tremblay}}]{Kowalski2021}%
  \BibitemOpen
  \bibfield  {author} {\bibinfo {author} {\bibfnamefont {N.}~\bibnamefont
  {Kowalski}}, \bibinfo {author} {\bibfnamefont {S.~S.}\ \bibnamefont {Dash}},
  \bibinfo {author} {\bibfnamefont {P.}~\bibnamefont {S{\'e}mon}}, \bibinfo
  {author} {\bibfnamefont {D.}~\bibnamefont {S{\'e}n{\'e}chal}},\ and\ \bibinfo
  {author} {\bibfnamefont {A.-M.}\ \bibnamefont {Tremblay}},\ }\bibfield
  {title} {\bibinfo {title} {Oxygen hole content, charge-transfer gap,
  covalency, and cuprate superconductivity},\ }\href
  {https://doi.org/10.1073/pnas.2106476118} {\bibfield  {journal} {\bibinfo
  {journal} {Proceedings of the National Academy of Sciences}\ }\textbf
  {\bibinfo {volume} {118}},\ \bibinfo {pages} {e2106476118} (\bibinfo {year}
  {2021})}\BibitemShut {NoStop}%
\bibitem [{\citenamefont {{Mai}}\ \emph {et~al.}(2021)\citenamefont {{Mai}},
  \citenamefont {{Balduzzi}}, \citenamefont {{Johnston}},\ and\ \citenamefont
  {{Maier}}}]{Mai2021}%
  \BibitemOpen
  \bibfield  {author} {\bibinfo {author} {\bibfnamefont {P.}~\bibnamefont
  {{Mai}}}, \bibinfo {author} {\bibfnamefont {G.}~\bibnamefont {{Balduzzi}}},
  \bibinfo {author} {\bibfnamefont {S.}~\bibnamefont {{Johnston}}},\ and\
  \bibinfo {author} {\bibfnamefont {T.~A.}\ \bibnamefont {{Maier}}},\
  }\bibfield  {title} {\bibinfo {title} {{Orbital structure of the effective
  pairing interaction in the high-temperature superconducting cuprates}},\
  }\href {https://doi.org/10.1038/s41535-021-00326-5} {\bibfield  {journal}
  {\bibinfo  {journal} {npj Quantum Materials}\ }\textbf {\bibinfo {volume}
  {6}},\ \bibinfo {eid} {26} (\bibinfo {year} {2021})}\BibitemShut {NoStop}%
\bibitem [{\citenamefont {Mai}\ \emph {et~al.}(2021)\citenamefont {Mai},
  \citenamefont {Balduzzi}, \citenamefont {Johnston},\ and\ \citenamefont
  {Maier}}]{Mai2021b}%
  \BibitemOpen
  \bibfield  {author} {\bibinfo {author} {\bibfnamefont {P.}~\bibnamefont
  {Mai}}, \bibinfo {author} {\bibfnamefont {G.}~\bibnamefont {Balduzzi}},
  \bibinfo {author} {\bibfnamefont {S.}~\bibnamefont {Johnston}},\ and\
  \bibinfo {author} {\bibfnamefont {T.~A.}\ \bibnamefont {Maier}},\ }\bibfield
  {title} {\bibinfo {title} {{Pairing correlations in the cuprates: A numerical
  study of the three-band Hubbard model}},\ }\href
  {https://doi.org/10.1103/PhysRevB.103.144514} {\bibfield  {journal} {\bibinfo
   {journal} {Phys. Rev. B}\ }\textbf {\bibinfo {volume} {103}},\ \bibinfo
  {pages} {144514} (\bibinfo {year} {2021})}\BibitemShut {NoStop}%
\bibitem [{Sup()}]{Supplemental}%
  \BibitemOpen
  \href@noop {} {\bibinfo  {journal} {See Supplemental Material [url] for
  details on the technique and additional data, which includes
  Refs.~\cite{Altland2008, Metzner1989, Georges1992, Georges1996, Rubtsov2005,
  Gull2011, TRIQS, Rohringer2013, Rohringer2012, DelRe2021b, Maier2005,
  Katanin2009, Schaefer2015, Rohringer2011, Schaefer2017, Schaefer2019,
  Kitatani2018, Ortiz2022, Klett2022, Held2022, Rohringer2016, Ayral2015,
  Ayral2016, MORIYA1991, Stobbe2022, Del_Re2019, Ding1996, Marshall1996}}\
  }\BibitemShut {NoStop}%
\bibitem [{\citenamefont {Galler}\ \emph {et~al.}(2017)\citenamefont {Galler},
  \citenamefont {Thunstr\"om}, \citenamefont {Gunacker}, \citenamefont
  {Tomczak},\ and\ \citenamefont {Held}}]{Galler2017}%
  \BibitemOpen
\bibfield  {journal} {  }\bibfield  {author} {\bibinfo {author} {\bibfnamefont
  {A.}~\bibnamefont {Galler}}, \bibinfo {author} {\bibfnamefont
  {P.}~\bibnamefont {Thunstr\"om}}, \bibinfo {author} {\bibfnamefont
  {P.}~\bibnamefont {Gunacker}}, \bibinfo {author} {\bibfnamefont {J.~M.}\
  \bibnamefont {Tomczak}},\ and\ \bibinfo {author} {\bibfnamefont
  {K.}~\bibnamefont {Held}},\ }\bibfield  {title} {\bibinfo {title} {Ab initio
  dynamical vertex approximation},\ }\href
  {https://doi.org/10.1103/PhysRevB.95.115107} {\bibfield  {journal} {\bibinfo
  {journal} {Phys. Rev. B}\ }\textbf {\bibinfo {volume} {95}},\ \bibinfo
  {pages} {115107} (\bibinfo {year} {2017})}\BibitemShut {NoStop}%
\bibitem [{\citenamefont {Weber}\ \emph {et~al.}(2012)\citenamefont {Weber},
  \citenamefont {Yee}, \citenamefont {Haule},\ and\ \citenamefont
  {Kotliar}}]{Weber2012}%
  \BibitemOpen
  \bibfield  {author} {\bibinfo {author} {\bibfnamefont {C.}~\bibnamefont
  {Weber}}, \bibinfo {author} {\bibfnamefont {C.}~\bibnamefont {Yee}}, \bibinfo
  {author} {\bibfnamefont {K.}~\bibnamefont {Haule}},\ and\ \bibinfo {author}
  {\bibfnamefont {G.}~\bibnamefont {Kotliar}},\ }\bibfield  {title} {\bibinfo
  {title} {Scaling of the transition temperature of hole-doped cuprate
  superconductors with the charge-transfer energy},\ }\href
  {https://doi.org/10.1209/0295-5075/100/37001} {\bibfield  {journal} {\bibinfo
   {journal} {Europhysics Letters}\ }\textbf {\bibinfo {volume} {100}},\
  \bibinfo {pages} {37001} (\bibinfo {year} {2012})}\BibitemShut {NoStop}%
\bibitem [{\citenamefont {Bocquet}\ \emph {et~al.}(1992)\citenamefont
  {Bocquet}, \citenamefont {Mizokawa}, \citenamefont {Saitoh}, \citenamefont
  {Namatame},\ and\ \citenamefont {Fujimori}}]{XAS_1}%
  \BibitemOpen
  \bibfield  {author} {\bibinfo {author} {\bibfnamefont {A.~E.}\ \bibnamefont
  {Bocquet}}, \bibinfo {author} {\bibfnamefont {T.}~\bibnamefont {Mizokawa}},
  \bibinfo {author} {\bibfnamefont {T.}~\bibnamefont {Saitoh}}, \bibinfo
  {author} {\bibfnamefont {H.}~\bibnamefont {Namatame}},\ and\ \bibinfo
  {author} {\bibfnamefont {A.}~\bibnamefont {Fujimori}},\ }\bibfield  {title}
  {\bibinfo {title} {Electronic structure of 3d-transition-metal compounds by
  analysis of the 2p core-level photoemission spectra},\ }\href
  {https://doi.org/10.1103/PhysRevB.46.3771} {\bibfield  {journal} {\bibinfo
  {journal} {Phys. Rev. B}\ }\textbf {\bibinfo {volume} {46}},\ \bibinfo
  {pages} {3771} (\bibinfo {year} {1992})}\BibitemShut {NoStop}%
\bibitem [{\citenamefont {Bocquet}\ \emph {et~al.}(1996)\citenamefont
  {Bocquet}, \citenamefont {Mizokawa}, \citenamefont {Morikawa}, \citenamefont
  {Fujimori}, \citenamefont {Barman}, \citenamefont {Maiti}, \citenamefont
  {Sarma}, \citenamefont {Tokura},\ and\ \citenamefont {Onoda}}]{XAS_2}%
  \BibitemOpen
  \bibfield  {author} {\bibinfo {author} {\bibfnamefont {A.~E.}\ \bibnamefont
  {Bocquet}}, \bibinfo {author} {\bibfnamefont {T.}~\bibnamefont {Mizokawa}},
  \bibinfo {author} {\bibfnamefont {K.}~\bibnamefont {Morikawa}}, \bibinfo
  {author} {\bibfnamefont {A.}~\bibnamefont {Fujimori}}, \bibinfo {author}
  {\bibfnamefont {S.~R.}\ \bibnamefont {Barman}}, \bibinfo {author}
  {\bibfnamefont {K.}~\bibnamefont {Maiti}}, \bibinfo {author} {\bibfnamefont
  {D.~D.}\ \bibnamefont {Sarma}}, \bibinfo {author} {\bibfnamefont
  {Y.}~\bibnamefont {Tokura}},\ and\ \bibinfo {author} {\bibfnamefont
  {M.}~\bibnamefont {Onoda}},\ }\bibfield  {title} {\bibinfo {title}
  {Electronic structure of early 3d-transition-metal oxides by analysis of the
  2p core-level photoemission spectra},\ }\href
  {https://doi.org/10.1103/PhysRevB.53.1161} {\bibfield  {journal} {\bibinfo
  {journal} {Phys. Rev. B}\ }\textbf {\bibinfo {volume} {53}},\ \bibinfo
  {pages} {1161} (\bibinfo {year} {1996})}\BibitemShut {NoStop}%
\bibitem [{\citenamefont {Toschi}\ \emph {et~al.}(2007)\citenamefont {Toschi},
  \citenamefont {Katanin},\ and\ \citenamefont {Held}}]{Toschi2007}%
  \BibitemOpen
  \bibfield  {author} {\bibinfo {author} {\bibfnamefont {A.}~\bibnamefont
  {Toschi}}, \bibinfo {author} {\bibfnamefont {A.~A.}\ \bibnamefont
  {Katanin}},\ and\ \bibinfo {author} {\bibfnamefont {K.}~\bibnamefont
  {Held}},\ }\bibfield  {title} {\bibinfo {title} {{Dynamical vertex
  approximation; A step beyond dynamical mean-field theory}},\ }\href
  {https://doi.org/10.1103/PhysRevB.75.045118} {\bibfield  {journal} {\bibinfo
  {journal} {Phys Rev. B}\ }\textbf {\bibinfo {volume} {75}},\ \bibinfo {pages}
  {045118} (\bibinfo {year} {2007})}\BibitemShut {NoStop}%
\bibitem [{\citenamefont {Rohringer}\ \emph {et~al.}(2018)\citenamefont
  {Rohringer}, \citenamefont {Hafermann}, \citenamefont {Toschi}, \citenamefont
  {Katanin}, \citenamefont {Antipov}, \citenamefont {Katsnelson}, \citenamefont
  {Lichtenstein}, \citenamefont {Rubtsov},\ and\ \citenamefont
  {Held}}]{Rohringer2018}%
  \BibitemOpen
  \bibfield  {author} {\bibinfo {author} {\bibfnamefont {G.}~\bibnamefont
  {Rohringer}}, \bibinfo {author} {\bibfnamefont {H.}~\bibnamefont
  {Hafermann}}, \bibinfo {author} {\bibfnamefont {A.}~\bibnamefont {Toschi}},
  \bibinfo {author} {\bibfnamefont {A.~A.}\ \bibnamefont {Katanin}}, \bibinfo
  {author} {\bibfnamefont {A.~E.}\ \bibnamefont {Antipov}}, \bibinfo {author}
  {\bibfnamefont {M.~I.}\ \bibnamefont {Katsnelson}}, \bibinfo {author}
  {\bibfnamefont {A.~I.}\ \bibnamefont {Lichtenstein}}, \bibinfo {author}
  {\bibfnamefont {A.~N.}\ \bibnamefont {Rubtsov}},\ and\ \bibinfo {author}
  {\bibfnamefont {K.}~\bibnamefont {Held}},\ }\bibfield  {title} {\bibinfo
  {title} {Diagrammatic routes to nonlocal correlations beyond dynamical mean
  field theory},\ }\href {https://doi.org/10.1103/RevModPhys.90.025003}
  {\bibfield  {journal} {\bibinfo  {journal} {Rev. Mod. Phys.}\ }\textbf
  {\bibinfo {volume} {90}},\ \bibinfo {pages} {025003} (\bibinfo {year}
  {2018})}\BibitemShut {NoStop}%
\bibitem [{\citenamefont {Georges}\ \emph {et~al.}(1996)\citenamefont
  {Georges}, \citenamefont {Kotliar}, \citenamefont {Krauth},\ and\
  \citenamefont {Rozenberg}}]{Georges1996}%
  \BibitemOpen
  \bibfield  {author} {\bibinfo {author} {\bibfnamefont {A.}~\bibnamefont
  {Georges}}, \bibinfo {author} {\bibfnamefont {G.}~\bibnamefont {Kotliar}},
  \bibinfo {author} {\bibfnamefont {W.}~\bibnamefont {Krauth}},\ and\ \bibinfo
  {author} {\bibfnamefont {M.~J.}\ \bibnamefont {Rozenberg}},\ }\bibfield
  {title} {\bibinfo {title} {Dynamical mean-field theory of strongly correlated
  fermion systems and the limit of infinite dimensions},\ }\href
  {https://doi.org/10.1103/RevModPhys.68.13} {\bibfield  {journal} {\bibinfo
  {journal} {Rev. Mod. Phys.}\ }\textbf {\bibinfo {volume} {68}},\ \bibinfo
  {pages} {13} (\bibinfo {year} {1996})}\BibitemShut {NoStop}%
\bibitem [{\citenamefont {Maier}\ \emph {et~al.}(2005)\citenamefont {Maier},
  \citenamefont {Jarrell}, \citenamefont {Pruschke},\ and\ \citenamefont
  {Hettler}}]{Maier2005}%
  \BibitemOpen
  \bibfield  {author} {\bibinfo {author} {\bibfnamefont {T.}~\bibnamefont
  {Maier}}, \bibinfo {author} {\bibfnamefont {M.}~\bibnamefont {Jarrell}},
  \bibinfo {author} {\bibfnamefont {T.}~\bibnamefont {Pruschke}},\ and\
  \bibinfo {author} {\bibfnamefont {M.~H.}\ \bibnamefont {Hettler}},\
  }\bibfield  {title} {\bibinfo {title} {Quantum cluster theories},\ }\href
  {https://doi.org/10.1103/RevModPhys.77.1027} {\bibfield  {journal} {\bibinfo
  {journal} {Rev. Mod. Phys.}\ }\textbf {\bibinfo {volume} {77}},\ \bibinfo
  {pages} {1027} (\bibinfo {year} {2005})}\BibitemShut {NoStop}%
\bibitem [{\citenamefont {Mai}\ \emph {et~al.}(2024)\citenamefont {Mai},
  \citenamefont {Cohen-Stead}, \citenamefont {Maier},\ and\ \citenamefont
  {Johnston}}]{Mai2024}%
  \BibitemOpen
  \bibfield  {author} {\bibinfo {author} {\bibfnamefont {P.}~\bibnamefont
  {Mai}}, \bibinfo {author} {\bibfnamefont {B.}~\bibnamefont {Cohen-Stead}},
  \bibinfo {author} {\bibfnamefont {T.~A.}\ \bibnamefont {Maier}},\ and\
  \bibinfo {author} {\bibfnamefont {S.}~\bibnamefont {Johnston}},\ }\href@noop
  {} {\bibinfo {title} {{Fluctuating charge-density-wave correlations in the
  three-band Hubbard model}}} (\bibinfo {year} {2024}),\ \Eprint
  {https://arxiv.org/abs/2405.13164} {arXiv:2405.13164 [cond-mat.str-el]}
  \BibitemShut {NoStop}%
\bibitem [{\citenamefont {Reymbaut}\ \emph {et~al.}(2017)\citenamefont
  {Reymbaut}, \citenamefont {Gagnon}, \citenamefont {Bergeron},\ and\
  \citenamefont {Tremblay}}]{Reymbaut2017}%
  \BibitemOpen
  \bibfield  {author} {\bibinfo {author} {\bibfnamefont {A.}~\bibnamefont
  {Reymbaut}}, \bibinfo {author} {\bibfnamefont {A.-M.}\ \bibnamefont
  {Gagnon}}, \bibinfo {author} {\bibfnamefont {D.}~\bibnamefont {Bergeron}},\
  and\ \bibinfo {author} {\bibfnamefont {A.-M.~S.}\ \bibnamefont {Tremblay}},\
  }\bibfield  {title} {\bibinfo {title} {Maximum entropy analytic continuation
  for frequency-dependent transport coefficients with nonpositive spectral
  weight},\ }\href {https://doi.org/10.1103/PhysRevB.95.121104} {\bibfield
  {journal} {\bibinfo  {journal} {Phys. Rev. B}\ }\textbf {\bibinfo {volume}
  {95}},\ \bibinfo {pages} {121104} (\bibinfo {year} {2017})}\BibitemShut
  {NoStop}%
\bibitem [{\citenamefont {Wu}\ \emph {et~al.}(2018)\citenamefont {Wu},
  \citenamefont {Scheurer}, \citenamefont {Chatterjee}, \citenamefont
  {Sachdev}, \citenamefont {Georges},\ and\ \citenamefont {Ferrero}}]{Wu2018}%
  \BibitemOpen
  \bibfield  {author} {\bibinfo {author} {\bibfnamefont {W.}~\bibnamefont
  {Wu}}, \bibinfo {author} {\bibfnamefont {M.~S.}\ \bibnamefont {Scheurer}},
  \bibinfo {author} {\bibfnamefont {S.}~\bibnamefont {Chatterjee}}, \bibinfo
  {author} {\bibfnamefont {S.}~\bibnamefont {Sachdev}}, \bibinfo {author}
  {\bibfnamefont {A.}~\bibnamefont {Georges}},\ and\ \bibinfo {author}
  {\bibfnamefont {M.}~\bibnamefont {Ferrero}},\ }\bibfield  {title} {\bibinfo
  {title} {{Pseudogap and Fermi-Surface Topology in the Two-Dimensional Hubbard
  Model}},\ }\href {https://doi.org/10.1103/PhysRevX.8.021048} {\bibfield
  {journal} {\bibinfo  {journal} {Phys. Rev. X}\ }\textbf {\bibinfo {volume}
  {8}},\ \bibinfo {pages} {021048} (\bibinfo {year} {2018})}\BibitemShut
  {NoStop}%
\bibitem [{\citenamefont {Meixner}\ \emph {et~al.}(2024)\citenamefont
  {Meixner}, \citenamefont {Menke}, \citenamefont {Klett}, \citenamefont
  {Heinzelmann}, \citenamefont {Andergassen}, \citenamefont {Hansmann},\ and\
  \citenamefont {Schäfer}}]{Meixner2024}%
  \BibitemOpen
  \bibfield  {author} {\bibinfo {author} {\bibfnamefont {M.}~\bibnamefont
  {Meixner}}, \bibinfo {author} {\bibfnamefont {H.}~\bibnamefont {Menke}},
  \bibinfo {author} {\bibfnamefont {M.}~\bibnamefont {Klett}}, \bibinfo
  {author} {\bibfnamefont {S.}~\bibnamefont {Heinzelmann}}, \bibinfo {author}
  {\bibfnamefont {S.}~\bibnamefont {Andergassen}}, \bibinfo {author}
  {\bibfnamefont {P.}~\bibnamefont {Hansmann}},\ and\ \bibinfo {author}
  {\bibfnamefont {T.}~\bibnamefont {Schäfer}},\ }\bibfield  {title} {\bibinfo
  {title} {{Mott transition and pseudogap of the square-lattice Hubbard model:
  Results from center-focused cellular dynamical mean-field theory}},\ }\href
  {https://doi.org/10.21468/SciPostPhys.16.2.059} {\bibfield  {journal}
  {\bibinfo  {journal} {SciPost Phys.}\ }\textbf {\bibinfo {volume} {16}},\
  \bibinfo {pages} {059} (\bibinfo {year} {2024})}\BibitemShut {NoStop}%
\bibitem [{\citenamefont {Kitatani}\ \emph {et~al.}(2023)\citenamefont
  {Kitatani}, \citenamefont {Si}, \citenamefont {Worm}, \citenamefont
  {Tomczak}, \citenamefont {Arita},\ and\ \citenamefont {Held}}]{Kitatani2023}%
  \BibitemOpen
  \bibfield  {author} {\bibinfo {author} {\bibfnamefont {M.}~\bibnamefont
  {Kitatani}}, \bibinfo {author} {\bibfnamefont {L.}~\bibnamefont {Si}},
  \bibinfo {author} {\bibfnamefont {P.}~\bibnamefont {Worm}}, \bibinfo {author}
  {\bibfnamefont {J.~M.}\ \bibnamefont {Tomczak}}, \bibinfo {author}
  {\bibfnamefont {R.}~\bibnamefont {Arita}},\ and\ \bibinfo {author}
  {\bibfnamefont {K.}~\bibnamefont {Held}},\ }\bibfield  {title} {\bibinfo
  {title} {Optimizing superconductivity: From cuprates via nickelates to
  palladates},\ }\href {https://doi.org/10.1103/PhysRevLett.130.166002}
  {\bibfield  {journal} {\bibinfo  {journal} {Phys. Rev. Lett.}\ }\textbf
  {\bibinfo {volume} {130}},\ \bibinfo {pages} {166002} (\bibinfo {year}
  {2023})}\BibitemShut {NoStop}%
\bibitem [{\citenamefont {Ye}\ \emph {et~al.}(2023)\citenamefont {Ye},
  \citenamefont {Wang}, \citenamefont {Fernandes},\ and\ \citenamefont
  {Chubukov}}]{Ye2023}%
  \BibitemOpen
  \bibfield  {author} {\bibinfo {author} {\bibfnamefont {M.}~\bibnamefont
  {Ye}}, \bibinfo {author} {\bibfnamefont {Z.}~\bibnamefont {Wang}}, \bibinfo
  {author} {\bibfnamefont {R.~M.}\ \bibnamefont {Fernandes}},\ and\ \bibinfo
  {author} {\bibfnamefont {A.~V.}\ \bibnamefont {Chubukov}},\ }\bibfield
  {title} {\bibinfo {title} {{Location and thermal evolution of the pseudogap
  due to spin fluctuations}},\ }\href
  {https://doi.org/10.1103/PhysRevB.108.115156} {\bibfield  {journal} {\bibinfo
   {journal} {Phys. Rev. B}\ }\textbf {\bibinfo {volume} {108}},\ \bibinfo
  {pages} {115156} (\bibinfo {year} {2023})}\BibitemShut {NoStop}%
\bibitem [{\citenamefont {Ye}\ and\ \citenamefont {Chubukov}(2023)}]{Ye2023b}%
  \BibitemOpen
  \bibfield  {author} {\bibinfo {author} {\bibfnamefont {M.}~\bibnamefont
  {Ye}}\ and\ \bibinfo {author} {\bibfnamefont {A.~V.}\ \bibnamefont
  {Chubukov}},\ }\bibfield  {title} {\bibinfo {title} {{Crucial role of thermal
  fluctuations and vertex corrections for the magnetic pseudogap}},\ }\href
  {https://doi.org/10.1103/PhysRevB.108.L081118} {\bibfield  {journal}
  {\bibinfo  {journal} {Phys. Rev. B}\ }\textbf {\bibinfo {volume} {108}},\
  \bibinfo {pages} {L081118} (\bibinfo {year} {2023})}\BibitemShut {NoStop}%
\bibitem [{\citenamefont {Gunnarsson}\ \emph {et~al.}(2015)\citenamefont
  {Gunnarsson}, \citenamefont {Sch\"afer}, \citenamefont {LeBlanc},
  \citenamefont {Gull}, \citenamefont {Merino}, \citenamefont {Sangiovanni},
  \citenamefont {Rohringer},\ and\ \citenamefont {Toschi}}]{Gunnarsson2015}%
  \BibitemOpen
  \bibfield  {author} {\bibinfo {author} {\bibfnamefont {O.}~\bibnamefont
  {Gunnarsson}}, \bibinfo {author} {\bibfnamefont {T.}~\bibnamefont
  {Sch\"afer}}, \bibinfo {author} {\bibfnamefont {J.~P.~F.}\ \bibnamefont
  {LeBlanc}}, \bibinfo {author} {\bibfnamefont {E.}~\bibnamefont {Gull}},
  \bibinfo {author} {\bibfnamefont {J.}~\bibnamefont {Merino}}, \bibinfo
  {author} {\bibfnamefont {G.}~\bibnamefont {Sangiovanni}}, \bibinfo {author}
  {\bibfnamefont {G.}~\bibnamefont {Rohringer}},\ and\ \bibinfo {author}
  {\bibfnamefont {A.}~\bibnamefont {Toschi}},\ }\bibfield  {title} {\bibinfo
  {title} {{Fluctuation Diagnostics of the Electron Self-Energy: Origin of the
  Pseudogap Physics}},\ }\href {https://doi.org/10.1103/PhysRevLett.114.236402}
  {\bibfield  {journal} {\bibinfo  {journal} {Phys. Rev. Lett.}\ }\textbf
  {\bibinfo {volume} {114}},\ \bibinfo {pages} {236402} (\bibinfo {year}
  {2015})}\BibitemShut {NoStop}%
\bibitem [{\citenamefont {Wu}\ \emph {et~al.}(2017)\citenamefont {Wu},
  \citenamefont {Ferrero}, \citenamefont {Georges},\ and\ \citenamefont
  {Kozik}}]{Wu2016}%
  \BibitemOpen
  \bibfield  {author} {\bibinfo {author} {\bibfnamefont {W.}~\bibnamefont
  {Wu}}, \bibinfo {author} {\bibfnamefont {M.}~\bibnamefont {Ferrero}},
  \bibinfo {author} {\bibfnamefont {A.}~\bibnamefont {Georges}},\ and\ \bibinfo
  {author} {\bibfnamefont {E.}~\bibnamefont {Kozik}},\ }\bibfield  {title}
  {\bibinfo {title} {{Controlling Feynman diagrammatic expansions: Physical
  nature of the pseudogap in the two-dimensional Hubbard model}},\ }\href
  {https://doi.org/10.1103/PhysRevB.96.041105} {\bibfield  {journal} {\bibinfo
  {journal} {Phys. Rev. B}\ }\textbf {\bibinfo {volume} {96}},\ \bibinfo
  {pages} {041105(R)} (\bibinfo {year} {2017})}\BibitemShut {NoStop}%
\bibitem [{\citenamefont {Schäfer}\ and\ \citenamefont
  {Toschi}(2021)}]{Schaefer2021b}%
  \BibitemOpen
  \bibfield  {author} {\bibinfo {author} {\bibfnamefont {T.}~\bibnamefont
  {Schäfer}}\ and\ \bibinfo {author} {\bibfnamefont {A.}~\bibnamefont
  {Toschi}},\ }\bibfield  {title} {\bibinfo {title} {How to read between the
  lines of electronic spectra: the diagnostics of fluctuations in strongly
  correlated electron systems},\ }\href
  {https://doi.org/10.1088/1361-648x/abeb44} {\bibfield  {journal} {\bibinfo
  {journal} {Journal of Physics: Condensed Matter}\ }\textbf {\bibinfo {volume}
  {33}},\ \bibinfo {pages} {214001} (\bibinfo {year} {2021})}\BibitemShut
  {NoStop}%
\bibitem [{\citenamefont {Mermin}\ and\ \citenamefont
  {Wagner}(1966)}]{Mermin1966}%
  \BibitemOpen
  \bibfield  {author} {\bibinfo {author} {\bibfnamefont {N.~D.}\ \bibnamefont
  {Mermin}}\ and\ \bibinfo {author} {\bibfnamefont {H.}~\bibnamefont
  {Wagner}},\ }\bibfield  {title} {\bibinfo {title} {{Absence of Ferromagnetism
  or Antiferromagnetism in One- or Two-Dimensional Isotropic Heisenberg
  Models}},\ }\href {https://doi.org/10.1103/PhysRevLett.17.1307} {\bibfield
  {journal} {\bibinfo  {journal} {Phys. Rev. Lett.}\ }\textbf {\bibinfo
  {volume} {17}},\ \bibinfo {pages} {1307} (\bibinfo {year}
  {1966})}\BibitemShut {NoStop}%
\bibitem [{\citenamefont {Alloul}(2024)}]{Alloul2024}%
  \BibitemOpen
  \bibfield  {author} {\bibinfo {author} {\bibfnamefont {H.}~\bibnamefont
  {Alloul}},\ }\bibfield  {title} {\bibinfo {title} {{What do we learn from
  impurities and disorder in high-$T_c$ cuprates?}},\ }\bibfield  {journal}
  {\bibinfo  {journal} {Frontiers in Physics}\ }\textbf {\bibinfo {volume}
  {12}},\ \href {https://doi.org/10.3389/fphy.2024.1406242}
  {10.3389/fphy.2024.1406242} (\bibinfo {year} {2024})\BibitemShut {NoStop}%
\bibitem [{\citenamefont {Yoshida}\ \emph {et~al.}(2009)\citenamefont
  {Yoshida}, \citenamefont {Hashimoto}, \citenamefont {Ideta}, \citenamefont
  {Fujimori}, \citenamefont {Tanaka}, \citenamefont {Mannella}, \citenamefont
  {Hussain}, \citenamefont {Shen}, \citenamefont {Kubota}, \citenamefont {Ono},
  \citenamefont {Komiya}, \citenamefont {Ando}, \citenamefont {Eisaki},\ and\
  \citenamefont {Uchida}}]{Yoshida2009}%
  \BibitemOpen
  \bibfield  {author} {\bibinfo {author} {\bibfnamefont {T.}~\bibnamefont
  {Yoshida}}, \bibinfo {author} {\bibfnamefont {M.}~\bibnamefont {Hashimoto}},
  \bibinfo {author} {\bibfnamefont {S.}~\bibnamefont {Ideta}}, \bibinfo
  {author} {\bibfnamefont {A.}~\bibnamefont {Fujimori}}, \bibinfo {author}
  {\bibfnamefont {K.}~\bibnamefont {Tanaka}}, \bibinfo {author} {\bibfnamefont
  {N.}~\bibnamefont {Mannella}}, \bibinfo {author} {\bibfnamefont
  {Z.}~\bibnamefont {Hussain}}, \bibinfo {author} {\bibfnamefont {Z.-X.}\
  \bibnamefont {Shen}}, \bibinfo {author} {\bibfnamefont {M.}~\bibnamefont
  {Kubota}}, \bibinfo {author} {\bibfnamefont {K.}~\bibnamefont {Ono}},
  \bibinfo {author} {\bibfnamefont {S.}~\bibnamefont {Komiya}}, \bibinfo
  {author} {\bibfnamefont {Y.}~\bibnamefont {Ando}}, \bibinfo {author}
  {\bibfnamefont {H.}~\bibnamefont {Eisaki}},\ and\ \bibinfo {author}
  {\bibfnamefont {S.}~\bibnamefont {Uchida}},\ }\bibfield  {title} {\bibinfo
  {title} {{Universal versus Material-Dependent Two-Gap Behaviors of the
  High-${T}_{c}$ Cuprate Superconductors: Angle-Resolved Photoemission Study of
  ${\mathrm{La}}_{2\ensuremath{-}x}{\mathrm{Sr}}_{x}{\mathrm{CuO}}_{4}$}},\
  }\href {https://doi.org/10.1103/PhysRevLett.103.037004} {\bibfield  {journal}
  {\bibinfo  {journal} {Phys. Rev. Lett.}\ }\textbf {\bibinfo {volume} {103}},\
  \bibinfo {pages} {037004} (\bibinfo {year} {2009})}\BibitemShut {NoStop}%
\bibitem [{\citenamefont {Cyr-Choini\`ere}\ \emph {et~al.}(2018)\citenamefont
  {Cyr-Choini\`ere}, \citenamefont {Daou}, \citenamefont {Lalibert\'e},
  \citenamefont {Collignon}, \citenamefont {Badoux}, \citenamefont {LeBoeuf},
  \citenamefont {Chang}, \citenamefont {Ramshaw}, \citenamefont {Bonn},
  \citenamefont {Hardy}, \citenamefont {Liang}, \citenamefont {Yan},
  \citenamefont {Cheng}, \citenamefont {Zhou}, \citenamefont {Goodenough},
  \citenamefont {Pyon}, \citenamefont {Takayama}, \citenamefont {Takagi},
  \citenamefont {Doiron-Leyraud},\ and\ \citenamefont {Taillefer}}]{Cyr2018}%
  \BibitemOpen
  \bibfield  {author} {\bibinfo {author} {\bibfnamefont {O.}~\bibnamefont
  {Cyr-Choini\`ere}}, \bibinfo {author} {\bibfnamefont {R.}~\bibnamefont
  {Daou}}, \bibinfo {author} {\bibfnamefont {F.}~\bibnamefont {Lalibert\'e}},
  \bibinfo {author} {\bibfnamefont {C.}~\bibnamefont {Collignon}}, \bibinfo
  {author} {\bibfnamefont {S.}~\bibnamefont {Badoux}}, \bibinfo {author}
  {\bibfnamefont {D.}~\bibnamefont {LeBoeuf}}, \bibinfo {author} {\bibfnamefont
  {J.}~\bibnamefont {Chang}}, \bibinfo {author} {\bibfnamefont {B.~J.}\
  \bibnamefont {Ramshaw}}, \bibinfo {author} {\bibfnamefont {D.~A.}\
  \bibnamefont {Bonn}}, \bibinfo {author} {\bibfnamefont {W.~N.}\ \bibnamefont
  {Hardy}}, \bibinfo {author} {\bibfnamefont {R.}~\bibnamefont {Liang}},
  \bibinfo {author} {\bibfnamefont {J.-Q.}\ \bibnamefont {Yan}}, \bibinfo
  {author} {\bibfnamefont {J.-G.}\ \bibnamefont {Cheng}}, \bibinfo {author}
  {\bibfnamefont {J.-S.}\ \bibnamefont {Zhou}}, \bibinfo {author}
  {\bibfnamefont {J.~B.}\ \bibnamefont {Goodenough}}, \bibinfo {author}
  {\bibfnamefont {S.}~\bibnamefont {Pyon}}, \bibinfo {author} {\bibfnamefont
  {T.}~\bibnamefont {Takayama}}, \bibinfo {author} {\bibfnamefont
  {H.}~\bibnamefont {Takagi}}, \bibinfo {author} {\bibfnamefont
  {N.}~\bibnamefont {Doiron-Leyraud}},\ and\ \bibinfo {author} {\bibfnamefont
  {L.}~\bibnamefont {Taillefer}},\ }\bibfield  {title} {\bibinfo {title}
  {{Pseudogap temperature ${T}^{*}$ of cuprate superconductors from the Nernst
  effect}},\ }\href {https://doi.org/10.1103/PhysRevB.97.064502} {\bibfield
  {journal} {\bibinfo  {journal} {Phys. Rev. B}\ }\textbf {\bibinfo {volume}
  {97}},\ \bibinfo {pages} {064502} (\bibinfo {year} {2018})}\BibitemShut
  {NoStop}%
\bibitem [{\citenamefont {Birgeneau}\ \emph {et~al.}(1989)\citenamefont
  {Birgeneau}, \citenamefont {Endoh}, \citenamefont {Kakurai}, \citenamefont
  {Hidaka}, \citenamefont {Murakami}, \citenamefont {Kastner}, \citenamefont
  {Thurston}, \citenamefont {Shirane},\ and\ \citenamefont
  {Yamada}}]{Birgeneau1989}%
  \BibitemOpen
  \bibfield  {author} {\bibinfo {author} {\bibfnamefont {R.~J.}\ \bibnamefont
  {Birgeneau}}, \bibinfo {author} {\bibfnamefont {Y.}~\bibnamefont {Endoh}},
  \bibinfo {author} {\bibfnamefont {K.}~\bibnamefont {Kakurai}}, \bibinfo
  {author} {\bibfnamefont {Y.}~\bibnamefont {Hidaka}}, \bibinfo {author}
  {\bibfnamefont {T.}~\bibnamefont {Murakami}}, \bibinfo {author}
  {\bibfnamefont {M.~A.}\ \bibnamefont {Kastner}}, \bibinfo {author}
  {\bibfnamefont {T.~R.}\ \bibnamefont {Thurston}}, \bibinfo {author}
  {\bibfnamefont {G.}~\bibnamefont {Shirane}},\ and\ \bibinfo {author}
  {\bibfnamefont {K.}~\bibnamefont {Yamada}},\ }\bibfield  {title} {\bibinfo
  {title} {{Static and dynamic spin fluctuations in superconducting
  ${\mathrm{La}}_{2\ensuremath{-}x}{\mathrm{Sr}}_{x}\mathrm{Cu}{\mathrm{O}}_{4}$}},\
  }\href {https://doi.org/10.1103/PhysRevB.39.2868} {\bibfield  {journal}
  {\bibinfo  {journal} {Phys. Rev. B}\ }\textbf {\bibinfo {volume} {39}},\
  \bibinfo {pages} {2868} (\bibinfo {year} {1989})}\BibitemShut {NoStop}%
\bibitem [{\citenamefont {Yamada}\ \emph {et~al.}(1998)\citenamefont {Yamada},
  \citenamefont {Lee}, \citenamefont {Kurahashi}, \citenamefont {Wada},
  \citenamefont {Wakimoto}, \citenamefont {Ueki}, \citenamefont {Kimura},
  \citenamefont {Endoh}, \citenamefont {Hosoya}, \citenamefont {Shirane},
  \citenamefont {Birgeneau}, \citenamefont {Greven}, \citenamefont {Kastner},\
  and\ \citenamefont {Kim}}]{Yamada1998}%
  \BibitemOpen
  \bibfield  {author} {\bibinfo {author} {\bibfnamefont {K.}~\bibnamefont
  {Yamada}}, \bibinfo {author} {\bibfnamefont {C.~H.}\ \bibnamefont {Lee}},
  \bibinfo {author} {\bibfnamefont {K.}~\bibnamefont {Kurahashi}}, \bibinfo
  {author} {\bibfnamefont {J.}~\bibnamefont {Wada}}, \bibinfo {author}
  {\bibfnamefont {S.}~\bibnamefont {Wakimoto}}, \bibinfo {author}
  {\bibfnamefont {S.}~\bibnamefont {Ueki}}, \bibinfo {author} {\bibfnamefont
  {H.}~\bibnamefont {Kimura}}, \bibinfo {author} {\bibfnamefont
  {Y.}~\bibnamefont {Endoh}}, \bibinfo {author} {\bibfnamefont
  {S.}~\bibnamefont {Hosoya}}, \bibinfo {author} {\bibfnamefont
  {G.}~\bibnamefont {Shirane}}, \bibinfo {author} {\bibfnamefont {R.~J.}\
  \bibnamefont {Birgeneau}}, \bibinfo {author} {\bibfnamefont {M.}~\bibnamefont
  {Greven}}, \bibinfo {author} {\bibfnamefont {M.~A.}\ \bibnamefont
  {Kastner}},\ and\ \bibinfo {author} {\bibfnamefont {Y.~J.}\ \bibnamefont
  {Kim}},\ }\bibfield  {title} {\bibinfo {title} {{Doping dependence of the
  spatially modulated dynamical spin correlations and the
  superconducting-transition temperature in
  ${\mathrm{La}}_{2\mathrm{\ensuremath{-}}\mathit{x}}{\mathrm{Sr}}_{x}{\mathrm{CuO}}_{4}$}},\
  }\href {https://doi.org/10.1103/PhysRevB.57.6165} {\bibfield  {journal}
  {\bibinfo  {journal} {Phys. Rev. B}\ }\textbf {\bibinfo {volume} {57}},\
  \bibinfo {pages} {6165} (\bibinfo {year} {1998})}\BibitemShut {NoStop}%
\bibitem [{\citenamefont {Haug}\ \emph {et~al.}(2010)\citenamefont {Haug},
  \citenamefont {Hinkov}, \citenamefont {Sidis}, \citenamefont {Bourges},
  \citenamefont {Christensen}, \citenamefont {Ivanov}, \citenamefont {Keller},
  \citenamefont {Lin},\ and\ \citenamefont {Keimer}}]{Haug2010}%
  \BibitemOpen
  \bibfield  {author} {\bibinfo {author} {\bibfnamefont {D.}~\bibnamefont
  {Haug}}, \bibinfo {author} {\bibfnamefont {V.}~\bibnamefont {Hinkov}},
  \bibinfo {author} {\bibfnamefont {Y.}~\bibnamefont {Sidis}}, \bibinfo
  {author} {\bibfnamefont {P.}~\bibnamefont {Bourges}}, \bibinfo {author}
  {\bibfnamefont {N.~B.}\ \bibnamefont {Christensen}}, \bibinfo {author}
  {\bibfnamefont {A.}~\bibnamefont {Ivanov}}, \bibinfo {author} {\bibfnamefont
  {T.}~\bibnamefont {Keller}}, \bibinfo {author} {\bibfnamefont {C.~T.}\
  \bibnamefont {Lin}},\ and\ \bibinfo {author} {\bibfnamefont {B.}~\bibnamefont
  {Keimer}},\ }\bibfield  {title} {\bibinfo {title} {{Neutron scattering study
  of the magnetic phase diagram of underdoped YBa$_2$Cu$_3$O$_{6+x}$}},\ }\href
  {https://doi.org/10.1088/1367-2630/12/10/105006} {\bibfield  {journal}
  {\bibinfo  {journal} {New Journal of Physics}\ }\textbf {\bibinfo {volume}
  {12}},\ \bibinfo {pages} {105006} (\bibinfo {year} {2010})}\BibitemShut
  {NoStop}%
\bibitem [{\citenamefont {Anderson}\ \emph {et~al.}(2024)\citenamefont
  {Anderson}, \citenamefont {Tang}, \citenamefont {Nagarajan}, \citenamefont
  {Chan}, \citenamefont {Dorow}, \citenamefont {Yu}, \citenamefont {Abernathy},
  \citenamefont {Christianson}, \citenamefont {Mangin-Thro}, \citenamefont
  {Steffens}, \citenamefont {Sterling}, \citenamefont {Reznik}, \citenamefont
  {Bounoua}, \citenamefont {Sidis}, \citenamefont {Bourges},\ and\
  \citenamefont {Greven}}]{Anderson2024}%
  \BibitemOpen
  \bibfield  {author} {\bibinfo {author} {\bibfnamefont {Z.~W.}\ \bibnamefont
  {Anderson}}, \bibinfo {author} {\bibfnamefont {Y.}~\bibnamefont {Tang}},
  \bibinfo {author} {\bibfnamefont {V.}~\bibnamefont {Nagarajan}}, \bibinfo
  {author} {\bibfnamefont {M.~K.}\ \bibnamefont {Chan}}, \bibinfo {author}
  {\bibfnamefont {C.~J.}\ \bibnamefont {Dorow}}, \bibinfo {author}
  {\bibfnamefont {G.}~\bibnamefont {Yu}}, \bibinfo {author} {\bibfnamefont
  {D.~L.}\ \bibnamefont {Abernathy}}, \bibinfo {author} {\bibfnamefont {A.~D.}\
  \bibnamefont {Christianson}}, \bibinfo {author} {\bibfnamefont
  {L.}~\bibnamefont {Mangin-Thro}}, \bibinfo {author} {\bibfnamefont
  {P.}~\bibnamefont {Steffens}}, \bibinfo {author} {\bibfnamefont
  {T.}~\bibnamefont {Sterling}}, \bibinfo {author} {\bibfnamefont
  {D.}~\bibnamefont {Reznik}}, \bibinfo {author} {\bibfnamefont
  {D.}~\bibnamefont {Bounoua}}, \bibinfo {author} {\bibfnamefont
  {Y.}~\bibnamefont {Sidis}}, \bibinfo {author} {\bibfnamefont
  {P.}~\bibnamefont {Bourges}},\ and\ \bibinfo {author} {\bibfnamefont
  {M.}~\bibnamefont {Greven}},\ }\href {https://arxiv.org/abs/2412.03524}
  {\bibinfo {title} {Gapped commensurate antiferromagnetic response in a
  strongly underdoped model cuprate superconductor}} (\bibinfo {year} {2024}),\
  \Eprint {https://arxiv.org/abs/2412.03524} {arXiv:2412.03524
  [cond-mat.supr-con]} \BibitemShut {NoStop}%
\bibitem [{\citenamefont {Birgeneau}\ \emph {et~al.}(1988)\citenamefont
  {Birgeneau}, \citenamefont {Gabbe}, \citenamefont {Jenssen}, \citenamefont
  {Kastner}, \citenamefont {Picone}, \citenamefont {Thurston}, \citenamefont
  {Shirane}, \citenamefont {Endoh}, \citenamefont {Sato}, \citenamefont
  {Yamada}, \citenamefont {Hidaka}, \citenamefont {Oda}, \citenamefont
  {Enomoto}, \citenamefont {Suzuki},\ and\ \citenamefont
  {Murakami}}]{Birgeneau1988}%
  \BibitemOpen
  \bibfield  {author} {\bibinfo {author} {\bibfnamefont {R.~J.}\ \bibnamefont
  {Birgeneau}}, \bibinfo {author} {\bibfnamefont {D.~R.}\ \bibnamefont
  {Gabbe}}, \bibinfo {author} {\bibfnamefont {H.~P.}\ \bibnamefont {Jenssen}},
  \bibinfo {author} {\bibfnamefont {M.~A.}\ \bibnamefont {Kastner}}, \bibinfo
  {author} {\bibfnamefont {P.~J.}\ \bibnamefont {Picone}}, \bibinfo {author}
  {\bibfnamefont {T.~R.}\ \bibnamefont {Thurston}}, \bibinfo {author}
  {\bibfnamefont {G.}~\bibnamefont {Shirane}}, \bibinfo {author} {\bibfnamefont
  {Y.}~\bibnamefont {Endoh}}, \bibinfo {author} {\bibfnamefont
  {M.}~\bibnamefont {Sato}}, \bibinfo {author} {\bibfnamefont {K.}~\bibnamefont
  {Yamada}}, \bibinfo {author} {\bibfnamefont {Y.}~\bibnamefont {Hidaka}},
  \bibinfo {author} {\bibfnamefont {M.}~\bibnamefont {Oda}}, \bibinfo {author}
  {\bibfnamefont {Y.}~\bibnamefont {Enomoto}}, \bibinfo {author} {\bibfnamefont
  {M.}~\bibnamefont {Suzuki}},\ and\ \bibinfo {author} {\bibfnamefont
  {T.}~\bibnamefont {Murakami}},\ }\bibfield  {title} {\bibinfo {title}
  {{Antiferromagnetic spin correlations in insulating, metallic, and
  superconducting
  ${\mathrm{La}}_{2\ensuremath{-}x}{\mathrm{Sr}}_{x}\mathrm{Cu}{\mathrm{O}}_{4}$}},\
  }\href {https://doi.org/10.1103/PhysRevB.38.6614} {\bibfield  {journal}
  {\bibinfo  {journal} {Phys. Rev. B}\ }\textbf {\bibinfo {volume} {38}},\
  \bibinfo {pages} {6614} (\bibinfo {year} {1988})}\BibitemShut {NoStop}%
\bibitem [{\citenamefont {Altland}\ and\ \citenamefont
  {Simons}(2008)}]{Altland2008}%
  \BibitemOpen
  \bibfield  {author} {\bibinfo {author} {\bibfnamefont {A.}~\bibnamefont
  {Altland}}\ and\ \bibinfo {author} {\bibfnamefont {B.}~\bibnamefont
  {Simons}},\ }\href@noop {} {\emph {\bibinfo {title} {Condensed Matter Field
  Theory}}}\ (\bibinfo  {publisher} {Cambridge University Press},\ \bibinfo
  {year} {2008})\BibitemShut {NoStop}%
\bibitem [{\citenamefont {Metzner}\ and\ \citenamefont
  {Vollhardt}(1989)}]{Metzner1989}%
  \BibitemOpen
  \bibfield  {author} {\bibinfo {author} {\bibfnamefont {W.}~\bibnamefont
  {Metzner}}\ and\ \bibinfo {author} {\bibfnamefont {D.}~\bibnamefont
  {Vollhardt}},\ }\bibfield  {title} {\bibinfo {title} {{Correlated Lattice
  Fermions in $d=\ensuremath{\infty}$ Dimensions}},\ }\href
  {https://doi.org/10.1103/PhysRevLett.62.324} {\bibfield  {journal} {\bibinfo
  {journal} {Phys. Rev. Lett.}\ }\textbf {\bibinfo {volume} {62}},\ \bibinfo
  {pages} {324} (\bibinfo {year} {1989})}\BibitemShut {NoStop}%
\bibitem [{\citenamefont {Georges}\ and\ \citenamefont
  {Kotliar}(1992)}]{Georges1992}%
  \BibitemOpen
  \bibfield  {author} {\bibinfo {author} {\bibfnamefont {A.}~\bibnamefont
  {Georges}}\ and\ \bibinfo {author} {\bibfnamefont {G.}~\bibnamefont
  {Kotliar}},\ }\bibfield  {title} {\bibinfo {title} {Hubbard model in infinite
  dimensions},\ }\href {https://doi.org/10.1103/PhysRevB.45.6479} {\bibfield
  {journal} {\bibinfo  {journal} {Phys. Rev. B}\ }\textbf {\bibinfo {volume}
  {45}},\ \bibinfo {pages} {6479} (\bibinfo {year} {1992})}\BibitemShut
  {NoStop}%
\bibitem [{\citenamefont {Rubtsov}\ \emph {et~al.}(2005)\citenamefont
  {Rubtsov}, \citenamefont {Savkin},\ and\ \citenamefont
  {Lichtenstein}}]{Rubtsov2005}%
  \BibitemOpen
  \bibfield  {author} {\bibinfo {author} {\bibfnamefont {A.~N.}\ \bibnamefont
  {Rubtsov}}, \bibinfo {author} {\bibfnamefont {V.~V.}\ \bibnamefont
  {Savkin}},\ and\ \bibinfo {author} {\bibfnamefont {A.~I.}\ \bibnamefont
  {Lichtenstein}},\ }\bibfield  {title} {\bibinfo {title} {{Continuous-time
  quantum Monte Carlo method for fermions}},\ }\href
  {https://doi.org/10.1103/PhysRevB.72.035122} {\bibfield  {journal} {\bibinfo
  {journal} {Phys. Rev. B}\ }\textbf {\bibinfo {volume} {72}},\ \bibinfo
  {pages} {035122} (\bibinfo {year} {2005})}\BibitemShut {NoStop}%
\bibitem [{\citenamefont {Gull}\ \emph {et~al.}(2011)\citenamefont {Gull},
  \citenamefont {Millis}, \citenamefont {Lichtenstein}, \citenamefont
  {Rubtsov}, \citenamefont {Troyer},\ and\ \citenamefont {Werner}}]{Gull2011}%
  \BibitemOpen
  \bibfield  {author} {\bibinfo {author} {\bibfnamefont {E.}~\bibnamefont
  {Gull}}, \bibinfo {author} {\bibfnamefont {A.~J.}\ \bibnamefont {Millis}},
  \bibinfo {author} {\bibfnamefont {A.~I.}\ \bibnamefont {Lichtenstein}},
  \bibinfo {author} {\bibfnamefont {A.~N.}\ \bibnamefont {Rubtsov}}, \bibinfo
  {author} {\bibfnamefont {M.}~\bibnamefont {Troyer}},\ and\ \bibinfo {author}
  {\bibfnamefont {P.}~\bibnamefont {Werner}},\ }\bibfield  {title} {\bibinfo
  {title} {{Continuous-time Monte Carlo methods for quantum impurity models}},\
  }\href {https://doi.org/10.1103/RevModPhys.83.349} {\bibfield  {journal}
  {\bibinfo  {journal} {Rev. Mod. Phys.}\ }\textbf {\bibinfo {volume} {83}},\
  \bibinfo {pages} {349} (\bibinfo {year} {2011})}\BibitemShut {NoStop}%
\bibitem [{\citenamefont {Parcollet}\ \emph {et~al.}(2015)\citenamefont
  {Parcollet}, \citenamefont {Ferrero}, \citenamefont {Ayral}, \citenamefont
  {Hafermann}, \citenamefont {Krivenko}, \citenamefont {Messio},\ and\
  \citenamefont {Seth}}]{TRIQS}%
  \BibitemOpen
  \bibfield  {author} {\bibinfo {author} {\bibfnamefont {O.}~\bibnamefont
  {Parcollet}}, \bibinfo {author} {\bibfnamefont {M.}~\bibnamefont {Ferrero}},
  \bibinfo {author} {\bibfnamefont {T.}~\bibnamefont {Ayral}}, \bibinfo
  {author} {\bibfnamefont {H.}~\bibnamefont {Hafermann}}, \bibinfo {author}
  {\bibfnamefont {I.}~\bibnamefont {Krivenko}}, \bibinfo {author}
  {\bibfnamefont {L.}~\bibnamefont {Messio}},\ and\ \bibinfo {author}
  {\bibfnamefont {P.}~\bibnamefont {Seth}},\ }\bibfield  {title} {\bibinfo
  {title} {Triqs: A toolbox for research on interacting quantum systems},\
  }\href {https://doi.org/https://doi.org/10.1016/j.cpc.2015.04.023} {\bibfield
   {journal} {\bibinfo  {journal} {Computer Physics Communications}\ }\textbf
  {\bibinfo {volume} {196}},\ \bibinfo {pages} {398 } (\bibinfo {year}
  {2015})}\BibitemShut {NoStop}%
\bibitem [{\citenamefont {Rohringer}(2013)}]{Rohringer2013}%
  \BibitemOpen
  \bibfield  {author} {\bibinfo {author} {\bibfnamefont {G.}~\bibnamefont
  {Rohringer}},\ }\emph {\bibinfo {title} {New routes towards a theoretical
  treatment of nonlocal electronic correlations}},\ \href
  {http://digital.obvsg.at/download/pdf/1631831} {Ph.D. thesis},\ \bibinfo
  {school} {Vienna University of Technology} (\bibinfo {year}
  {2013})\BibitemShut {NoStop}%
\bibitem [{\citenamefont {Rohringer}\ \emph {et~al.}(2012)\citenamefont
  {Rohringer}, \citenamefont {Valli},\ and\ \citenamefont
  {Toschi}}]{Rohringer2012}%
  \BibitemOpen
  \bibfield  {author} {\bibinfo {author} {\bibfnamefont {G.}~\bibnamefont
  {Rohringer}}, \bibinfo {author} {\bibfnamefont {A.}~\bibnamefont {Valli}},\
  and\ \bibinfo {author} {\bibfnamefont {A.}~\bibnamefont {Toschi}},\
  }\bibfield  {title} {\bibinfo {title} {Local electronic correlation at the
  two-particle level},\ }\href {https://doi.org/10.1103/PhysRevB.86.125114}
  {\bibfield  {journal} {\bibinfo  {journal} {Phys. Rev. B}\ }\textbf {\bibinfo
  {volume} {86}},\ \bibinfo {pages} {125114} (\bibinfo {year}
  {2012})}\BibitemShut {NoStop}%
\bibitem [{\citenamefont {Del~Re}\ and\ \citenamefont
  {Toschi}(2021)}]{DelRe2021b}%
  \BibitemOpen
  \bibfield  {author} {\bibinfo {author} {\bibfnamefont {L.}~\bibnamefont
  {Del~Re}}\ and\ \bibinfo {author} {\bibfnamefont {A.}~\bibnamefont
  {Toschi}},\ }\bibfield  {title} {\bibinfo {title} {{Dynamical vertex
  approximation for many-electron systems with spontaneously broken SU(2)
  symmetry}},\ }\href {https://doi.org/10.1103/PhysRevB.104.085120} {\bibfield
  {journal} {\bibinfo  {journal} {Phys. Rev. B}\ }\textbf {\bibinfo {volume}
  {104}},\ \bibinfo {pages} {085120} (\bibinfo {year} {2021})}\BibitemShut
  {NoStop}%
\bibitem [{\citenamefont {Katanin}\ \emph {et~al.}(2009)\citenamefont
  {Katanin}, \citenamefont {Toschi},\ and\ \citenamefont {Held}}]{Katanin2009}%
  \BibitemOpen
  \bibfield  {author} {\bibinfo {author} {\bibfnamefont {A.~A.}\ \bibnamefont
  {Katanin}}, \bibinfo {author} {\bibfnamefont {A.}~\bibnamefont {Toschi}},\
  and\ \bibinfo {author} {\bibfnamefont {K.}~\bibnamefont {Held}},\ }\bibfield
  {title} {\bibinfo {title} {{Comparing pertinent effects of antiferromagnetic
  fluctuations in the two- and three-dimensional Hubbard model}},\ }\href
  {https://doi.org/10.1103/PhysRevB.80.075104} {\bibfield  {journal} {\bibinfo
  {journal} {Phys. Rev. B}\ }\textbf {\bibinfo {volume} {80}},\ \bibinfo
  {pages} {075104} (\bibinfo {year} {2009})}\BibitemShut {NoStop}%
\bibitem [{\citenamefont {Sch\"afer}\ \emph {et~al.}(2015)\citenamefont
  {Sch\"afer}, \citenamefont {Geles}, \citenamefont {Rost}, \citenamefont
  {Rohringer}, \citenamefont {Arrigoni}, \citenamefont {Held}, \citenamefont
  {Bl\"umer}, \citenamefont {Aichhorn},\ and\ \citenamefont
  {Toschi}}]{Schaefer2015}%
  \BibitemOpen
  \bibfield  {author} {\bibinfo {author} {\bibfnamefont {T.}~\bibnamefont
  {Sch\"afer}}, \bibinfo {author} {\bibfnamefont {F.}~\bibnamefont {Geles}},
  \bibinfo {author} {\bibfnamefont {D.}~\bibnamefont {Rost}}, \bibinfo {author}
  {\bibfnamefont {G.}~\bibnamefont {Rohringer}}, \bibinfo {author}
  {\bibfnamefont {E.}~\bibnamefont {Arrigoni}}, \bibinfo {author}
  {\bibfnamefont {K.}~\bibnamefont {Held}}, \bibinfo {author} {\bibfnamefont
  {N.}~\bibnamefont {Bl\"umer}}, \bibinfo {author} {\bibfnamefont
  {M.}~\bibnamefont {Aichhorn}},\ and\ \bibinfo {author} {\bibfnamefont
  {A.}~\bibnamefont {Toschi}},\ }\bibfield  {title} {\bibinfo {title} {{Fate of
  the false Mott-Hubbard transition in two dimensions}},\ }\href
  {https://doi.org/10.1103/PhysRevB.91.125109} {\bibfield  {journal} {\bibinfo
  {journal} {Phys. Rev. B}\ }\textbf {\bibinfo {volume} {91}},\ \bibinfo
  {pages} {125109} (\bibinfo {year} {2015})}\BibitemShut {NoStop}%
\bibitem [{\citenamefont {Rohringer}\ \emph {et~al.}(2011)\citenamefont
  {Rohringer}, \citenamefont {Toschi}, \citenamefont {Katanin},\ and\
  \citenamefont {Held}}]{Rohringer2011}%
  \BibitemOpen
  \bibfield  {author} {\bibinfo {author} {\bibfnamefont {G.}~\bibnamefont
  {Rohringer}}, \bibinfo {author} {\bibfnamefont {A.}~\bibnamefont {Toschi}},
  \bibinfo {author} {\bibfnamefont {A.}~\bibnamefont {Katanin}},\ and\ \bibinfo
  {author} {\bibfnamefont {K.}~\bibnamefont {Held}},\ }\bibfield  {title}
  {\bibinfo {title} {{Critical Properties of the Half-Filled Hubbard Model in
  Three Dimensions}},\ }\href {https://doi.org/10.1103/PhysRevLett.107.256402}
  {\bibfield  {journal} {\bibinfo  {journal} {Phys. Rev. Lett.}\ }\textbf
  {\bibinfo {volume} {107}},\ \bibinfo {pages} {256402} (\bibinfo {year}
  {2011})}\BibitemShut {NoStop}%
\bibitem [{\citenamefont {Sch\"afer}\ \emph {et~al.}(2017)\citenamefont
  {Sch\"afer}, \citenamefont {Katanin}, \citenamefont {Held},\ and\
  \citenamefont {Toschi}}]{Schaefer2017}%
  \BibitemOpen
  \bibfield  {author} {\bibinfo {author} {\bibfnamefont {T.}~\bibnamefont
  {Sch\"afer}}, \bibinfo {author} {\bibfnamefont {A.~A.}\ \bibnamefont
  {Katanin}}, \bibinfo {author} {\bibfnamefont {K.}~\bibnamefont {Held}},\ and\
  \bibinfo {author} {\bibfnamefont {A.}~\bibnamefont {Toschi}},\ }\bibfield
  {title} {\bibinfo {title} {{Interplay of Correlations and Kohn Anomalies in
  Three Dimensions: Quantum Criticality with a Twist}},\ }\href
  {https://doi.org/10.1103/PhysRevLett.119.046402} {\bibfield  {journal}
  {\bibinfo  {journal} {Phys. Rev. Lett.}\ }\textbf {\bibinfo {volume} {119}},\
  \bibinfo {pages} {046402} (\bibinfo {year} {2017})}\BibitemShut {NoStop}%
\bibitem [{\citenamefont {Sch\"afer}\ \emph {et~al.}(2019)\citenamefont
  {Sch\"afer}, \citenamefont {Katanin}, \citenamefont {Kitatani}, \citenamefont
  {Toschi},\ and\ \citenamefont {Held}}]{Schaefer2019}%
  \BibitemOpen
  \bibfield  {author} {\bibinfo {author} {\bibfnamefont {T.}~\bibnamefont
  {Sch\"afer}}, \bibinfo {author} {\bibfnamefont {A.~A.}\ \bibnamefont
  {Katanin}}, \bibinfo {author} {\bibfnamefont {M.}~\bibnamefont {Kitatani}},
  \bibinfo {author} {\bibfnamefont {A.}~\bibnamefont {Toschi}},\ and\ \bibinfo
  {author} {\bibfnamefont {K.}~\bibnamefont {Held}},\ }\bibfield  {title}
  {\bibinfo {title} {{Quantum Criticality in the Two-Dimensional Periodic
  Anderson Model}},\ }\href {https://doi.org/10.1103/PhysRevLett.122.227201}
  {\bibfield  {journal} {\bibinfo  {journal} {Phys. Rev. Lett.}\ }\textbf
  {\bibinfo {volume} {122}},\ \bibinfo {pages} {227201} (\bibinfo {year}
  {2019})}\BibitemShut {NoStop}%
\bibitem [{\citenamefont {Kitatani}\ \emph {et~al.}(2019)\citenamefont
  {Kitatani}, \citenamefont {Sch\"afer}, \citenamefont {Aoki},\ and\
  \citenamefont {Held}}]{Kitatani2018}%
  \BibitemOpen
  \bibfield  {author} {\bibinfo {author} {\bibfnamefont {M.}~\bibnamefont
  {Kitatani}}, \bibinfo {author} {\bibfnamefont {T.}~\bibnamefont {Sch\"afer}},
  \bibinfo {author} {\bibfnamefont {H.}~\bibnamefont {Aoki}},\ and\ \bibinfo
  {author} {\bibfnamefont {K.}~\bibnamefont {Held}},\ }\bibfield  {title}
  {\bibinfo {title} {Why the critical temperature of high-${T}_{c}$ cuprate
  superconductors is so low: The importance of the dynamical vertex
  structure},\ }\href {https://doi.org/10.1103/PhysRevB.99.041115} {\bibfield
  {journal} {\bibinfo  {journal} {Phys. Rev. B}\ }\textbf {\bibinfo {volume}
  {99}},\ \bibinfo {pages} {041115} (\bibinfo {year} {2019})}\BibitemShut
  {NoStop}%
\bibitem [{\citenamefont {Ortiz}\ \emph {et~al.}(2022)\citenamefont {Ortiz},
  \citenamefont {Puphal}, \citenamefont {Klett}, \citenamefont {Hotz},
  \citenamefont {Kremer}, \citenamefont {Trepka}, \citenamefont {Hemmida},
  \citenamefont {von Nidda}, \citenamefont {Isobe}, \citenamefont {Khasanov},
  \citenamefont {Luetkens}, \citenamefont {Hansmann}, \citenamefont {Keimer},
  \citenamefont {Sch\"afer},\ and\ \citenamefont {Hepting}}]{Ortiz2022}%
  \BibitemOpen
  \bibfield  {author} {\bibinfo {author} {\bibfnamefont {R.~A.}\ \bibnamefont
  {Ortiz}}, \bibinfo {author} {\bibfnamefont {P.}~\bibnamefont {Puphal}},
  \bibinfo {author} {\bibfnamefont {M.}~\bibnamefont {Klett}}, \bibinfo
  {author} {\bibfnamefont {F.}~\bibnamefont {Hotz}}, \bibinfo {author}
  {\bibfnamefont {R.~K.}\ \bibnamefont {Kremer}}, \bibinfo {author}
  {\bibfnamefont {H.}~\bibnamefont {Trepka}}, \bibinfo {author} {\bibfnamefont
  {M.}~\bibnamefont {Hemmida}}, \bibinfo {author} {\bibfnamefont {H.-A.~K.}\
  \bibnamefont {von Nidda}}, \bibinfo {author} {\bibfnamefont {M.}~\bibnamefont
  {Isobe}}, \bibinfo {author} {\bibfnamefont {R.}~\bibnamefont {Khasanov}},
  \bibinfo {author} {\bibfnamefont {H.}~\bibnamefont {Luetkens}}, \bibinfo
  {author} {\bibfnamefont {P.}~\bibnamefont {Hansmann}}, \bibinfo {author}
  {\bibfnamefont {B.}~\bibnamefont {Keimer}}, \bibinfo {author} {\bibfnamefont
  {T.}~\bibnamefont {Sch\"afer}},\ and\ \bibinfo {author} {\bibfnamefont
  {M.}~\bibnamefont {Hepting}},\ }\bibfield  {title} {\bibinfo {title}
  {{Magnetic correlations in infinite-layer nickelates: An experimental and
  theoretical multimethod study}},\ }\href
  {https://doi.org/10.1103/PhysRevResearch.4.023093} {\bibfield  {journal}
  {\bibinfo  {journal} {Phys. Rev. Research}\ }\textbf {\bibinfo {volume}
  {4}},\ \bibinfo {pages} {023093} (\bibinfo {year} {2022})}\BibitemShut
  {NoStop}%
\bibitem [{\citenamefont {Held}\ \emph {et~al.}(2022)\citenamefont {Held},
  \citenamefont {Si}, \citenamefont {Worm}, \citenamefont {Janson},
  \citenamefont {Arita}, \citenamefont {Zhong}, \citenamefont {Tomczak},\ and\
  \citenamefont {Kitatani}}]{Held2022}%
  \BibitemOpen
  \bibfield  {author} {\bibinfo {author} {\bibfnamefont {K.}~\bibnamefont
  {Held}}, \bibinfo {author} {\bibfnamefont {L.}~\bibnamefont {Si}}, \bibinfo
  {author} {\bibfnamefont {P.}~\bibnamefont {Worm}}, \bibinfo {author}
  {\bibfnamefont {O.}~\bibnamefont {Janson}}, \bibinfo {author} {\bibfnamefont
  {R.}~\bibnamefont {Arita}}, \bibinfo {author} {\bibfnamefont
  {Z.}~\bibnamefont {Zhong}}, \bibinfo {author} {\bibfnamefont {J.~M.}\
  \bibnamefont {Tomczak}},\ and\ \bibinfo {author} {\bibfnamefont
  {M.}~\bibnamefont {Kitatani}},\ }\bibfield  {title} {\bibinfo {title} {{Phase
  Diagram of Nickelate Superconductors Calculated by Dynamical Vertex
  Approximation}},\ }\href {https://doi.org/10.3389%2Ffphy.2021.810394}
  {\bibfield  {journal} {\bibinfo  {journal} {Frontiers in Physics}\ }\textbf
  {\bibinfo {volume} {9}} (\bibinfo {year} {2022})}\BibitemShut {NoStop}%
\bibitem [{\citenamefont {Rohringer}\ and\ \citenamefont
  {Toschi}(2016)}]{Rohringer2016}%
  \BibitemOpen
  \bibfield  {author} {\bibinfo {author} {\bibfnamefont {G.}~\bibnamefont
  {Rohringer}}\ and\ \bibinfo {author} {\bibfnamefont {A.}~\bibnamefont
  {Toschi}},\ }\bibfield  {title} {\bibinfo {title} {Impact of nonlocal
  correlations over different energy scales: A dynamical vertex approximation
  study},\ }\href {https://doi.org/10.1103%2Fphysrevb.94.125144} {\bibfield
  {journal} {\bibinfo  {journal} {Physical Review B}\ }\textbf {\bibinfo
  {volume} {94}} (\bibinfo {year} {2016})}\BibitemShut {NoStop}%
\bibitem [{\citenamefont {Ayral}\ and\ \citenamefont
  {Parcollet}(2015)}]{Ayral2015}%
  \BibitemOpen
  \bibfield  {author} {\bibinfo {author} {\bibfnamefont {T.}~\bibnamefont
  {Ayral}}\ and\ \bibinfo {author} {\bibfnamefont {O.}~\bibnamefont
  {Parcollet}},\ }\bibfield  {title} {\bibinfo {title} {Mott physics and spin
  fluctuations: A unified framework},\ }\href
  {https://doi.org/10.1103/PhysRevB.92.115109} {\bibfield  {journal} {\bibinfo
  {journal} {Phys. Rev. B}\ }\textbf {\bibinfo {volume} {92}},\ \bibinfo
  {pages} {115109} (\bibinfo {year} {2015})}\BibitemShut {NoStop}%
\bibitem [{\citenamefont {Ayral}\ and\ \citenamefont
  {Parcollet}(2016)}]{Ayral2016}%
  \BibitemOpen
  \bibfield  {author} {\bibinfo {author} {\bibfnamefont {T.}~\bibnamefont
  {Ayral}}\ and\ \bibinfo {author} {\bibfnamefont {O.}~\bibnamefont
  {Parcollet}},\ }\bibfield  {title} {\bibinfo {title} {Mott physics and spin
  fluctuations: A functional viewpoint},\ }\href
  {https://doi.org/10.1103/PhysRevB.93.235124} {\bibfield  {journal} {\bibinfo
  {journal} {Phys. Rev. B}\ }\textbf {\bibinfo {volume} {93}},\ \bibinfo
  {pages} {235124} (\bibinfo {year} {2016})}\BibitemShut {NoStop}%
\bibitem [{\citenamefont {Moriya}(1991)}]{MORIYA1991}%
  \BibitemOpen
  \bibfield  {author} {\bibinfo {author} {\bibfnamefont {T.}~\bibnamefont
  {Moriya}},\ }\bibfield  {title} {\bibinfo {title} {Theory of itinerant
  electron magnetism},\ }\href
  {https://doi.org/https://doi.org/10.1016/0304-8853(91)90824-T} {\bibfield
  {journal} {\bibinfo  {journal} {Journal of Magnetism and Magnetic Materials}\
  }\textbf {\bibinfo {volume} {100}},\ \bibinfo {pages} {261} (\bibinfo {year}
  {1991})}\BibitemShut {NoStop}%
\bibitem [{\citenamefont {Stobbe}\ and\ \citenamefont
  {Rohringer}(2022)}]{Stobbe2022}%
  \BibitemOpen
  \bibfield  {author} {\bibinfo {author} {\bibfnamefont {J.}~\bibnamefont
  {Stobbe}}\ and\ \bibinfo {author} {\bibfnamefont {G.}~\bibnamefont
  {Rohringer}},\ }\bibfield  {title} {\bibinfo {title} {Consistency of
  potential energy in the dynamical vertex approximation},\ }\href
  {https://doi.org/10.1103/PhysRevB.106.205101} {\bibfield  {journal} {\bibinfo
   {journal} {Phys. Rev. B}\ }\textbf {\bibinfo {volume} {106}},\ \bibinfo
  {pages} {205101} (\bibinfo {year} {2022})}\BibitemShut {NoStop}%
\bibitem [{\citenamefont {Del~Re}\ \emph {et~al.}(2019)\citenamefont {Del~Re},
  \citenamefont {Capone},\ and\ \citenamefont {Toschi}}]{Del_Re2019}%
  \BibitemOpen
  \bibfield  {author} {\bibinfo {author} {\bibfnamefont {L.}~\bibnamefont
  {Del~Re}}, \bibinfo {author} {\bibfnamefont {M.}~\bibnamefont {Capone}},\
  and\ \bibinfo {author} {\bibfnamefont {A.}~\bibnamefont {Toschi}},\
  }\bibfield  {title} {\bibinfo {title} {{Dynamical vertex approximation for
  the attractive Hubbard model}},\ }\href
  {https://doi.org/10.1103/PhysRevB.99.045137} {\bibfield  {journal} {\bibinfo
  {journal} {Phys. Rev. B}\ }\textbf {\bibinfo {volume} {99}},\ \bibinfo
  {pages} {045137} (\bibinfo {year} {2019})}\BibitemShut {NoStop}%
\bibitem [{\citenamefont {Ding}\ \emph {et~al.}(1996)\citenamefont {Ding},
  \citenamefont {Norman}, \citenamefont {Campuzano}, \citenamefont {Randeria},
  \citenamefont {Bellman}, \citenamefont {Yokoya}, \citenamefont {Takahashi},
  \citenamefont {Mochiku},\ and\ \citenamefont {Kadowaki}}]{Ding1996}%
  \BibitemOpen
  \bibfield  {author} {\bibinfo {author} {\bibfnamefont {H.}~\bibnamefont
  {Ding}}, \bibinfo {author} {\bibfnamefont {M.~R.}\ \bibnamefont {Norman}},
  \bibinfo {author} {\bibfnamefont {J.~C.}\ \bibnamefont {Campuzano}}, \bibinfo
  {author} {\bibfnamefont {M.}~\bibnamefont {Randeria}}, \bibinfo {author}
  {\bibfnamefont {A.~F.}\ \bibnamefont {Bellman}}, \bibinfo {author}
  {\bibfnamefont {T.}~\bibnamefont {Yokoya}}, \bibinfo {author} {\bibfnamefont
  {T.}~\bibnamefont {Takahashi}}, \bibinfo {author} {\bibfnamefont
  {T.}~\bibnamefont {Mochiku}},\ and\ \bibinfo {author} {\bibfnamefont
  {K.}~\bibnamefont {Kadowaki}},\ }\bibfield  {title} {\bibinfo {title}
  {{Angle-resolved photoemission spectroscopy study of the superconducting gap
  anisotropy in
  ${\mathrm{Bi}}_{2}{\mathrm{Sr}}_{2}\mathrm{Ca}{\mathrm{Cu}}_{2}{\mathrm{O}}_{8+x}$}},\
  }\href {https://doi.org/10.1103/PhysRevB.54.R9678} {\bibfield  {journal}
  {\bibinfo  {journal} {Phys. Rev. B}\ }\textbf {\bibinfo {volume} {54}},\
  \bibinfo {pages} {R9678} (\bibinfo {year} {1996})}\BibitemShut {NoStop}%
\bibitem [{\citenamefont {Marshall}\ \emph {et~al.}(1996)\citenamefont
  {Marshall}, \citenamefont {Dessau}, \citenamefont {Loeser}, \citenamefont
  {Park}, \citenamefont {Matsuura}, \citenamefont {Eckstein}, \citenamefont
  {Bozovic}, \citenamefont {Fournier}, \citenamefont {Kapitulnik},
  \citenamefont {Spicer},\ and\ \citenamefont {Shen}}]{Marshall1996}%
  \BibitemOpen
  \bibfield  {author} {\bibinfo {author} {\bibfnamefont {D.~S.}\ \bibnamefont
  {Marshall}}, \bibinfo {author} {\bibfnamefont {D.~S.}\ \bibnamefont
  {Dessau}}, \bibinfo {author} {\bibfnamefont {A.~G.}\ \bibnamefont {Loeser}},
  \bibinfo {author} {\bibfnamefont {C.-H.}\ \bibnamefont {Park}}, \bibinfo
  {author} {\bibfnamefont {A.~Y.}\ \bibnamefont {Matsuura}}, \bibinfo {author}
  {\bibfnamefont {J.~N.}\ \bibnamefont {Eckstein}}, \bibinfo {author}
  {\bibfnamefont {I.}~\bibnamefont {Bozovic}}, \bibinfo {author} {\bibfnamefont
  {P.}~\bibnamefont {Fournier}}, \bibinfo {author} {\bibfnamefont
  {A.}~\bibnamefont {Kapitulnik}}, \bibinfo {author} {\bibfnamefont {W.~E.}\
  \bibnamefont {Spicer}},\ and\ \bibinfo {author} {\bibfnamefont {Z.-X.}\
  \bibnamefont {Shen}},\ }\bibfield  {title} {\bibinfo {title} {{Unconventional
  Electronic Structure Evolution with Hole Doping in
  ${\mathrm{Bi}}_{2}{\mathrm{Sr}}_{2}{\mathrm{CaCu}}_{2}{O}_{8+\ensuremath{\delta}}$:
  Angle-Resolved Photoemission Results}},\ }\href
  {https://doi.org/10.1103/PhysRevLett.76.4841} {\bibfield  {journal} {\bibinfo
   {journal} {Phys. Rev. Lett.}\ }\textbf {\bibinfo {volume} {76}},\ \bibinfo
  {pages} {4841} (\bibinfo {year} {1996})}\BibitemShut {NoStop}%
\end{thebibliography}%


\begin{thebibliography}{39}%
\makeatletter
\providecommand \@ifxundefined [1]{%
 \@ifx{#1\undefined}
}%
\providecommand \@ifnum [1]{%
 \ifnum #1\expandafter \@firstoftwo
 \else \expandafter \@secondoftwo
 \fi
}%
\providecommand \@ifx [1]{%
 \ifx #1\expandafter \@firstoftwo
 \else \expandafter \@secondoftwo
 \fi
}%
\providecommand \natexlab [1]{#1}%
\providecommand \enquote  [1]{``#1''}%
\providecommand \bibnamefont  [1]{#1}%
\providecommand \bibfnamefont [1]{#1}%
\providecommand \citenamefont [1]{#1}%
\providecommand \href@noop [0]{\@secondoftwo}%
\providecommand \href [0]{\begingroup \@sanitize@url \@href}%
\providecommand \@href[1]{\@@startlink{#1}\@@href}%
\providecommand \@@href[1]{\endgroup#1\@@endlink}%
\providecommand \@sanitize@url [0]{\catcode `\\12\catcode `\$12\catcode
  `\&12\catcode `\#12\catcode `\^12\catcode `\_12\catcode `\%12\relax}%
\providecommand \@@startlink[1]{}%
\providecommand \@@endlink[0]{}%
\providecommand \url  [0]{\begingroup\@sanitize@url \@url }%
\providecommand \@url [1]{\endgroup\@href {#1}{\urlprefix }}%
\providecommand \urlprefix  [0]{URL }%
\providecommand \Eprint [0]{\href }%
\providecommand \doibase [0]{https://doi.org/}%
\providecommand \selectlanguage [0]{\@gobble}%
\providecommand \bibinfo  [0]{\@secondoftwo}%
\providecommand \bibfield  [0]{\@secondoftwo}%
\providecommand \translation [1]{[#1]}%
\providecommand \BibitemOpen [0]{}%
\providecommand \bibitemStop [0]{}%
\providecommand \bibitemNoStop [0]{.\EOS\space}%
\providecommand \EOS [0]{\spacefactor3000\relax}%
\providecommand \BibitemShut  [1]{\csname bibitem#1\endcsname}%
\let\auto@bib@innerbib\@empty
\bibitem [{\citenamefont {Metzner}\ and\ \citenamefont
  {Vollhardt}(1989)}]{Metzner1989}%
  \BibitemOpen
  \bibfield  {author} {\bibinfo {author} {\bibfnamefont {W.}~\bibnamefont
  {Metzner}}\ and\ \bibinfo {author} {\bibfnamefont {D.}~\bibnamefont
  {Vollhardt}},\ }\bibfield  {title} {\bibinfo {title} {{Correlated Lattice
  Fermions in $d=\ensuremath{\infty}$ Dimensions}},\ }\href
  {https://doi.org/10.1103/PhysRevLett.62.324} {\bibfield  {journal} {\bibinfo
  {journal} {Phys. Rev. Lett.}\ }\textbf {\bibinfo {volume} {62}},\ \bibinfo
  {pages} {324} (\bibinfo {year} {1989})}\BibitemShut {NoStop}%
\bibitem [{\citenamefont {Georges}\ and\ \citenamefont
  {Kotliar}(1992)}]{Georges1992}%
  \BibitemOpen
  \bibfield  {author} {\bibinfo {author} {\bibfnamefont {A.}~\bibnamefont
  {Georges}}\ and\ \bibinfo {author} {\bibfnamefont {G.}~\bibnamefont
  {Kotliar}},\ }\bibfield  {title} {\bibinfo {title} {Hubbard model in infinite
  dimensions},\ }\href {https://doi.org/10.1103/PhysRevB.45.6479} {\bibfield
  {journal} {\bibinfo  {journal} {Phys. Rev. B}\ }\textbf {\bibinfo {volume}
  {45}},\ \bibinfo {pages} {6479} (\bibinfo {year} {1992})}\BibitemShut
  {NoStop}%
\bibitem [{\citenamefont {Georges}\ \emph {et~al.}(1996)\citenamefont
  {Georges}, \citenamefont {Kotliar}, \citenamefont {Krauth},\ and\
  \citenamefont {Rozenberg}}]{Georges1996}%
  \BibitemOpen
  \bibfield  {author} {\bibinfo {author} {\bibfnamefont {A.}~\bibnamefont
  {Georges}}, \bibinfo {author} {\bibfnamefont {G.}~\bibnamefont {Kotliar}},
  \bibinfo {author} {\bibfnamefont {W.}~\bibnamefont {Krauth}},\ and\ \bibinfo
  {author} {\bibfnamefont {M.~J.}\ \bibnamefont {Rozenberg}},\ }\bibfield
  {title} {\bibinfo {title} {Dynamical mean-field theory of strongly correlated
  fermion systems and the limit of infinite dimensions},\ }\href
  {https://doi.org/10.1103/RevModPhys.68.13} {\bibfield  {journal} {\bibinfo
  {journal} {Rev. Mod. Phys.}\ }\textbf {\bibinfo {volume} {68}},\ \bibinfo
  {pages} {13} (\bibinfo {year} {1996})}\BibitemShut {NoStop}%
\bibitem [{\citenamefont {Rubtsov}\ \emph {et~al.}(2005)\citenamefont
  {Rubtsov}, \citenamefont {Savkin},\ and\ \citenamefont
  {Lichtenstein}}]{Rubtsov2005}%
  \BibitemOpen
  \bibfield  {author} {\bibinfo {author} {\bibfnamefont {A.~N.}\ \bibnamefont
  {Rubtsov}}, \bibinfo {author} {\bibfnamefont {V.~V.}\ \bibnamefont
  {Savkin}},\ and\ \bibinfo {author} {\bibfnamefont {A.~I.}\ \bibnamefont
  {Lichtenstein}},\ }\bibfield  {title} {\bibinfo {title} {{Continuous-time
  quantum Monte Carlo method for fermions}},\ }\href
  {https://doi.org/10.1103/PhysRevB.72.035122} {\bibfield  {journal} {\bibinfo
  {journal} {Phys. Rev. B}\ }\textbf {\bibinfo {volume} {72}},\ \bibinfo
  {pages} {035122} (\bibinfo {year} {2005})}\BibitemShut {NoStop}%
\bibitem [{\citenamefont {Gull}\ \emph {et~al.}(2011)\citenamefont {Gull},
  \citenamefont {Millis}, \citenamefont {Lichtenstein}, \citenamefont
  {Rubtsov}, \citenamefont {Troyer},\ and\ \citenamefont {Werner}}]{Gull2011}%
  \BibitemOpen
  \bibfield  {author} {\bibinfo {author} {\bibfnamefont {E.}~\bibnamefont
  {Gull}}, \bibinfo {author} {\bibfnamefont {A.~J.}\ \bibnamefont {Millis}},
  \bibinfo {author} {\bibfnamefont {A.~I.}\ \bibnamefont {Lichtenstein}},
  \bibinfo {author} {\bibfnamefont {A.~N.}\ \bibnamefont {Rubtsov}}, \bibinfo
  {author} {\bibfnamefont {M.}~\bibnamefont {Troyer}},\ and\ \bibinfo {author}
  {\bibfnamefont {P.}~\bibnamefont {Werner}},\ }\bibfield  {title} {\bibinfo
  {title} {{Continuous-time Monte Carlo methods for quantum impurity models}},\
  }\href {https://doi.org/10.1103/RevModPhys.83.349} {\bibfield  {journal}
  {\bibinfo  {journal} {Rev. Mod. Phys.}\ }\textbf {\bibinfo {volume} {83}},\
  \bibinfo {pages} {349} (\bibinfo {year} {2011})}\BibitemShut {NoStop}%
\bibitem [{\citenamefont {Parcollet}\ \emph {et~al.}(2015)\citenamefont
  {Parcollet}, \citenamefont {Ferrero}, \citenamefont {Ayral}, \citenamefont
  {Hafermann}, \citenamefont {Krivenko}, \citenamefont {Messio},\ and\
  \citenamefont {Seth}}]{TRIQS}%
  \BibitemOpen
  \bibfield  {author} {\bibinfo {author} {\bibfnamefont {O.}~\bibnamefont
  {Parcollet}}, \bibinfo {author} {\bibfnamefont {M.}~\bibnamefont {Ferrero}},
  \bibinfo {author} {\bibfnamefont {T.}~\bibnamefont {Ayral}}, \bibinfo
  {author} {\bibfnamefont {H.}~\bibnamefont {Hafermann}}, \bibinfo {author}
  {\bibfnamefont {I.}~\bibnamefont {Krivenko}}, \bibinfo {author}
  {\bibfnamefont {L.}~\bibnamefont {Messio}},\ and\ \bibinfo {author}
  {\bibfnamefont {P.}~\bibnamefont {Seth}},\ }\bibfield  {title} {\bibinfo
  {title} {Triqs: A toolbox for research on interacting quantum systems},\
  }\href {https://doi.org/https://doi.org/10.1016/j.cpc.2015.04.023} {\bibfield
   {journal} {\bibinfo  {journal} {Computer Physics Communications}\ }\textbf
  {\bibinfo {volume} {196}},\ \bibinfo {pages} {398 } (\bibinfo {year}
  {2015})}\BibitemShut {NoStop}%
\bibitem [{\citenamefont {Rohringer}(2013)}]{Rohringer2013}%
  \BibitemOpen
  \bibfield  {author} {\bibinfo {author} {\bibfnamefont {G.}~\bibnamefont
  {Rohringer}},\ }\emph {\bibinfo {title} {New routes towards a theoretical
  treatment of nonlocal electronic correlations}},\ \href
  {http://digital.obvsg.at/download/pdf/1631831} {Ph.D. thesis},\ \bibinfo
  {school} {Vienna University of Technology} (\bibinfo {year}
  {2013})\BibitemShut {NoStop}%
\bibitem [{\citenamefont {Rohringer}\ \emph {et~al.}(2012)\citenamefont
  {Rohringer}, \citenamefont {Valli},\ and\ \citenamefont
  {Toschi}}]{Rohringer2012}%
  \BibitemOpen
  \bibfield  {author} {\bibinfo {author} {\bibfnamefont {G.}~\bibnamefont
  {Rohringer}}, \bibinfo {author} {\bibfnamefont {A.}~\bibnamefont {Valli}},\
  and\ \bibinfo {author} {\bibfnamefont {A.}~\bibnamefont {Toschi}},\
  }\bibfield  {title} {\bibinfo {title} {Local electronic correlation at the
  two-particle level},\ }\href {https://doi.org/10.1103/PhysRevB.86.125114}
  {\bibfield  {journal} {\bibinfo  {journal} {Phys. Rev. B}\ }\textbf {\bibinfo
  {volume} {86}},\ \bibinfo {pages} {125114} (\bibinfo {year}
  {2012})}\BibitemShut {NoStop}%
\bibitem [{\citenamefont {Del~Re}\ and\ \citenamefont
  {Toschi}(2021)}]{DelRe2021b}%
  \BibitemOpen
  \bibfield  {author} {\bibinfo {author} {\bibfnamefont {L.}~\bibnamefont
  {Del~Re}}\ and\ \bibinfo {author} {\bibfnamefont {A.}~\bibnamefont
  {Toschi}},\ }\bibfield  {title} {\bibinfo {title} {{Dynamical vertex
  approximation for many-electron systems with spontaneously broken SU(2)
  symmetry}},\ }\href {https://doi.org/10.1103/PhysRevB.104.085120} {\bibfield
  {journal} {\bibinfo  {journal} {Phys. Rev. B}\ }\textbf {\bibinfo {volume}
  {104}},\ \bibinfo {pages} {085120} (\bibinfo {year} {2021})}\BibitemShut
  {NoStop}%
\bibitem [{\citenamefont {Maier}\ \emph {et~al.}(2005)\citenamefont {Maier},
  \citenamefont {Jarrell}, \citenamefont {Pruschke},\ and\ \citenamefont
  {Hettler}}]{Maier2005}%
  \BibitemOpen
  \bibfield  {author} {\bibinfo {author} {\bibfnamefont {T.}~\bibnamefont
  {Maier}}, \bibinfo {author} {\bibfnamefont {M.}~\bibnamefont {Jarrell}},
  \bibinfo {author} {\bibfnamefont {T.}~\bibnamefont {Pruschke}},\ and\
  \bibinfo {author} {\bibfnamefont {M.~H.}\ \bibnamefont {Hettler}},\
  }\bibfield  {title} {\bibinfo {title} {Quantum cluster theories},\ }\href
  {https://doi.org/10.1103/RevModPhys.77.1027} {\bibfield  {journal} {\bibinfo
  {journal} {Rev. Mod. Phys.}\ }\textbf {\bibinfo {volume} {77}},\ \bibinfo
  {pages} {1027} (\bibinfo {year} {2005})}\BibitemShut {NoStop}%
\bibitem [{\citenamefont {Rohringer}\ \emph {et~al.}(2018)\citenamefont
  {Rohringer}, \citenamefont {Hafermann}, \citenamefont {Toschi}, \citenamefont
  {Katanin}, \citenamefont {Antipov}, \citenamefont {Katsnelson}, \citenamefont
  {Lichtenstein}, \citenamefont {Rubtsov},\ and\ \citenamefont
  {Held}}]{Rohringer2018}%
  \BibitemOpen
  \bibfield  {author} {\bibinfo {author} {\bibfnamefont {G.}~\bibnamefont
  {Rohringer}}, \bibinfo {author} {\bibfnamefont {H.}~\bibnamefont
  {Hafermann}}, \bibinfo {author} {\bibfnamefont {A.}~\bibnamefont {Toschi}},
  \bibinfo {author} {\bibfnamefont {A.~A.}\ \bibnamefont {Katanin}}, \bibinfo
  {author} {\bibfnamefont {A.~E.}\ \bibnamefont {Antipov}}, \bibinfo {author}
  {\bibfnamefont {M.~I.}\ \bibnamefont {Katsnelson}}, \bibinfo {author}
  {\bibfnamefont {A.~I.}\ \bibnamefont {Lichtenstein}}, \bibinfo {author}
  {\bibfnamefont {A.~N.}\ \bibnamefont {Rubtsov}},\ and\ \bibinfo {author}
  {\bibfnamefont {K.}~\bibnamefont {Held}},\ }\bibfield  {title} {\bibinfo
  {title} {Diagrammatic routes to nonlocal correlations beyond dynamical mean
  field theory},\ }\href {https://doi.org/10.1103/RevModPhys.90.025003}
  {\bibfield  {journal} {\bibinfo  {journal} {Rev. Mod. Phys.}\ }\textbf
  {\bibinfo {volume} {90}},\ \bibinfo {pages} {025003} (\bibinfo {year}
  {2018})}\BibitemShut {NoStop}%
\bibitem [{\citenamefont {Toschi}\ \emph {et~al.}(2007)\citenamefont {Toschi},
  \citenamefont {Katanin},\ and\ \citenamefont {Held}}]{Toschi2007}%
  \BibitemOpen
  \bibfield  {author} {\bibinfo {author} {\bibfnamefont {A.}~\bibnamefont
  {Toschi}}, \bibinfo {author} {\bibfnamefont {A.~A.}\ \bibnamefont
  {Katanin}},\ and\ \bibinfo {author} {\bibfnamefont {K.}~\bibnamefont
  {Held}},\ }\bibfield  {title} {\bibinfo {title} {{Dynamical vertex
  approximation; A step beyond dynamical mean-field theory}},\ }\href
  {https://doi.org/10.1103/PhysRevB.75.045118} {\bibfield  {journal} {\bibinfo
  {journal} {Phys Rev. B}\ }\textbf {\bibinfo {volume} {75}},\ \bibinfo {pages}
  {045118} (\bibinfo {year} {2007})}\BibitemShut {NoStop}%
\bibitem [{\citenamefont {Katanin}\ \emph {et~al.}(2009)\citenamefont
  {Katanin}, \citenamefont {Toschi},\ and\ \citenamefont {Held}}]{Katanin2009}%
  \BibitemOpen
  \bibfield  {author} {\bibinfo {author} {\bibfnamefont {A.~A.}\ \bibnamefont
  {Katanin}}, \bibinfo {author} {\bibfnamefont {A.}~\bibnamefont {Toschi}},\
  and\ \bibinfo {author} {\bibfnamefont {K.}~\bibnamefont {Held}},\ }\bibfield
  {title} {\bibinfo {title} {{Comparing pertinent effects of antiferromagnetic
  fluctuations in the two- and three-dimensional Hubbard model}},\ }\href
  {https://doi.org/10.1103/PhysRevB.80.075104} {\bibfield  {journal} {\bibinfo
  {journal} {Phys. Rev. B}\ }\textbf {\bibinfo {volume} {80}},\ \bibinfo
  {pages} {075104} (\bibinfo {year} {2009})}\BibitemShut {NoStop}%
\bibitem [{\citenamefont {Sch\"afer}\ \emph {et~al.}(2021)\citenamefont
  {Sch\"afer}, \citenamefont {Wentzell}, \citenamefont {\ifmmode~\check{S}\else
  \v{S}\fi{}imkovic}, \citenamefont {He}, \citenamefont {Hille}, \citenamefont
  {Klett}, \citenamefont {Eckhardt}, \citenamefont {Arzhang}, \citenamefont
  {Harkov}, \citenamefont {Le~R\'egent}, \citenamefont {Kirsch}, \citenamefont
  {Wang}, \citenamefont {Kim}, \citenamefont {Kozik}, \citenamefont {Stepanov},
  \citenamefont {Kauch}, \citenamefont {Andergassen}, \citenamefont {Hansmann},
  \citenamefont {Rohe}, \citenamefont {Vilk}, \citenamefont {LeBlanc},
  \citenamefont {Zhang}, \citenamefont {Tremblay}, \citenamefont {Ferrero},
  \citenamefont {Parcollet},\ and\ \citenamefont {Georges}}]{Schaefer2021}%
  \BibitemOpen
  \bibfield  {author} {\bibinfo {author} {\bibfnamefont {T.}~\bibnamefont
  {Sch\"afer}}, \bibinfo {author} {\bibfnamefont {N.}~\bibnamefont {Wentzell}},
  \bibinfo {author} {\bibfnamefont {F.}~\bibnamefont {\ifmmode~\check{S}\else
  \v{S}\fi{}imkovic}}, \bibinfo {author} {\bibfnamefont {Y.-Y.}\ \bibnamefont
  {He}}, \bibinfo {author} {\bibfnamefont {C.}~\bibnamefont {Hille}}, \bibinfo
  {author} {\bibfnamefont {M.}~\bibnamefont {Klett}}, \bibinfo {author}
  {\bibfnamefont {C.~J.}\ \bibnamefont {Eckhardt}}, \bibinfo {author}
  {\bibfnamefont {B.}~\bibnamefont {Arzhang}}, \bibinfo {author} {\bibfnamefont
  {V.}~\bibnamefont {Harkov}}, \bibinfo {author} {\bibfnamefont
  {F.}~\bibnamefont {Le~R\'egent}}, \bibinfo {author} {\bibfnamefont
  {A.}~\bibnamefont {Kirsch}}, \bibinfo {author} {\bibfnamefont
  {Y.}~\bibnamefont {Wang}}, \bibinfo {author} {\bibfnamefont {A.~J.}\
  \bibnamefont {Kim}}, \bibinfo {author} {\bibfnamefont {E.}~\bibnamefont
  {Kozik}}, \bibinfo {author} {\bibfnamefont {E.~A.}\ \bibnamefont {Stepanov}},
  \bibinfo {author} {\bibfnamefont {A.}~\bibnamefont {Kauch}}, \bibinfo
  {author} {\bibfnamefont {S.}~\bibnamefont {Andergassen}}, \bibinfo {author}
  {\bibfnamefont {P.}~\bibnamefont {Hansmann}}, \bibinfo {author}
  {\bibfnamefont {D.}~\bibnamefont {Rohe}}, \bibinfo {author} {\bibfnamefont
  {Y.~M.}\ \bibnamefont {Vilk}}, \bibinfo {author} {\bibfnamefont {J.~P.~F.}\
  \bibnamefont {LeBlanc}}, \bibinfo {author} {\bibfnamefont {S.}~\bibnamefont
  {Zhang}}, \bibinfo {author} {\bibfnamefont {A.-M.~S.}\ \bibnamefont
  {Tremblay}}, \bibinfo {author} {\bibfnamefont {M.}~\bibnamefont {Ferrero}},
  \bibinfo {author} {\bibfnamefont {O.}~\bibnamefont {Parcollet}},\ and\
  \bibinfo {author} {\bibfnamefont {A.}~\bibnamefont {Georges}},\ }\bibfield
  {title} {\bibinfo {title} {{Tracking the Footprints of Spin Fluctuations: A
  MultiMethod, MultiMessenger Study of the Two-Dimensional Hubbard Model}},\
  }\href {https://doi.org/10.1103/PhysRevX.11.011058} {\bibfield  {journal}
  {\bibinfo  {journal} {Phys. Rev. X}\ }\textbf {\bibinfo {volume} {11}},\
  \bibinfo {pages} {011058} (\bibinfo {year} {2021})}\BibitemShut {NoStop}%
\bibitem [{\citenamefont {Sch\"afer}\ \emph {et~al.}(2015)\citenamefont
  {Sch\"afer}, \citenamefont {Geles}, \citenamefont {Rost}, \citenamefont
  {Rohringer}, \citenamefont {Arrigoni}, \citenamefont {Held}, \citenamefont
  {Bl\"umer}, \citenamefont {Aichhorn},\ and\ \citenamefont
  {Toschi}}]{Schaefer2015}%
  \BibitemOpen
  \bibfield  {author} {\bibinfo {author} {\bibfnamefont {T.}~\bibnamefont
  {Sch\"afer}}, \bibinfo {author} {\bibfnamefont {F.}~\bibnamefont {Geles}},
  \bibinfo {author} {\bibfnamefont {D.}~\bibnamefont {Rost}}, \bibinfo {author}
  {\bibfnamefont {G.}~\bibnamefont {Rohringer}}, \bibinfo {author}
  {\bibfnamefont {E.}~\bibnamefont {Arrigoni}}, \bibinfo {author}
  {\bibfnamefont {K.}~\bibnamefont {Held}}, \bibinfo {author} {\bibfnamefont
  {N.}~\bibnamefont {Bl\"umer}}, \bibinfo {author} {\bibfnamefont
  {M.}~\bibnamefont {Aichhorn}},\ and\ \bibinfo {author} {\bibfnamefont
  {A.}~\bibnamefont {Toschi}},\ }\bibfield  {title} {\bibinfo {title} {{Fate of
  the false Mott-Hubbard transition in two dimensions}},\ }\href
  {https://doi.org/10.1103/PhysRevB.91.125109} {\bibfield  {journal} {\bibinfo
  {journal} {Phys. Rev. B}\ }\textbf {\bibinfo {volume} {91}},\ \bibinfo
  {pages} {125109} (\bibinfo {year} {2015})}\BibitemShut {NoStop}%
\bibitem [{\citenamefont {Rohringer}\ \emph {et~al.}(2011)\citenamefont
  {Rohringer}, \citenamefont {Toschi}, \citenamefont {Katanin},\ and\
  \citenamefont {Held}}]{Rohringer2011}%
  \BibitemOpen
  \bibfield  {author} {\bibinfo {author} {\bibfnamefont {G.}~\bibnamefont
  {Rohringer}}, \bibinfo {author} {\bibfnamefont {A.}~\bibnamefont {Toschi}},
  \bibinfo {author} {\bibfnamefont {A.}~\bibnamefont {Katanin}},\ and\ \bibinfo
  {author} {\bibfnamefont {K.}~\bibnamefont {Held}},\ }\bibfield  {title}
  {\bibinfo {title} {{Critical Properties of the Half-Filled Hubbard Model in
  Three Dimensions}},\ }\href {https://doi.org/10.1103/PhysRevLett.107.256402}
  {\bibfield  {journal} {\bibinfo  {journal} {Phys. Rev. Lett.}\ }\textbf
  {\bibinfo {volume} {107}},\ \bibinfo {pages} {256402} (\bibinfo {year}
  {2011})}\BibitemShut {NoStop}%
\bibitem [{\citenamefont {Sch\"afer}\ \emph {et~al.}(2017)\citenamefont
  {Sch\"afer}, \citenamefont {Katanin}, \citenamefont {Held},\ and\
  \citenamefont {Toschi}}]{Schaefer2017}%
  \BibitemOpen
  \bibfield  {author} {\bibinfo {author} {\bibfnamefont {T.}~\bibnamefont
  {Sch\"afer}}, \bibinfo {author} {\bibfnamefont {A.~A.}\ \bibnamefont
  {Katanin}}, \bibinfo {author} {\bibfnamefont {K.}~\bibnamefont {Held}},\ and\
  \bibinfo {author} {\bibfnamefont {A.}~\bibnamefont {Toschi}},\ }\bibfield
  {title} {\bibinfo {title} {{Interplay of Correlations and Kohn Anomalies in
  Three Dimensions: Quantum Criticality with a Twist}},\ }\href
  {https://doi.org/10.1103/PhysRevLett.119.046402} {\bibfield  {journal}
  {\bibinfo  {journal} {Phys. Rev. Lett.}\ }\textbf {\bibinfo {volume} {119}},\
  \bibinfo {pages} {046402} (\bibinfo {year} {2017})}\BibitemShut {NoStop}%
\bibitem [{\citenamefont {Sch\"afer}\ \emph {et~al.}(2019)\citenamefont
  {Sch\"afer}, \citenamefont {Katanin}, \citenamefont {Kitatani}, \citenamefont
  {Toschi},\ and\ \citenamefont {Held}}]{Schaefer2019}%
  \BibitemOpen
  \bibfield  {author} {\bibinfo {author} {\bibfnamefont {T.}~\bibnamefont
  {Sch\"afer}}, \bibinfo {author} {\bibfnamefont {A.~A.}\ \bibnamefont
  {Katanin}}, \bibinfo {author} {\bibfnamefont {M.}~\bibnamefont {Kitatani}},
  \bibinfo {author} {\bibfnamefont {A.}~\bibnamefont {Toschi}},\ and\ \bibinfo
  {author} {\bibfnamefont {K.}~\bibnamefont {Held}},\ }\bibfield  {title}
  {\bibinfo {title} {{Quantum Criticality in the Two-Dimensional Periodic
  Anderson Model}},\ }\href {https://doi.org/10.1103/PhysRevLett.122.227201}
  {\bibfield  {journal} {\bibinfo  {journal} {Phys. Rev. Lett.}\ }\textbf
  {\bibinfo {volume} {122}},\ \bibinfo {pages} {227201} (\bibinfo {year}
  {2019})}\BibitemShut {NoStop}%
\bibitem [{\citenamefont {Kitatani}\ \emph {et~al.}(2019)\citenamefont
  {Kitatani}, \citenamefont {Sch\"afer}, \citenamefont {Aoki},\ and\
  \citenamefont {Held}}]{Kitatani2018}%
  \BibitemOpen
  \bibfield  {author} {\bibinfo {author} {\bibfnamefont {M.}~\bibnamefont
  {Kitatani}}, \bibinfo {author} {\bibfnamefont {T.}~\bibnamefont {Sch\"afer}},
  \bibinfo {author} {\bibfnamefont {H.}~\bibnamefont {Aoki}},\ and\ \bibinfo
  {author} {\bibfnamefont {K.}~\bibnamefont {Held}},\ }\bibfield  {title}
  {\bibinfo {title} {Why the critical temperature of high-${T}_{c}$ cuprate
  superconductors is so low: The importance of the dynamical vertex
  structure},\ }\href {https://doi.org/10.1103/PhysRevB.99.041115} {\bibfield
  {journal} {\bibinfo  {journal} {Phys. Rev. B}\ }\textbf {\bibinfo {volume}
  {99}},\ \bibinfo {pages} {041115} (\bibinfo {year} {2019})}\BibitemShut
  {NoStop}%
\bibitem [{\citenamefont {Ortiz}\ \emph {et~al.}(2022)\citenamefont {Ortiz},
  \citenamefont {Puphal}, \citenamefont {Klett}, \citenamefont {Hotz},
  \citenamefont {Kremer}, \citenamefont {Trepka}, \citenamefont {Hemmida},
  \citenamefont {von Nidda}, \citenamefont {Isobe}, \citenamefont {Khasanov},
  \citenamefont {Luetkens}, \citenamefont {Hansmann}, \citenamefont {Keimer},
  \citenamefont {Sch\"afer},\ and\ \citenamefont {Hepting}}]{Ortiz2022}%
  \BibitemOpen
  \bibfield  {author} {\bibinfo {author} {\bibfnamefont {R.~A.}\ \bibnamefont
  {Ortiz}}, \bibinfo {author} {\bibfnamefont {P.}~\bibnamefont {Puphal}},
  \bibinfo {author} {\bibfnamefont {M.}~\bibnamefont {Klett}}, \bibinfo
  {author} {\bibfnamefont {F.}~\bibnamefont {Hotz}}, \bibinfo {author}
  {\bibfnamefont {R.~K.}\ \bibnamefont {Kremer}}, \bibinfo {author}
  {\bibfnamefont {H.}~\bibnamefont {Trepka}}, \bibinfo {author} {\bibfnamefont
  {M.}~\bibnamefont {Hemmida}}, \bibinfo {author} {\bibfnamefont {H.-A.~K.}\
  \bibnamefont {von Nidda}}, \bibinfo {author} {\bibfnamefont {M.}~\bibnamefont
  {Isobe}}, \bibinfo {author} {\bibfnamefont {R.}~\bibnamefont {Khasanov}},
  \bibinfo {author} {\bibfnamefont {H.}~\bibnamefont {Luetkens}}, \bibinfo
  {author} {\bibfnamefont {P.}~\bibnamefont {Hansmann}}, \bibinfo {author}
  {\bibfnamefont {B.}~\bibnamefont {Keimer}}, \bibinfo {author} {\bibfnamefont
  {T.}~\bibnamefont {Sch\"afer}},\ and\ \bibinfo {author} {\bibfnamefont
  {M.}~\bibnamefont {Hepting}},\ }\bibfield  {title} {\bibinfo {title}
  {{Magnetic correlations in infinite-layer nickelates: An experimental and
  theoretical multimethod study}},\ }\href
  {https://doi.org/10.1103/PhysRevResearch.4.023093} {\bibfield  {journal}
  {\bibinfo  {journal} {Phys. Rev. Research}\ }\textbf {\bibinfo {volume}
  {4}},\ \bibinfo {pages} {023093} (\bibinfo {year} {2022})}\BibitemShut
  {NoStop}%
\bibitem [{\citenamefont {Klett}\ \emph {et~al.}(2022)\citenamefont {Klett},
  \citenamefont {Hansmann},\ and\ \citenamefont {Schäfer}}]{Klett2022}%
  \BibitemOpen
  \bibfield  {author} {\bibinfo {author} {\bibfnamefont {M.}~\bibnamefont
  {Klett}}, \bibinfo {author} {\bibfnamefont {P.}~\bibnamefont {Hansmann}},\
  and\ \bibinfo {author} {\bibfnamefont {T.}~\bibnamefont {Schäfer}},\
  }\bibfield  {title} {\bibinfo {title} {Magnetic properties and pseudogap
  formation in infinite-layer nickelates: Insights from the single-band hubbard
  model},\ }\href {https://doi.org/10.3389/fphy.2022.834682} {\bibfield
  {journal} {\bibinfo  {journal} {Frontiers in Physics}\ }\textbf {\bibinfo
  {volume} {10}},\ \bibinfo {pages} {834682} (\bibinfo {year}
  {2022})}\BibitemShut {NoStop}%
\bibitem [{\citenamefont {Kitatani}\ \emph {et~al.}(2020)\citenamefont
  {Kitatani}, \citenamefont {Si}, \citenamefont {Janson}, \citenamefont
  {Arita}, \citenamefont {Zhong},\ and\ \citenamefont {Held}}]{Kitatani2020}%
  \BibitemOpen
  \bibfield  {author} {\bibinfo {author} {\bibfnamefont {M.}~\bibnamefont
  {Kitatani}}, \bibinfo {author} {\bibfnamefont {L.}~\bibnamefont {Si}},
  \bibinfo {author} {\bibfnamefont {O.}~\bibnamefont {Janson}}, \bibinfo
  {author} {\bibfnamefont {R.}~\bibnamefont {Arita}}, \bibinfo {author}
  {\bibfnamefont {Z.}~\bibnamefont {Zhong}},\ and\ \bibinfo {author}
  {\bibfnamefont {K.}~\bibnamefont {Held}},\ }\bibfield  {title} {\bibinfo
  {title} {{Nickelate superconductors{\textemdash}a renaissance of the one-band
  Hubbard model}},\ }\bibfield  {journal} {\bibinfo  {journal} {npj Quantum
  Materials}\ }\textbf {\bibinfo {volume} {5}},\ \href
  {https://doi.org/10.1038/s41535-020-00260-y} {10.1038/s41535-020-00260-y}
  (\bibinfo {year} {2020})\BibitemShut {NoStop}%
\bibitem [{\citenamefont {Held}\ \emph {et~al.}(2022)\citenamefont {Held},
  \citenamefont {Si}, \citenamefont {Worm}, \citenamefont {Janson},
  \citenamefont {Arita}, \citenamefont {Zhong}, \citenamefont {Tomczak},\ and\
  \citenamefont {Kitatani}}]{Held2022}%
  \BibitemOpen
  \bibfield  {author} {\bibinfo {author} {\bibfnamefont {K.}~\bibnamefont
  {Held}}, \bibinfo {author} {\bibfnamefont {L.}~\bibnamefont {Si}}, \bibinfo
  {author} {\bibfnamefont {P.}~\bibnamefont {Worm}}, \bibinfo {author}
  {\bibfnamefont {O.}~\bibnamefont {Janson}}, \bibinfo {author} {\bibfnamefont
  {R.}~\bibnamefont {Arita}}, \bibinfo {author} {\bibfnamefont
  {Z.}~\bibnamefont {Zhong}}, \bibinfo {author} {\bibfnamefont {J.~M.}\
  \bibnamefont {Tomczak}},\ and\ \bibinfo {author} {\bibfnamefont
  {M.}~\bibnamefont {Kitatani}},\ }\bibfield  {title} {\bibinfo {title} {{Phase
  Diagram of Nickelate Superconductors Calculated by Dynamical Vertex
  Approximation}},\ }\href {https://doi.org/10.3389%2Ffphy.2021.810394}
  {\bibfield  {journal} {\bibinfo  {journal} {Frontiers in Physics}\ }\textbf
  {\bibinfo {volume} {9}} (\bibinfo {year} {2022})}\BibitemShut {NoStop}%
\bibitem [{\citenamefont {Kitatani}\ \emph {et~al.}(2023)\citenamefont
  {Kitatani}, \citenamefont {Si}, \citenamefont {Worm}, \citenamefont
  {Tomczak}, \citenamefont {Arita},\ and\ \citenamefont {Held}}]{Kitatani2023}%
  \BibitemOpen
  \bibfield  {author} {\bibinfo {author} {\bibfnamefont {M.}~\bibnamefont
  {Kitatani}}, \bibinfo {author} {\bibfnamefont {L.}~\bibnamefont {Si}},
  \bibinfo {author} {\bibfnamefont {P.}~\bibnamefont {Worm}}, \bibinfo {author}
  {\bibfnamefont {J.~M.}\ \bibnamefont {Tomczak}}, \bibinfo {author}
  {\bibfnamefont {R.}~\bibnamefont {Arita}},\ and\ \bibinfo {author}
  {\bibfnamefont {K.}~\bibnamefont {Held}},\ }\bibfield  {title} {\bibinfo
  {title} {Optimizing superconductivity: From cuprates via nickelates to
  palladates},\ }\href {https://doi.org/10.1103/PhysRevLett.130.166002}
  {\bibfield  {journal} {\bibinfo  {journal} {Phys. Rev. Lett.}\ }\textbf
  {\bibinfo {volume} {130}},\ \bibinfo {pages} {166002} (\bibinfo {year}
  {2023})}\BibitemShut {NoStop}%
\bibitem [{\citenamefont {Rohringer}\ and\ \citenamefont
  {Toschi}(2016)}]{Rohringer2016}%
  \BibitemOpen
  \bibfield  {author} {\bibinfo {author} {\bibfnamefont {G.}~\bibnamefont
  {Rohringer}}\ and\ \bibinfo {author} {\bibfnamefont {A.}~\bibnamefont
  {Toschi}},\ }\bibfield  {title} {\bibinfo {title} {Impact of nonlocal
  correlations over different energy scales: A dynamical vertex approximation
  study},\ }\href {https://doi.org/10.1103%2Fphysrevb.94.125144} {\bibfield
  {journal} {\bibinfo  {journal} {Physical Review B}\ }\textbf {\bibinfo
  {volume} {94}} (\bibinfo {year} {2016})}\BibitemShut {NoStop}%
\bibitem [{\citenamefont {Ayral}\ and\ \citenamefont
  {Parcollet}(2015)}]{Ayral2015}%
  \BibitemOpen
  \bibfield  {author} {\bibinfo {author} {\bibfnamefont {T.}~\bibnamefont
  {Ayral}}\ and\ \bibinfo {author} {\bibfnamefont {O.}~\bibnamefont
  {Parcollet}},\ }\bibfield  {title} {\bibinfo {title} {Mott physics and spin
  fluctuations: A unified framework},\ }\href
  {https://doi.org/10.1103/PhysRevB.92.115109} {\bibfield  {journal} {\bibinfo
  {journal} {Phys. Rev. B}\ }\textbf {\bibinfo {volume} {92}},\ \bibinfo
  {pages} {115109} (\bibinfo {year} {2015})}\BibitemShut {NoStop}%
\bibitem [{\citenamefont {Ayral}\ and\ \citenamefont
  {Parcollet}(2016)}]{Ayral2016}%
  \BibitemOpen
  \bibfield  {author} {\bibinfo {author} {\bibfnamefont {T.}~\bibnamefont
  {Ayral}}\ and\ \bibinfo {author} {\bibfnamefont {O.}~\bibnamefont
  {Parcollet}},\ }\bibfield  {title} {\bibinfo {title} {Mott physics and spin
  fluctuations: A functional viewpoint},\ }\href
  {https://doi.org/10.1103/PhysRevB.93.235124} {\bibfield  {journal} {\bibinfo
  {journal} {Phys. Rev. B}\ }\textbf {\bibinfo {volume} {93}},\ \bibinfo
  {pages} {235124} (\bibinfo {year} {2016})}\BibitemShut {NoStop}%
\bibitem [{\citenamefont {Moriya}(1991)}]{MORIYA1991}%
  \BibitemOpen
  \bibfield  {author} {\bibinfo {author} {\bibfnamefont {T.}~\bibnamefont
  {Moriya}},\ }\bibfield  {title} {\bibinfo {title} {Theory of itinerant
  electron magnetism},\ }\href
  {https://doi.org/https://doi.org/10.1016/0304-8853(91)90824-T} {\bibfield
  {journal} {\bibinfo  {journal} {Journal of Magnetism and Magnetic Materials}\
  }\textbf {\bibinfo {volume} {100}},\ \bibinfo {pages} {261} (\bibinfo {year}
  {1991})}\BibitemShut {NoStop}%
\bibitem [{\citenamefont {Stobbe}\ and\ \citenamefont
  {Rohringer}(2022)}]{Stobbe2022}%
  \BibitemOpen
  \bibfield  {author} {\bibinfo {author} {\bibfnamefont {J.}~\bibnamefont
  {Stobbe}}\ and\ \bibinfo {author} {\bibfnamefont {G.}~\bibnamefont
  {Rohringer}},\ }\bibfield  {title} {\bibinfo {title} {Consistency of
  potential energy in the dynamical vertex approximation},\ }\href
  {https://doi.org/10.1103/PhysRevB.106.205101} {\bibfield  {journal} {\bibinfo
   {journal} {Phys. Rev. B}\ }\textbf {\bibinfo {volume} {106}},\ \bibinfo
  {pages} {205101} (\bibinfo {year} {2022})}\BibitemShut {NoStop}%
\bibitem [{\citenamefont {Del~Re}\ \emph {et~al.}(2019)\citenamefont {Del~Re},
  \citenamefont {Capone},\ and\ \citenamefont {Toschi}}]{Del_Re2019}%
  \BibitemOpen
  \bibfield  {author} {\bibinfo {author} {\bibfnamefont {L.}~\bibnamefont
  {Del~Re}}, \bibinfo {author} {\bibfnamefont {M.}~\bibnamefont {Capone}},\
  and\ \bibinfo {author} {\bibfnamefont {A.}~\bibnamefont {Toschi}},\
  }\bibfield  {title} {\bibinfo {title} {{Dynamical vertex approximation for
  the attractive Hubbard model}},\ }\href
  {https://doi.org/10.1103/PhysRevB.99.045137} {\bibfield  {journal} {\bibinfo
  {journal} {Phys. Rev. B}\ }\textbf {\bibinfo {volume} {99}},\ \bibinfo
  {pages} {045137} (\bibinfo {year} {2019})}\BibitemShut {NoStop}%
\bibitem [{\citenamefont {Mermin}\ and\ \citenamefont
  {Wagner}(1966)}]{Mermin1966}%
  \BibitemOpen
  \bibfield  {author} {\bibinfo {author} {\bibfnamefont {N.~D.}\ \bibnamefont
  {Mermin}}\ and\ \bibinfo {author} {\bibfnamefont {H.}~\bibnamefont
  {Wagner}},\ }\bibfield  {title} {\bibinfo {title} {{Absence of Ferromagnetism
  or Antiferromagnetism in One- or Two-Dimensional Isotropic Heisenberg
  Models}},\ }\href {https://doi.org/10.1103/PhysRevLett.17.1307} {\bibfield
  {journal} {\bibinfo  {journal} {Phys. Rev. Lett.}\ }\textbf {\bibinfo
  {volume} {17}},\ \bibinfo {pages} {1307} (\bibinfo {year}
  {1966})}\BibitemShut {NoStop}%
\bibitem [{\citenamefont {Damascelli}\ \emph {et~al.}(2003)\citenamefont
  {Damascelli}, \citenamefont {Hussain},\ and\ \citenamefont
  {Shen}}]{Damascelli2003}%
  \BibitemOpen
  \bibfield  {author} {\bibinfo {author} {\bibfnamefont {A.}~\bibnamefont
  {Damascelli}}, \bibinfo {author} {\bibfnamefont {Z.}~\bibnamefont
  {Hussain}},\ and\ \bibinfo {author} {\bibfnamefont {Z.-X.}\ \bibnamefont
  {Shen}},\ }\bibfield  {title} {\bibinfo {title} {Angle-resolved photoemission
  studies of the cuprate superconductors},\ }\href
  {https://doi.org/10.1103/RevModPhys.75.473} {\bibfield  {journal} {\bibinfo
  {journal} {Rev. Mod. Phys.}\ }\textbf {\bibinfo {volume} {75}},\ \bibinfo
  {pages} {473} (\bibinfo {year} {2003})}\BibitemShut {NoStop}%
\bibitem [{\citenamefont {Ding}\ \emph {et~al.}(1996)\citenamefont {Ding},
  \citenamefont {Norman}, \citenamefont {Campuzano}, \citenamefont {Randeria},
  \citenamefont {Bellman}, \citenamefont {Yokoya}, \citenamefont {Takahashi},
  \citenamefont {Mochiku},\ and\ \citenamefont {Kadowaki}}]{Ding1996}%
  \BibitemOpen
  \bibfield  {author} {\bibinfo {author} {\bibfnamefont {H.}~\bibnamefont
  {Ding}}, \bibinfo {author} {\bibfnamefont {M.~R.}\ \bibnamefont {Norman}},
  \bibinfo {author} {\bibfnamefont {J.~C.}\ \bibnamefont {Campuzano}}, \bibinfo
  {author} {\bibfnamefont {M.}~\bibnamefont {Randeria}}, \bibinfo {author}
  {\bibfnamefont {A.~F.}\ \bibnamefont {Bellman}}, \bibinfo {author}
  {\bibfnamefont {T.}~\bibnamefont {Yokoya}}, \bibinfo {author} {\bibfnamefont
  {T.}~\bibnamefont {Takahashi}}, \bibinfo {author} {\bibfnamefont
  {T.}~\bibnamefont {Mochiku}},\ and\ \bibinfo {author} {\bibfnamefont
  {K.}~\bibnamefont {Kadowaki}},\ }\bibfield  {title} {\bibinfo {title}
  {{Angle-resolved photoemission spectroscopy study of the superconducting gap
  anisotropy in
  ${\mathrm{Bi}}_{2}{\mathrm{Sr}}_{2}\mathrm{Ca}{\mathrm{Cu}}_{2}{\mathrm{O}}_{8+x}$}},\
  }\href {https://doi.org/10.1103/PhysRevB.54.R9678} {\bibfield  {journal}
  {\bibinfo  {journal} {Phys. Rev. B}\ }\textbf {\bibinfo {volume} {54}},\
  \bibinfo {pages} {R9678} (\bibinfo {year} {1996})}\BibitemShut {NoStop}%
\bibitem [{\citenamefont {Marshall}\ \emph {et~al.}(1996)\citenamefont
  {Marshall}, \citenamefont {Dessau}, \citenamefont {Loeser}, \citenamefont
  {Park}, \citenamefont {Matsuura}, \citenamefont {Eckstein}, \citenamefont
  {Bozovic}, \citenamefont {Fournier}, \citenamefont {Kapitulnik},
  \citenamefont {Spicer},\ and\ \citenamefont {Shen}}]{Marshall1996}%
  \BibitemOpen
  \bibfield  {author} {\bibinfo {author} {\bibfnamefont {D.~S.}\ \bibnamefont
  {Marshall}}, \bibinfo {author} {\bibfnamefont {D.~S.}\ \bibnamefont
  {Dessau}}, \bibinfo {author} {\bibfnamefont {A.~G.}\ \bibnamefont {Loeser}},
  \bibinfo {author} {\bibfnamefont {C.-H.}\ \bibnamefont {Park}}, \bibinfo
  {author} {\bibfnamefont {A.~Y.}\ \bibnamefont {Matsuura}}, \bibinfo {author}
  {\bibfnamefont {J.~N.}\ \bibnamefont {Eckstein}}, \bibinfo {author}
  {\bibfnamefont {I.}~\bibnamefont {Bozovic}}, \bibinfo {author} {\bibfnamefont
  {P.}~\bibnamefont {Fournier}}, \bibinfo {author} {\bibfnamefont
  {A.}~\bibnamefont {Kapitulnik}}, \bibinfo {author} {\bibfnamefont {W.~E.}\
  \bibnamefont {Spicer}},\ and\ \bibinfo {author} {\bibfnamefont {Z.-X.}\
  \bibnamefont {Shen}},\ }\bibfield  {title} {\bibinfo {title} {{Unconventional
  Electronic Structure Evolution with Hole Doping in
  ${\mathrm{Bi}}_{2}{\mathrm{Sr}}_{2}{\mathrm{CaCu}}_{2}{O}_{8+\ensuremath{\delta}}$:
  Angle-Resolved Photoemission Results}},\ }\href
  {https://doi.org/10.1103/PhysRevLett.76.4841} {\bibfield  {journal} {\bibinfo
   {journal} {Phys. Rev. Lett.}\ }\textbf {\bibinfo {volume} {76}},\ \bibinfo
  {pages} {4841} (\bibinfo {year} {1996})}\BibitemShut {NoStop}%
\bibitem [{\citenamefont {{Kanigel}}\ \emph {et~al.}(2006)\citenamefont
  {{Kanigel}}, \citenamefont {{Norman}}, \citenamefont {{Randeria}},
  \citenamefont {{Chatterjee}}, \citenamefont {{Souma}}, \citenamefont
  {{Kaminski}}, \citenamefont {{Fretwell}}, \citenamefont {{Rosenkranz}},
  \citenamefont {{Shi}}, \citenamefont {{Sato}}, \citenamefont {{Takahashi}},
  \citenamefont {{Li}}, \citenamefont {{Raffy}}, \citenamefont {{Kadowaki}},
  \citenamefont {{Hinks}}, \citenamefont {{Ozyuzer}},\ and\ \citenamefont
  {{Campuzano}}}]{Kanigel2006}%
  \BibitemOpen
  \bibfield  {author} {\bibinfo {author} {\bibfnamefont {A.}~\bibnamefont
  {{Kanigel}}}, \bibinfo {author} {\bibfnamefont {M.~R.}\ \bibnamefont
  {{Norman}}}, \bibinfo {author} {\bibfnamefont {M.}~\bibnamefont
  {{Randeria}}}, \bibinfo {author} {\bibfnamefont {U.}~\bibnamefont
  {{Chatterjee}}}, \bibinfo {author} {\bibfnamefont {S.}~\bibnamefont
  {{Souma}}}, \bibinfo {author} {\bibfnamefont {A.}~\bibnamefont {{Kaminski}}},
  \bibinfo {author} {\bibfnamefont {H.~M.}\ \bibnamefont {{Fretwell}}},
  \bibinfo {author} {\bibfnamefont {S.}~\bibnamefont {{Rosenkranz}}}, \bibinfo
  {author} {\bibfnamefont {M.}~\bibnamefont {{Shi}}}, \bibinfo {author}
  {\bibfnamefont {T.}~\bibnamefont {{Sato}}}, \bibinfo {author} {\bibfnamefont
  {T.}~\bibnamefont {{Takahashi}}}, \bibinfo {author} {\bibfnamefont {Z.~Z.}\
  \bibnamefont {{Li}}}, \bibinfo {author} {\bibfnamefont {H.}~\bibnamefont
  {{Raffy}}}, \bibinfo {author} {\bibfnamefont {K.}~\bibnamefont {{Kadowaki}}},
  \bibinfo {author} {\bibfnamefont {D.}~\bibnamefont {{Hinks}}}, \bibinfo
  {author} {\bibfnamefont {L.}~\bibnamefont {{Ozyuzer}}},\ and\ \bibinfo
  {author} {\bibfnamefont {J.~C.}\ \bibnamefont {{Campuzano}}},\ }\bibfield
  {title} {\bibinfo {title} {{Evolution of the pseudogap from Fermi arcs to the
  nodal liquid}},\ }\href {https://doi.org/10.1038/nphys334} {\bibfield
  {journal} {\bibinfo  {journal} {Nature Physics}\ }\textbf {\bibinfo {volume}
  {2}},\ \bibinfo {pages} {447} (\bibinfo {year} {2006})}\BibitemShut {NoStop}%
\bibitem [{\citenamefont {Yoshida}\ \emph {et~al.}(2006)\citenamefont
  {Yoshida}, \citenamefont {Zhou}, \citenamefont {Tanaka}, \citenamefont
  {Yang}, \citenamefont {Hussain}, \citenamefont {Shen}, \citenamefont
  {Fujimori}, \citenamefont {Sahrakorpi}, \citenamefont {Lindroos},
  \citenamefont {Markiewicz}, \citenamefont {Bansil}, \citenamefont {Komiya},
  \citenamefont {Ando}, \citenamefont {Eisaki}, \citenamefont {Kakeshita},\
  and\ \citenamefont {Uchida}}]{Yoshida2006}%
  \BibitemOpen
  \bibfield  {author} {\bibinfo {author} {\bibfnamefont {T.}~\bibnamefont
  {Yoshida}}, \bibinfo {author} {\bibfnamefont {X.~J.}\ \bibnamefont {Zhou}},
  \bibinfo {author} {\bibfnamefont {K.}~\bibnamefont {Tanaka}}, \bibinfo
  {author} {\bibfnamefont {W.~L.}\ \bibnamefont {Yang}}, \bibinfo {author}
  {\bibfnamefont {Z.}~\bibnamefont {Hussain}}, \bibinfo {author} {\bibfnamefont
  {Z.-X.}\ \bibnamefont {Shen}}, \bibinfo {author} {\bibfnamefont
  {A.}~\bibnamefont {Fujimori}}, \bibinfo {author} {\bibfnamefont
  {S.}~\bibnamefont {Sahrakorpi}}, \bibinfo {author} {\bibfnamefont
  {M.}~\bibnamefont {Lindroos}}, \bibinfo {author} {\bibfnamefont {R.~S.}\
  \bibnamefont {Markiewicz}}, \bibinfo {author} {\bibfnamefont
  {A.}~\bibnamefont {Bansil}}, \bibinfo {author} {\bibfnamefont
  {S.}~\bibnamefont {Komiya}}, \bibinfo {author} {\bibfnamefont
  {Y.}~\bibnamefont {Ando}}, \bibinfo {author} {\bibfnamefont {H.}~\bibnamefont
  {Eisaki}}, \bibinfo {author} {\bibfnamefont {T.}~\bibnamefont {Kakeshita}},\
  and\ \bibinfo {author} {\bibfnamefont {S.}~\bibnamefont {Uchida}},\
  }\bibfield  {title} {\bibinfo {title} {{Systematic doping evolution of the
  underlying Fermi surface of
  ${\mathrm{La}}_{2\ensuremath{-}x}{\mathrm{Sr}}_{x}\mathrm{Cu}{\mathrm{O}}_{4}$}},\
  }\href {https://doi.org/10.1103/PhysRevB.74.224510} {\bibfield  {journal}
  {\bibinfo  {journal} {Phys. Rev. B}\ }\textbf {\bibinfo {volume} {74}},\
  \bibinfo {pages} {224510} (\bibinfo {year} {2006})}\BibitemShut {NoStop}%
\bibitem [{\citenamefont {Shen}\ \emph {et~al.}(2005)\citenamefont {Shen},
  \citenamefont {Ronning}, \citenamefont {Lu}, \citenamefont {Baumberger},
  \citenamefont {Ingle}, \citenamefont {Lee}, \citenamefont {Meevasana},
  \citenamefont {Kohsaka}, \citenamefont {Azuma}, \citenamefont {Takano},
  \citenamefont {Takagi},\ and\ \citenamefont {Shen}}]{Shen2005}%
  \BibitemOpen
  \bibfield  {author} {\bibinfo {author} {\bibfnamefont {K.~M.}\ \bibnamefont
  {Shen}}, \bibinfo {author} {\bibfnamefont {F.}~\bibnamefont {Ronning}},
  \bibinfo {author} {\bibfnamefont {D.~H.}\ \bibnamefont {Lu}}, \bibinfo
  {author} {\bibfnamefont {F.}~\bibnamefont {Baumberger}}, \bibinfo {author}
  {\bibfnamefont {N.~J.~C.}\ \bibnamefont {Ingle}}, \bibinfo {author}
  {\bibfnamefont {W.~S.}\ \bibnamefont {Lee}}, \bibinfo {author} {\bibfnamefont
  {W.}~\bibnamefont {Meevasana}}, \bibinfo {author} {\bibfnamefont
  {Y.}~\bibnamefont {Kohsaka}}, \bibinfo {author} {\bibfnamefont
  {M.}~\bibnamefont {Azuma}}, \bibinfo {author} {\bibfnamefont
  {M.}~\bibnamefont {Takano}}, \bibinfo {author} {\bibfnamefont
  {H.}~\bibnamefont {Takagi}},\ and\ \bibinfo {author} {\bibfnamefont {Z.-X.}\
  \bibnamefont {Shen}},\ }\bibfield  {title} {\bibinfo {title} {{Nodal
  Quasiparticles and Antinodal Charge Ordering in
  Ca$_{2-x}$Na$_x$CuO$_2$Cl$_2$}},\ }\href
  {https://doi.org/10.1126/science.1103627} {\bibfield  {journal} {\bibinfo
  {journal} {Science}\ }\textbf {\bibinfo {volume} {307}},\ \bibinfo {pages}
  {901} (\bibinfo {year} {2005})}\BibitemShut {NoStop}%
\bibitem [{\citenamefont {Reymbaut}\ \emph {et~al.}(2017)\citenamefont
  {Reymbaut}, \citenamefont {Gagnon}, \citenamefont {Bergeron},\ and\
  \citenamefont {Tremblay}}]{Reymbaut2017}%
  \BibitemOpen
  \bibfield  {author} {\bibinfo {author} {\bibfnamefont {A.}~\bibnamefont
  {Reymbaut}}, \bibinfo {author} {\bibfnamefont {A.-M.}\ \bibnamefont
  {Gagnon}}, \bibinfo {author} {\bibfnamefont {D.}~\bibnamefont {Bergeron}},\
  and\ \bibinfo {author} {\bibfnamefont {A.-M.~S.}\ \bibnamefont {Tremblay}},\
  }\bibfield  {title} {\bibinfo {title} {Maximum entropy analytic continuation
  for frequency-dependent transport coefficients with nonpositive spectral
  weight},\ }\href {https://doi.org/10.1103/PhysRevB.95.121104} {\bibfield
  {journal} {\bibinfo  {journal} {Phys. Rev. B}\ }\textbf {\bibinfo {volume}
  {95}},\ \bibinfo {pages} {121104} (\bibinfo {year} {2017})}\BibitemShut
  {NoStop}%
\bibitem [{\citenamefont {Altland}\ and\ \citenamefont
  {Simons}(2008)}]{Altland2008}%
  \BibitemOpen
  \bibfield  {author} {\bibinfo {author} {\bibfnamefont {A.}~\bibnamefont
  {Altland}}\ and\ \bibinfo {author} {\bibfnamefont {B.}~\bibnamefont
  {Simons}},\ }\href@noop {} {\emph {\bibinfo {title} {Condensed Matter Field
  Theory}}}\ (\bibinfo  {publisher} {Cambridge University Press},\ \bibinfo
  {year} {2008})\BibitemShut {NoStop}%
\end{thebibliography}%

\end{document}


\title{Rise and Fall of the Pseudogap in the Emery model: Insights for Cuprates\\
\textit{-- Supplemental Material --}}

\author{M.~O.~Malcolms}
\email{m.deoliveira@fkf.mpg.de}
\affiliation{Max-Planck-Institut für Festkörperforschung, Heisenbergstraße 1, 70569 Stuttgart, Germany}

\author{Henri Menke}
\affiliation{Max-Planck-Institut für Festkörperforschung, Heisenbergstraße 1, 70569 Stuttgart, Germany}
\affiliation{Department of Physics, University of Erlangen-N\"{u}rnberg, 91058 Erlangen, Germany}

\author{Yi-Ting Tseng}
\affiliation{Department of Physics, University of Erlangen-N\"{u}rnberg, 91058 Erlangen, Germany}

\author{\\Eric Jacob}
\affiliation{Institute of Solid State Physics, TU Wien, 1040 Vienna, Austria}

\author{Karsten Held}
\affiliation{Institute of Solid State Physics, TU Wien, 1040 Vienna, Austria}

\author{Philipp Hansmann}
\affiliation{Department of Physics, University of Erlangen-N\"{u}rnberg, 91058 Erlangen, Germany}

\author{Thomas~Sch{\"a}fer}
\email{t.schaefer@fkf.mpg.de}
\affiliation{Max-Planck-Institut für Festkörperforschung, Heisenbergstraße 1, 70569 Stuttgart, Germany}

\maketitle

In this Supplemental Material, we provide the technical details on the dynamical mean-field theory (DMFT) applied to the Emery model (Sec.~\ref{sec:dmft_emery}) as well as its extension and main equations of the dynamical vertex approximation (D$\Gamma$A, Sec.~\ref{sec:dga_emery}). In addition, we provide the doping evolution of the low-frequency behavior of Matsubara Green function at the nodal and antinodal point (Sec.~\ref{sec:gf}), spectral function evolution over the Fermi surface (Sec.~\ref{sec:w_specFunc_FS}), local spectral function (Sec.~\ref{sec:spectrum}), the static lattice susceptibility (Sec.~\ref{sec:susc}), and DMFT spectral intensities over the Brillouin zone (Sec.~\ref{sec:dmft_spec}).

\section{DMFT for the Emery model}
\label{sec:dmft_emery}
Dynamical mean-field theory (DMFT) \cite{Metzner1989, Georges1992, Georges1996} is a successful scheme that accounts for many-body correlation effects beyond the static Hartree-Fock mean-field theory and density functional theory (DFT). In DMFT, the interacting lattice problem is mapped onto an interacting impurity coupled to a self-consistent bath. In this approach, the full self-consistent scheme imposes that the local lattice Green function coincides with the impurity one by replacing the lattice self-energy by the impurity self-energy. Such approximation implies that a converged DMFT solution accounts only for all local self-energy diagrams, i.e, all single site temporal (dynamical) fluctuations, while neglecting non-local ones. 

\subsection{Single-particle quantities}
We performed the paramagnetic restricted DMFT calculation of the on-site/impurity quantities using the continuous-time interaction expansion quantum Monte Carlo impurity solver (CT-INT) \cite{Rubtsov2005, Gull2011} implemented in the TRIQS library \cite{TRIQS}. We adopt the following workflow to obtain a converged DMFT solution:

1. We initialize the Weiss field from a non-interacting reference point $\left[\Sigma_{\nu}^{\left(j=0\right)}=0\right]$
\begin{equation}
\bar{\mathcal{G}}^{(j=0)}\left(i\nu\right)=\int\frac{d\mathbf{k}}{V_{BZ}}\left[i\nu-\bar{h}_{0}\left(\mathbf{k}\right)\right]^{-1},
\end{equation}
where $\nu=\left(2n+1\right)\pi/\beta$ with $n\,\epsilon\,\mathbb{Z}$ denotes the fermionic Matsubara frequencies, $\bar{h}_{0}\left(\mathbf{k}\right)$ is the non-interacting Hamiltonian, and superscript j denotes the iteration step. Here it is important to emphasize that we express all matrix quantities in the same basis as $\bar{h}_{0}\left(\mathbf{k}\right)$, where the first diagonal element $\left(\left[\bar{\mathcal{G}}^{(j)}\left(i\nu\right)\right]_{00}\right)$ corresponds to the subspace of the interacting orbital. 

2. Since in the three-band Emery model we assume that the Coulomb repulsion acts only in the $d_{x^{2}-y^{2}}$-orbital, for the impurity solver we give as input $\mathcal{G}_{dd}^{(j)}\left(i\nu\right)=\left[\bar{\mathcal{G}}^{(j)}\left(i\nu\right)\right]_{00}$, and as output we get the impurity self-energy $\Sigma_{\text{dd}}^{\left(j+1\right)}\left(i\nu\right)$ that is also restricted to then $d_{x^{2}-y^{2}}$ orbital.

3. By means of the lattice Dyson equation, we compute the new local lattice Green function 
\begin{equation}
    \bar{G}^{\left(j+1\right)}\left(i\nu\right)=\int\frac{d\mathbf{k}}{V_{BZ}}\left[i\nu-\bar{h}_{0}\left(\mathbf{k}\right)-\bar{\Sigma}^{j+1}(i\nu)\right]^{-1},
\end{equation}
where 
\begin{equation}
\bar{\Sigma}^{j+1}(i\nu) = \left(\begin{array}{ccc}
\Sigma_{\text{dd}}^{\left(j+1\right)}\left(i\nu\right) & 0 & 0\\
0 & 0 & 0\\
0 & 0 & 0
\end{array}\right).    
\end{equation}

4. From the local lattice Green function in step 3 and the local self-energy from step 2, we compute via the local Dyson equation the new Weiss field in the subspace of the interacting orbital
\begin{equation}
\mathcal{G}_{\text{dd}}^{(j+1)}\left(i\nu\right)=\left[\left[\bar{G}^{\left(j+1\right)}\left(i\nu\right)\right]_{00}^{-1}+\Sigma_{\text{dd}}^{\left(j+1\right)}\left(i\nu\right)\right]^{-1}.
\end{equation}

5. We repeat the steps 2 to 4 until the convergence is reached.

In this work, we focus only on paramagnetic SU$\left(2\right)$-symmetric solutions. In all the steps described above we omitted the spin index of the Green function, Weiss field and self-energy since we always consider the spin symmetrized quantities, i.e., $G=\left(G_{\uparrow}+G_{\downarrow}\right)/2$.

\subsection{DMFT lattice susceptibility and two-particle local irreducible vertex}

Under the assumption that the local interaction is restricted to the $d_{x^{2}-y^{2}}$-orbital, the local vertex corrections are also restricted to the subspace of this orbital. Then, the DMFT expression for the momentum dependent magnetic/charge susceptibility \cite{Georges1996, Rohringer2013} for the $d_{x^{2}-y^{2}}$ orbital reads 
\begin{equation}
\chi_{\text{dd, m/ch}}^{\textbf{q},\Omega}=T^{2}\sum_{\nu\nu'}\left[\left(\chi_{0,\text{dd}}^{\mathbf{q}\Omega\nu'}\right)^{-1}\delta_{\nu\nu'}-\Gamma_{\text{m/ch},\text{dd}}^{\Omega\nu\nu'}\right]^{-1},
\label{eq:BSE}
\end{equation}
where $\chi_{0,\text{dd}}^{\mathbf{q}\Omega\nu'}$ is the interacting $d_{x^{2}-y^{2}}$ orbital bubble term defined as  
\begin{equation}
\chi_{0,\text{dd}}^{\mathbf{q}\Omega\nu'}=T^{-1}\sum_{\mathbf{k}}G_{\text{dd}}^{\text{DMFT}}\left(\mathbf{k},\nu'\right)G_{\text{dd}}^{\text{DMFT}}\left(\mathbf{k}+\mathbf{q},\nu'+\Omega\right),
\label{eq:dmft_bubble}
\end{equation}
$T$ is the temperature, $\mathbf{k}$, $\mathbf{q}$ denote two-dimensional momentum vectors in the $1^{\text{st}}$ Brillouin zone, and $\nu$, $\nu'$ denote fermionic frequencies while $\Omega$ denotes bosonic ones. Please note that Eq.~(\ref{eq:BSE}) is a matrix inversion. The interacting bubble susceptibility is defined in terms of the one-particle DMFT lattice Green function
\begin{equation}
    \bar{G}^{\text{DMFT}}\left(\mathbf{k},i\nu\right)=\left[i\omega_{n}-\bar{h}_{0}\left(\mathbf{k}\right)-\bar{\Sigma}(i\nu)\right]^{-1},
\label{eq:DMFT_latt_GF}
\end{equation}
restricted to the $d_{x^{2}-y^{2}}$ orbital, i.e., $G_{\text{dd}}^{\text{DMFT}}\left(\mathbf{k},i\nu\right)=\left[\bar{G}\left(\mathbf{k},i\nu\right)\right]_{00}$, where
\begin{equation}
    \bar{\Sigma}(i\nu)=\left(\begin{array}{ccc}
\Sigma_{\text{dd}}\left(i\nu\right) & 0 & 0\\
0 & 0 & 0\\
0 & 0 & 0
\end{array}\right).
\end{equation}
$\Gamma_{\text{dd, m/ch}}^{\Omega\nu\nu'}$ represents the local irreducible vertex of DMFT \cite{Georges1996, Rohringer2013, Rohringer2012} in the magnetic/charge channel for $d_{x^{2}-y^{2}}$ orbital.

In order to obtain the irreducible vertex $\Gamma_{\text{dd,m/ch}}^{\Omega\nu\nu'}$, from a converged DMFT solution, one can compute via the impurity solver (e.g., in our case the continuous-time quantum Monte Carlo) the generalized impurity susceptibility \cite{Rohringer2012}, which explicitly reads
\begin{align}
    \chi_{\text{dd},\sigma\sigma^{'}}^{\Omega\nu\nu'} &=\int_{0}^{\beta}\text{d}\tau_{1}\text{d}\tau_{2}\text{d}\tau_{3}e^{-i\nu\tau_{1}}e^{i(\nu+\Omega)\tau_{2}}e^{-i(\nu'+\Omega)\tau_{3}}\nonumber\\
    &\left[\left\langle T_{\tau}d_{\sigma}(\tau_{1})d_{\sigma}^{\dagger}(\tau_{2})d_{\sigma^{'}}(\tau_{3})d_{\sigma^{'}}^{\dagger}(0)\right\rangle \right.\nonumber\\
    &\left.-\left\langle T_{\tau}d_{\sigma}(\tau_{1})d_{\sigma}^{\dagger}(\tau_{2})\right\rangle \left\langle T_{\tau}d_{\sigma}(\tau_{3})d_{\sigma}^{\dagger}(0)\right\rangle \right],
    \label{eq:generalized_imp_suscep}
\end{align}
where $\tau_{i}$ are imaginary times, $\beta=1/T$ and $d_{\sigma}^{\left(\dagger\right)}$ annihilates (creates) an electron on the $d_{x^{2}-y^{2}}$ impurity site with spin $\sigma=\uparrow,\downarrow$. In the SU$\left(2\right)$-symmetric phase, the impurity magnetic and charge susceptibilities are defined as \cite{Rohringer2012, Rohringer2013}
\begin{equation}
    \chi_{\text{\text{dd},m}}^{\Omega\nu\nu'}=\chi_{\text{dd},\uparrow\uparrow}^{\Omega\nu\nu'}-\chi_{\text{dd},\uparrow\downarrow}^{\Omega\nu\nu'},
    \label{eq:magnetic_imp_suscept}
\end{equation}
\begin{equation}
    \chi_{\text{\text{dd},ch}}^{\Omega\nu\nu'}=\chi_{\text{dd},\uparrow\uparrow}^{\Omega\nu\nu'}+\chi_{\text{dd},\uparrow\downarrow}^{\Omega\nu\nu'}.
    \label{eq:charge_imp_suscept}
\end{equation}
From the inversion of the Bethe Salpeter equation (BSE) for the impurity $d_{x^{2}-y^{2}}$ site \cite{Georges1996,DelRe2021b, Rohringer2012, Rohringer2013}, the irreducible vertex of DMFT in the corresponding channel (magnetic or charge) is given by
\begin{equation}
    \Gamma_{\text{dd,m/ch}}^{\Omega\nu\nu'}=\left[\chi_{\text{dd,m/ch}}^{\Omega\nu\nu'}\right]^{-1}-\left[\chi_{0,\text{dd}}^{\Omega\nu'}\delta_{\nu\nu'}\right]^{-1},
    \label{eq:Irreducible_vextex}
\end{equation}
where $\chi_{0,\text{dd}}^{\Omega\nu'}=T^{-1}G_{\text{dd}}\left(i\nu'\right)G_{\text{dd}}\left(i\nu'+i\Omega\right)$ is the local generalized DMFT dressed bubble term for the $d_{x^{2}-y^{2}}$-orbital with $G_{\text{dd}}\left(i\nu'\right)$ being the local DMFT Green function restricted to the same orbital.

\section{Ladder D$\Gamma$A for the Emery model}
\label{sec:dga_emery}
Strongly correlated electronic systems present a plethora of interesting phenoma, among these, the ones in which the electronic degrees of freedom are confined at low spatial dimensions or in the proximity of phase transitions are the hardest to describe. In these regimes, the electronic correlations emerge simultaneously in space and time, resulting in an interwined frequency and momentum dependence of electronic quantities like the self-energy. The accurate description of such regimes demands approaches beyond the DMFT framework, e.g., cluster extensions \cite{Maier2005} or diagrammatic extensions of DMFT \cite{Rohringer2018}. Although cluster approaches are capable of treating correlations confined to the size of the cluster, diagrammatic extensions are able to incorporate correlations on all length scales. 

In this work, we focus on the ladder version of the dynamical vertex approximation (D$\Gamma$A) \cite{Toschi2007, Katanin2009, Rohringer2018}, which accurately describe spin fluctuations in the parameter regime when compared to quantum Monte Carlo simulations (where they are still possible) \cite{Schaefer2021}. In single-band models such an approach was successful in describing both the emergence of the pseudogap phase \cite{Schaefer2015, Schaefer2021} and the manifestation of (quantum) critical effects \cite{Rohringer2011, Schaefer2017, Schaefer2019}, while in single-band real-material models, this technique satisfactorily described the origin of the low superconducting critical temperatures for cuprates \cite{Kitatani2018} as well as the magnetic \cite{Ortiz2022, Klett2022} and superconducting \cite{Kitatani2020, Held2022} properties of nickelates and palladates \cite{Kitatani2023}. 

The dynamical vertex approximation (D$\Gamma$A) was originally derived in Refs.~\cite{Toschi2007, Katanin2009} for the paramagnetic SU$(2)$-symmetric phase of the single-band Hubbard model. Since in the Emery model described in Eq.~(1) in the main text the Coulomb repulsion is restricted to the $d_{x^{2}-y^{2}}$-orbital and we only analyze SU$(2)$-symmetric paramagnetic solutions, the same single-band formalism to compute the frequency and momentum-dependent self-energy can be applied. In this section, we review the main equations of the ladder version of D$\Gamma$A, focusing on the algorithmic aspects.

\subsection{Ladder D$\Gamma$A self-energy in the particle-hole channel}

The main idea of D$\Gamma$A is to introduce non-local and temporal correlations beyond the local (temporal) ones of DMFT in the self-energy of the system by means of the Schwinger-Dyson equation of motion (SD-EOM) \cite{Toschi2007, Katanin2009}

\begin{align}
    \Sigma_{\text{dd}}\left(\mathbf{k},i\nu\right)&=\frac{U_{\text{dd}}n_{\text{dd}}}{2}\\
    &-\frac{U_{\text{dd}}}{\beta}\sum_{\mathbf{q}\nu'\Omega}F_{\text{dd},\uparrow\downarrow}^{\mathbf{qkk'},\Omega\nu\nu'}\chi_{0,\text{dd}}^{\mathbf{q}\Omega\nu'}G_{\text{dd}}^{\mathbf{k}+\mathbf{q},i\nu+i\Omega},
    \label{eq:Schwinger_Dyson}
\end{align}
where $U_{\text{dd}}$ is the local Coulomb repulsion restricted to the $d_{x^{2}-y^{2}}$ orbital, $n_{\text{dd}}$ is the average $d_{x^{2}-y^{2}}$-orbital occupation, $F_{\text{dd},\uparrow\downarrow}^{\mathbf{qkk'},\Omega\nu\nu'}$ is the $\uparrow\downarrow$-component of the frequency and momentum dependent $d_{x^{2}-y^{2}}$-orbital full vertex, $G_{\text{dd}}\left(\mathbf{k}',i\nu\right)$ is the interacting $d_{x^{2}-y^{2}}$-orbital Green function, and $\chi_{0,\text{dd}}^{\mathbf{q}\Omega\nu'}=T^{-1}\sum_{\mathbf{k}}G_{\text{dd}}\left(\mathbf{k},\nu'\right)G_{\text{dd}}\left(\mathbf{k}+\mathbf{q},\nu'+\Omega\right)$ is the generalized particle-hole bubble. 

In the ladder version of D$\Gamma$A, it is assumed that if we know which type of fluctuations prevail in the system, we can restrict ourselves, in the language of the Bethe-Salpeter equation \cite{Rohringer2012, DelRe2021b}, to the channel generating such fluctuations. In this approximation, the Green function in Eq.~(\ref{eq:Schwinger_Dyson}) is replaced by the DMFT one restricted to the $d_{x^{2}-y^{2}}$-orbital [see Eq.~(\ref{eq:DMFT_latt_GF})] and the $d_{x^{2}-y^{2}}$-orbital generalized susceptibility $\chi_{\text{dd},\eta}^{\mathbf{q},\Omega\nu\nu'}$ and the full vertex $F_{\text{dd},\eta}^{\mathbf{q},\Omega\nu\nu'}$ are constructed as a ladder summation via the Bethe-Salpeter equation in the given channels. In this work, we analyze the role of dynamical spin fluctuations in the pseudogap phase in the Emery model; for that, we restrict our analysis to the two particle-hole channels, charge ($\eta$ = ch) and magnetic ($\eta$ = m), in which the $d_{x^{2}-y^{2}}$ generalized susceptibilities reads

\begin{align}
    \chi_{\text{dd},\eta}^{\mathbf{q},\Omega\nu\nu'}&=\beta\delta_{\nu\nu'}\chi_{0,\text{dd}}^{\mathbf{q},\Omega\nu}-\frac{1}{\beta}\chi_{0,\text{dd}}^{\mathbf{q},\Omega\nu}\sum_{\nu_{1}}\Gamma_{\text{dd},\eta}^{\Omega\nu\nu_{1}}\chi_{\text{dd},\eta}^{\mathbf{q},\Omega\nu_{1}\nu'}\\&=\beta\delta_{\nu\nu'}\chi_{0,\text{dd}}^{\mathbf{q},\Omega\nu}-\chi_{0,\text{dd}}^{\mathbf{q},\Omega\nu}F_{\text{dd},\eta}^{\mathbf{q},\Omega\nu\nu'}\chi_{0,\text{dd}}^{\mathbf{q},\Omega\nu'},
    \label{eq:gen_latt_suscep}
\end{align}
where $\Gamma_{\text{dd},\eta}^{\Omega\nu\nu'}$ is the local DMFT $d_{x^{2}-y^{2}}$ irreducible vertex in the channel $\eta$ [see Eq.~(\ref{eq:Irreducible_vextex})], $\chi_{0,\text{dd}}^{\mathbf{q},\Omega\nu}$ is the DMFT $d_{x^{2}-y^{2}}$ bare susceptibility [see Eq.~(\ref{eq:dmft_bubble})] and $F_{\text{dd},\eta}^{\mathbf{q},\Omega\nu\nu'}$ is the ladder D$\Gamma$A $d_{x^{2}-y^{2}}$ full vertex. From Eq.~(\ref{eq:gen_latt_suscep}) we notice that in the ladder aprroximation, both the generalized lattice susceptibility $\chi_{\text{dd},\eta}^{\mathbf{q},\Omega\nu\nu'}$ and the full reducible vertex $F_{\text{dd},\eta}^{\mathbf{q},\Omega\nu\nu'}$ depend only on the transfer momentum $\mathbf{q}$ since these quantities are constructed as ladder series in the DMFT local irreducible vertex $\Gamma_{\text{dd},\eta}^{\Omega\nu\nu'}$.
Inserting the vertex $F_{\text{dd},\eta}^{\mathbf{q},\Omega\nu\nu'}$ into the SD-EOM, Eq.~(\ref{eq:Schwinger_Dyson}), we obtain the ladder D$\Gamma$A self-energy
\begin{align}
    \Sigma_{\text{D}\Gamma\text{A,\text{dd}}}^{\text{ladder}}\left(\mathbf{k},i\nu\right)&=\frac{U_{\text{dd}}n_{\text{dd}}}{2}\nonumber\\
    &-\frac{U_{\text{dd}}}{\beta^{2}}\sum_{\mathbf{q}\mathbf{k}'\nu'\Omega}F_{\text{dd},\uparrow\downarrow}^{\mathbf{q},\Omega\nu\nu'}\chi_{0,\text{dd}}^{\mathbf{q}\Omega\nu'}G_{\text{dd,DMFT}}^{\mathbf{k}+\mathbf{q},i\nu+i\Omega}
    \label{eq:ladder_sig}
\end{align}
for the $d_{x^{2}-y^{2}}$-orbital, where $\chi_{0,\text{dd}}^{\mathbf{q}\Omega\nu'}$ is given by Eq.~(\ref{eq:dmft_bubble}). 

Following the Ref. \cite{Katanin2009}, we can rewrite the Eq.~(\ref{eq:ladder_sig}) in a more physical way in terms of the magnetic and charge DMFT lattice susceptibilities [see Eq.~(\ref{eq:BSE})]. For that, we split the D$\Gamma$A ladders, Eq.~(\ref{eq:gen_latt_suscep}), by the bare interaction vertex $U_{\eta}=+/-U_{\text{dd}}$ for $\eta=\text{ch/m}$, which allows us to define the ladder quantity $\Phi_{\text{\text{dd},\ensuremath{\eta}}}^{\mathbf{q},\Omega\nu\nu'}$ by a Bethe-Salpeter-like equation

\begin{equation}
    \chi_{\text{dd},\eta}^{\mathbf{q},\Omega\nu\nu'}=\Phi_{\text{dd},\eta}^{\mathbf{q},\Omega\nu\nu'}-\frac{1}{\beta^{2}}\sum_{\nu_{1},\nu_{2}}\Phi_{\text{dd},\eta}^{\mathbf{q},\Omega\nu\nu_{1}}U_{\eta}\chi_{\text{dd},\eta}^{\mathbf{q},\Omega\nu_{2}\nu'}.
\end{equation}
Expressing the full vertex $F_{\text{dd},\eta}^{\mathbf{q},\Omega\nu\nu'}$ in terms of $\Phi_{\text{dd},\eta}^{\mathbf{q},\Omega\nu\nu'}$ allows us to rewrite the SD-EOM, Eq.~(\ref{eq:ladder_sig}), like \cite{Katanin2009} 
\begin{align}
    \Sigma_{\text{D}\Gamma\text{A},\text{dd}}^{\text{ladder}}\left(\mathbf{k},i\nu\right)
    &=\frac{Un}{2}-\frac{U_{\text{dd}}}{2\beta}\sum_{\mathbf{q}\Omega}\left[2+\gamma_{\text{dd,ch}}^{\mathbf{q}\Omega\nu}\left(1-U_{\text{dd}}\chi_{\text{dd,ch}}^{\mathbf{q}\Omega}\right)\right.\nonumber\\&-3\gamma_{\text{dd,m}}^{\mathbf{q}\Omega\nu}\left(1+U_{\text{dd}}\chi_{\text{dd,m}}^{\mathbf{q}\Omega}\right)\nonumber\\&\left.+\frac{1}{\beta}\sum_{\mathbf{q}\Omega\nu'}\left(F_{\text{dd,ch}}^{\Omega\nu\nu'}-F_{\text{dd,m}}^{\Omega\nu\nu'}\right)\chi_{0,\text{dd}}^{\mathbf{q},\nu'}\right]G_{\text{dd, \text{DMFT}}}^{\mathbf{k}+\mathbf{q},\nu+\Omega},
    \label{eq:sig_ladder_final}
\end{align}
where $\gamma_{\text{dd},\eta}^{\mathbf{q},\Omega\nu}$ is the so-called DMFT lattice three-leg Hedin vertex \cite{Katanin2009, Rohringer2016, Ayral2015, Ayral2016} defined as
\begin{align}
    \gamma_{\text{dd},\eta}^{\mathbf{q},\Omega\nu}&=\left(\chi_{0,\text{dd}}^{\mathbf{q},\Omega\nu}\right)^{-1}\frac{1}{\beta}\sum_{\nu'}\Phi_{\text{\text{dd}},\eta}^{\mathbf{q},\Omega\nu\nu'},\nonumber\\&=\frac{\left(\chi_{0,\text{dd}}^{\mathbf{q},\Omega\nu}\right)^{-1}T\sum_{\nu'}\chi_{\text{dd},\eta}^{\mathbf{q},\Omega\nu\nu'}}{1-U_{\eta}\chi_{\text{dd},\eta}^{\mathbf{q}\Omega}}.
    \label{eq:Hedin_vertex}
\end{align}
$\chi_{\text{dd,ch/m}}^{\mathbf{q}\Omega}$ is the DMFT $d_{x^{2}-y^{2}}$ lattice charge/magnetic susceptibility, and $F_{\text{ch/m,dd}}^{\Omega\nu\nu'}$ is the local DMFT $d_{x^{2}-y^{2}}$ full vertex in the charge/magnetic channel that is obtained from the local version (no $\mathbf{q}$-dependence) of the Eq.~(\ref{eq:gen_latt_suscep}). The last term in Eq.~(\ref{eq:sig_ladder_final}) avoids the double counting of local DMFT diagrams.

\subsection{$\lambda$-corrected (D$\Gamma$A) magnetic lattice susceptibility}

The computation of the self-energy by means of the ladder D$\Gamma$A diagrams, Eq.~(\ref{eq:sig_ladder_final}), violates the correct asymptotic behavior of a self-energy \cite{Rohringer2013, Rohringer2016}, i.e., for large fermionic Matsubara frequencies the 1/i$\nu$-part of Eq.~(\ref{eq:sig_ladder_final}) does not have the correct prefactor $U_{\text{dd}}^{2}\frac{n_{\text{d}}}{2}\left(1-\frac{n_{\text{d}}}{2}\right)$. Such incorrect asymptotic behavior arises as a direct consequence of the violation of the following sum rule by the DMFT lattice susceptibilities (a more detailed disccusion of this issue is reported in Refs.~\cite{Rohringer2013, Rohringer2016}):
\begin{align}
    \frac{T}{N_{\mathbf{q}}}\sum_{\mathbf{q},\Omega}\chi_{\text{dd,}\uparrow\uparrow}^{\mathbf{q}\Omega}=\frac{T}{2N_{\mathbf{q}}}\sum_{\mathbf{q},\Omega}\left[\chi_{\text{dd,ch}}^{\mathbf{q}\Omega}+\chi_{\text{dd,m}}^{\mathbf{q}\Omega}\right]\neq\frac{n_{d}}{2}\left(1-\frac{n_{d}}{2}\right).
\end{align}
However, this violation does not occur for the DMFT solution since the impurity susceptibilities, Eqs.~(\ref{eq:magnetic_imp_suscept}) and (\ref{eq:charge_imp_suscept}), fulfill the sum rule \cite{Rohringer2013}.

In order to restore the correct asymptotic behavior of the self-energy, following the Moriya theory of itinerant magnetism \cite{MORIYA1991}, we introduce the so-called $\lambda$-correction scheme for the magnetic lattice susceptibility. In principle the $\lambda$-correction scheme can also be applied to the charge susceptibility \cite{Rohringer2016,Stobbe2022}, however, the benchmark against the numerical exact quantum Monte Carlo results \cite{Schaefer2021} has shown that the correction of the magnetic susceptibility only accurately describes the spin fluctuations in the weak coupling regime of the square lattice Hubbard model. In practice, the $\lambda$-correction scheme consists of defining the $\lambda$-corrected magnetic lattice susceptibility as 

\begin{equation}
    \chi_{\text{dd,m}}^{\lambda_{\text{m}},\mathbf{q}\Omega}=\left[\left(\chi_{\text{\text{dd,m}}}^{\mathbf{q}\Omega}\right)^{-1}+\lambda_{\text{m}}\right]^{-1},
    \label{eq:lambda_corr_chi}
\end{equation}
where the determinations of the scalar parameter $\lambda_{\text{m}}$ is obtained by imposing that the $\lambda$-corrected magnetic lattice susceptibility fulfills the same sum rule as the impurity one, which mathematically translates as

\begin{equation}
    \frac{T}{2N_{\mathbf{q}}}\sum_{\mathbf{q},\Omega}\left[\chi_{\text{dd,ch}}^{\mathbf{q}\Omega}+\chi_{\text{dd,m}}^{\lambda_{\text{m}},\mathbf{q}\Omega}\right]\equiv\frac{n_{d}}{2}\left(1-\frac{n_{d}}{2}\right).
    \label{eq:sum_rule}
\end{equation}

To get a physical intuition about the role of the $\lambda$-correction, let us assume that the static DMFT magnetic lattice susceptibility is described by an Ornstein-Zernike form peaked around a leading vector $\mathbf{Q}$, and with magnetic correlation length $\xi_{\text{m}}$. In this case the Eq.~(\ref{eq:lambda_corr_chi}) reads

\begin{align}
    \left[\chi_{\text{dd,m}}^{\lambda_{\text{m}},\mathbf{q},i\Omega_{0}}\right]^{-1}&=\left[\chi_{\text{OZ}}^{\mathbf{q},i\Omega_{0}}\right]^{-1}+\lambda_{\text{m}}\nonumber\\&=\frac{1}{A}\left[\left(\mathbf{q}-\mathbf{Q}\right)^{-2}+\text{\ensuremath{\xi_{\text{m}}^{-2}}}+\lambda_{\text{m}}\right],
    \label{eq:lambda_chi}
\end{align}
which allows us to define the $\lambda$-corrected magnetic correlation length as 
\begin{equation}
    \xi_{\text{m}}^{\text{D}\Gamma\text{A}}=\frac{\xi_{\text{m}}}{\sqrt{1+\lambda\xi_{\text{m}}^{2}}},
    \label{eq:renorm_corr_len}
\end{equation}
leading us to the conclusion that the $\lambda$-correction in fact corresponds to a renormalization of the DMFT magnetic correlation length. In particular, for a positive value of $\lambda$, the Eq.~(\ref{eq:renorm_corr_len}) describes an effective reduction in the magnetic correlation length of DMFT $\xi_{\text{m}}$. In addition, the $\lambda$-correction renders the static lattice susceptibility of the system positive in a temperature range where the DMFT one, Eq.~(\ref{eq:BSE}), has already become thermodynamically unstable. In 3D systems this correction leads to a reduction of the transition temperatures \cite{Rohringer2011, Rohringer2016, Del_Re2019}, while for two-dimensional systems such correction avoids the emergence of long range orders at finte temperatures \cite{Schaefer2015,Schaefer2021}, fulfilling the Mermin-Wagner theorem \cite{Mermin1966} (see next subsection for a more detailed discussion).

We introduce the $\lambda$-corrected magnetic lattice susceptibility in the ladder D$\Gamma$A self-energy by simply replacing $\chi_{\text{dd,m}}^{\mathbf{q}\Omega}$ by $\chi_{\text{dd,m}}^{\lambda_{\text{m}},\mathbf{q}\Omega}$ in Eq.~(\ref{eq:sig_ladder_final}). 

\subsubsection{Fulfillment of the Mermin-Wagner theorem}
The absence of spatial fluctuations in DMFT leads to the emergence of spurious mean-field-like long range orders (e.g., magnetic order) in two-dimensional systems at finite temperatures, which violates the Mermin-Wagner theorem \cite{Mermin1966}. Such theorem states that long-range order is forbidden at finite temperatures in a two-dimensional system. It turned out that in addition to guarantee the correct asymptotic behavior of the ladder D$\Gamma$A self-energy, Eq.~(\ref{eq:sig_ladder_final}), the implementation of the $\lambda$-correction scheme via the imposition of the sum rule in Eq.~(\ref{eq:sum_rule}) also guarantees the fulfillment of the Mermin-Wagner theorem.
In the $\lambda$-correction scheme, Eq.~(\ref{eq:sum_rule}), the DMFT lattice charge susceptibility remains unchanged and the trace over its momentum and frequency degrees of freedom leads a constant contribution that decreases the right hand side of the Eq.~(\ref{eq:sum_rule}). Here we assume that no single site DMFT charge instability is present in the system, which is the case for the parameter regime where we analyze the Emery model in this work. Since the right-hand side of Eq.~(\ref{eq:sum_rule}) is always constant for a fixed $d_{x^{2}-y^{2}}$ local average density, if we isolate the $\lambda$-corrected magnetic lattice susceptibility in Eq.~(\ref{eq:sum_rule})
\begin{equation}
    \frac{1}{2\beta}\sum_{\mathbf{q}\Omega}\chi_{\text{dd,m}}^{\lambda,\mathbf{q}\Omega}<\infty,
    \label{eq:local_equal_time_spin_suscep}
\end{equation}
we conclude that the trace over its momentum and bosonic frequencies degrees of freedom always gives rise to a finite constant value. 
In order to understand the implications of the Eq.~(\ref{eq:local_equal_time_spin_suscep}), we restrict our analysis to the static mode ($\Omega=0$) of the $\lambda$-corrected magnetic lattice susceptibility, and we assume that it can be described by Ornstein-Zernike form around the the leading vector $\mathbf{Q}$ with a magnetic correlation lengh $\xi_{\lambda}$, as in Eq.~(\ref{eq:lambda_chi}). For convinience, in Eq.~(\ref{eq:local_equal_time_spin_suscep}) we replace the summation over the vectors belonging to the 1$^{\text{st.}}$ Brillouin zone by an integral over it that reads 
\begin{equation}
    \frac{1}{\left(2\pi\right)^{s}}\int\frac{\text{d}^{s}\tilde{q}}{\tilde{\mathbf{q}}^{2}+\xi_{\lambda}^{-2}}<\infty,
    \label{eq:mermin}
\end{equation}
where $s$ denotes the spatial dimensionality of the system, and $\tilde{\mathbf{q}}=\mathbf{q}-\mathbf{Q}$. Now we are able to study the behavior of the Eq.~(\ref{eq:mermin}) with respect to the spatial dimensionality of the system

We first analyze the three-dimensional (3D) system in which the volume element is $\text{d}^{s}\tilde{q}\propto\tilde{q}^{2}\text{d}\tilde{q}$ (for our discussion the angular variable is irrelevant). If we assume a divergent $\lambda$-corrected magnetic correlation length $\left(\xi_{\lambda}^{-2}\rightarrow0\right)$, we obtain that
\begin{equation}
    \frac{1}{\left(2\pi\right)^{s}}\int\frac{\text{d}^{s}\tilde{q}}{\tilde{\mathbf{q}}^{2}+\xi_{\lambda}^{-2}}\propto\int_{0}^{q^{*}}\text{d}\tilde{q}<\infty,
    \label{eq:3D_div}
\end{equation}
where the upper limit value $q^{*}$ of the integral is a natural finite cutoff since the integration is performed over a finite Brillouin zone. From the Eq.~(\ref{eq:3D_div}) we conclude that in 3D a divergent $\lambda$-corrected magnetic correlation length still guarantees the fulfillment of the Eq.~(\ref{eq:sum_rule}). In fact, previous works have shown that in 3D the $\lambda$-correction procedure reduces both the N{\'e}el temperature \cite{Rohringer2011, Schaefer2017} as well as superconducting critical temperature \cite{Del_Re2019}, when applied in the particle-particle channel, in comparison to DMFT.

Now, if we assume a two-dimensional (2D) system $\left(\text{d}^{s}\tilde{q}\propto\tilde{q}\text{d}\tilde{q}\right)$ with a divergent $\lambda$-corrected magnetic correlation length in Eq.~(\ref{eq:mermin}), we in fact obtain a divergent quantity 

\begin{equation}
    \frac{1}{\left(2\pi\right)^{s}}\int\frac{\text{d}^{s}\tilde{q}}{\tilde{\mathbf{q}}^{2}+\xi_{\lambda}^{-2}}\propto\int_{0}^{q^{*}}\frac{\text{d}\tilde{q}}{\tilde{q}}=\lim_{q_{0}\rightarrow0}\ln\left(\frac{q^{*}}{q_{0}}\right)\rightarrow\infty.
    \label{eq:2D_div}
\end{equation}
Hence, from Eq.~(\ref{eq:2D_div}) we conclude that in 2D systems, once the $\lambda$-correction is determined from the Eq.~(\ref{eq:sum_rule}), it leads to a finite correlation length which avoids the emergence of a long-range magnetic order in the system. Previous studies have shown that on both the square lattice Hubbard model \cite{Schaefer2015, Rohringer2016, Schaefer2021} as well as in square lattice periodic Anderson model \cite{Schaefer2019}, upon cooling the $\lambda$-corrected magnetic correlation length remains finite besides it increases monotonically when the ground state of the system is magnetically ordered.

\section{Doping evolution of the Matsubara Green function}
\label{sec:gf}

\begin{figure}[h!]
\centering
\includegraphics[width=.8\columnwidth]{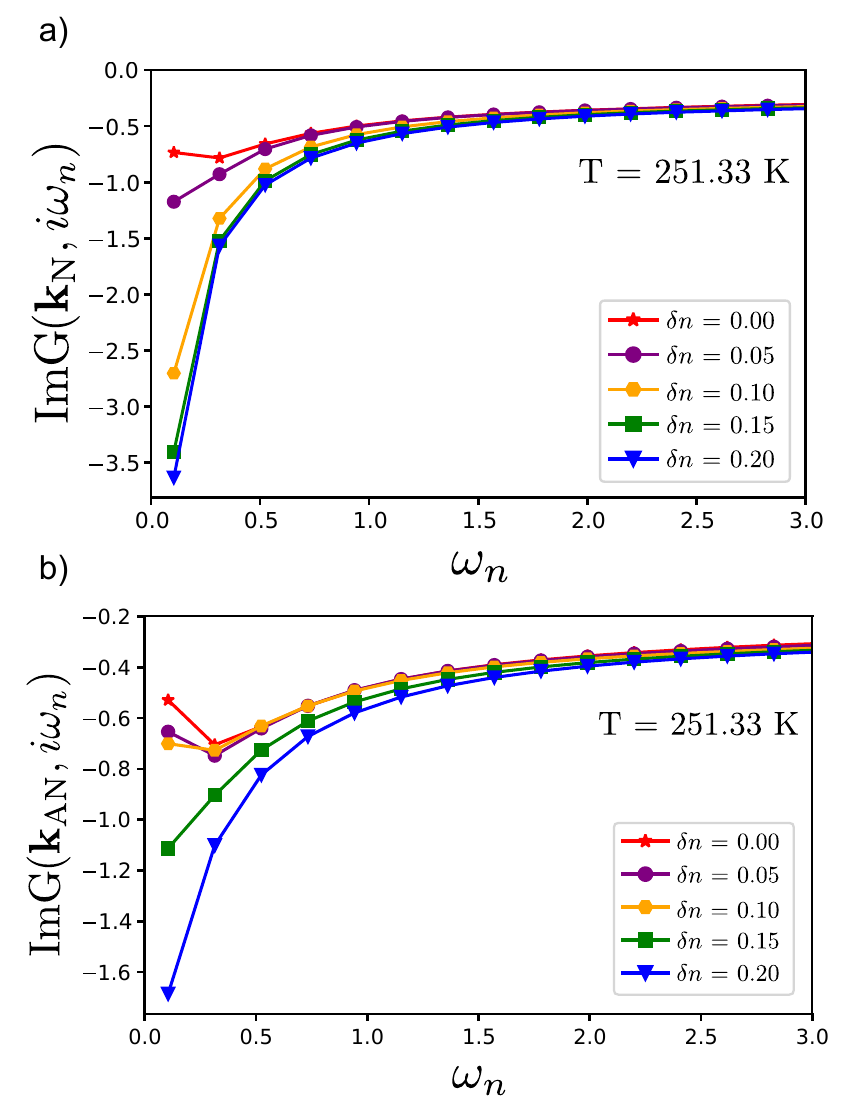}
\caption{Hole-doping evolution of the imaginary part of the Matsubara Green function at the lowest temperature, T = 251.33 K ($\beta t_{\text{pp}}$ = 30), at the a) nodal and b) antinodal points.}  
\label{fig:ImG_N_AN}
\end{figure}

In the main text, we have shown that the inclusion of the non-local correlations on all length scales in the ladder D$\Gamma$A approach reveals that the Emery model, upon hole-doping, evolves from an undoped insulating ($\delta n$ = 0.0) regime into a pseudogap regime ($0.05\leq\delta n<0.15$) and then eventually develops into a conventional metal ($\delta n\geq0.15$). The fingerprints of such crossovers are also present in the low-frequency behavior of the imaginary part of the Matsubara Green function. In particular, the slope of this quantity, $\text{Im}\left[G(\mathbf{k},i\omega_{0})-G(\mathbf{k},i\omega_{1})\right]$, is a proxy that when negative (positive) hints for a metallic (insulating) behavior.

Fig.~\ref{fig:ImG_N_AN} displays the hole-doping evolution of the imaginary part of the Matsubara Green function at the nodal and antinodal points at $T=251.33$ K ($\beta t_{\text{pp}}=30.0$). For the undoped system ($\delta n = 0.0$), the slope of the respective Green function at both nodal and antinodal points indicates an insulating-like behavior. However, at $\delta n = 0.05$ and $\delta n = 0.1$, while the slope of the Green function  at the antinodal point remains insulating-like, the one of the nodal point turns negative, which indicates a metallic-like behavior. Such dichotomy between the nodal and the antinodal points is a signature of the pseudogap regime. Upon further doping ($\delta n\geq0.15$), the system falls into a metallic regime where the slope of the Green functions turns negative for both nodal and antinodal points. 

In addition, Fig.~\ref{fig:ImG_N_AN} also shows that non-local effects induce a sizable momentum differentiation on the Fermi surface of the hole-doped metallic regime ($\delta n\geq0.15$) since the amplitude of the Green function at the nodal point is larger than at the antinodal one.   

\section{Spectral function over the Fermi surface}
\label{sec:w_specFunc_FS}
\begin{figure}[b!]
\centering
\includegraphics[width=1.\columnwidth]{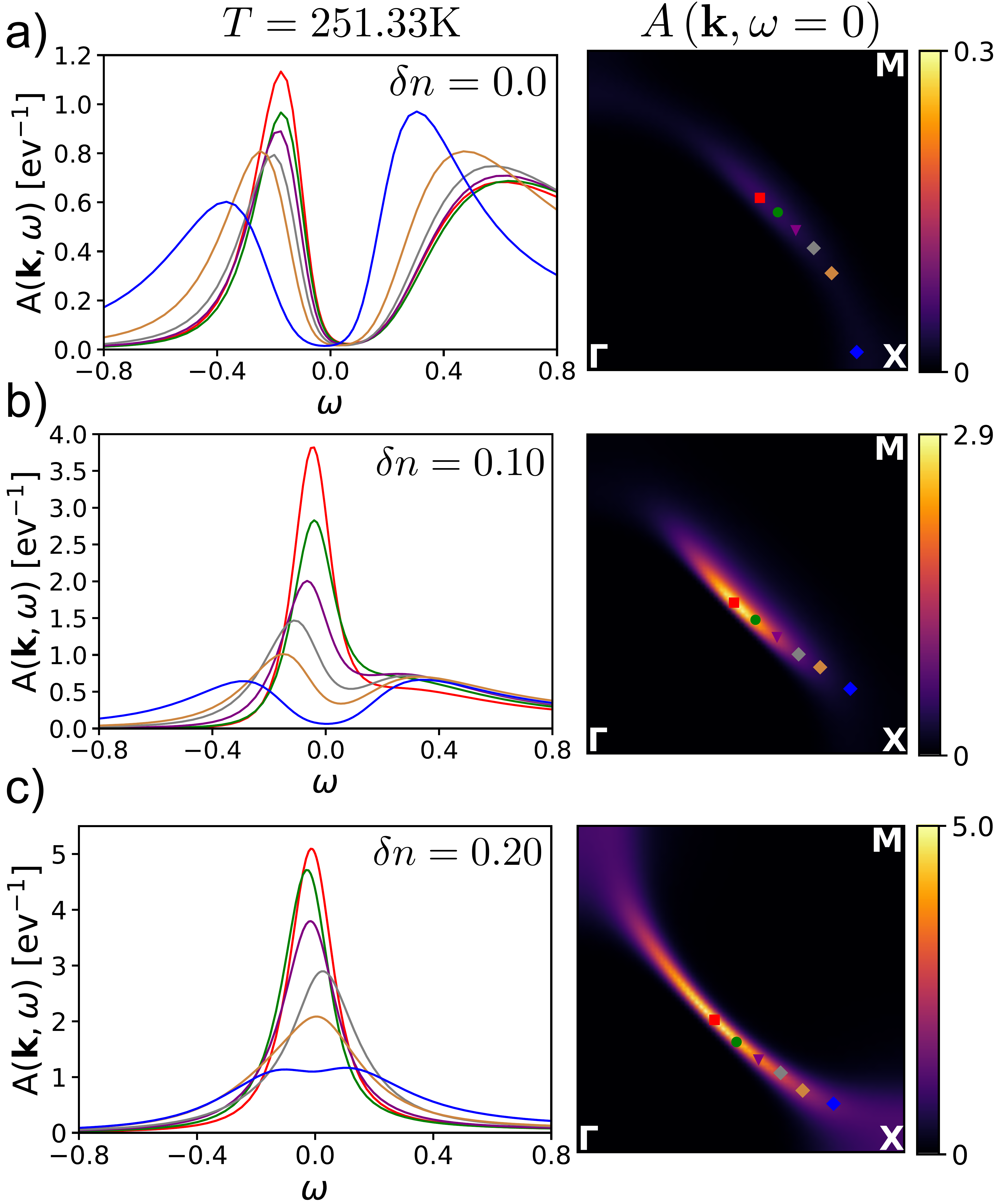}
\caption{Hole-doping evolution of the low-frequency behavior of the electronic spectral function (left panel) at points along the Fermi surface (right panel) at the lowest temperature, T = 251.33 K ($\beta t_{\text{pp}}$ = 30), in the a) insulating ($\delta n =0.00$), b) pseudogap ($\delta n =0.10$), and c) metallic ($\delta n =0.20$) regime. The local spectral functions were obtained by MaxEnt analytical continuation \cite{Reymbaut2017}.}  
\label{fig:spec_func_FS}
\end{figure}

Rigorously, the Fermi surface exists only at zero temperature. However, as observed in angle-resolved photoemission (ARPES) experiments \cite{Damascelli2003}, at finite temperatures, we can define the Fermi surface as the surface in momentum space where the spectral intensity reaches its maximum value. In Fig.~\ref{fig:spec_func_FS}, we show the electronic spectral function, obtained by MaxEnt analytical continuation at T = $251.33$ K, over one-quarter of the Brillouin zone at $\omega=0$  and its low-frequency dependence along a couple of points over the Fermi surface.

In the undoped regime ($\delta n = 0.00$),  Fig.~\ref{fig:spec_func_FS}(a),  the frequency-dependent spectral function over the Fermi surface (left panel) presents a clear gap at $\omega = 0.00$, which is an emblematic feature of the insulating state. The presence of the low-energy gap reflects the absence of spectral weight over one-quarter of the Brillouin zone at $\omega = 0$ (right panel). At finite temperatures, thermal excitations always populate the insulating gap, leading to a finite spectral weight at $\omega = 0$. In the data shown in Fig.~\ref{fig:spec_func_FS}(a), the actual value of the spectral weight at $\omega = 0.0$ is $0.05$ eV$^{-1}$.

Deep inside the pseudogap regime ($\delta n = 0.10$), the spectral 
function over one-quarter of the Brillouin zone displays a Fermi arc [see the right panel of Fig.~\ref{fig:spec_func_FS}(b)],  characterized by a high spectral weight along the $\Gamma$M-direction 
(nodal point) and a low spectral weight one 
along the $\Gamma$X-direction (antinodal point). This Fermi arc structure is one of the hallmark features of the pseudogap regime in cuprate superconductors 
\cite{Ding1996, Marshall1996, Damascelli2003, Kanigel2006,Yoshida2006,Shen2005}. 
In addition, from the left panel in Fig.~\ref{fig:spec_func_FS}(b), we observe that the low-frequency dependence of the electronic spectral function over the Fermi surface evolves from a coherent quasiparticle peak at the nodal point (red curve) into a gapped structure at the antinodal one (blue curve). 

Unlike the pseudogap regime ($\delta n = 0.10$), within the metallic regime ($\delta n = 0.20$), the spectral function displays a finite value along the $\Gamma$X-direction, as seen in the right panel of Fig.~\ref{fig:spec_func_FS}(c). Furthermore, in the left panel of  Fig.~\ref{fig:spec_func_FS}(c), we observe that along the Fermi surface, the low-frequency  spectral function exhibits a coherent quasiparticle peak at the nodal point along the $\Gamma$M-direction (red curve) and evolves into a less coherent quasiparticle peak in the proximity of the $\Gamma$X-direction (blue curve). In the metallic state, we also notice that the feedback of the short-ranged dynamical spin fluctuation introduced via the ladder D$\Gamma$A self-energy leads to a momentum differentiation of the Fermi surface states.

\section{Local spectral function}
\label{sec:spectrum}

\subsection{Temperature dependence}

\begin{figure}[b!]
\centering
\includegraphics[width=.7\columnwidth]{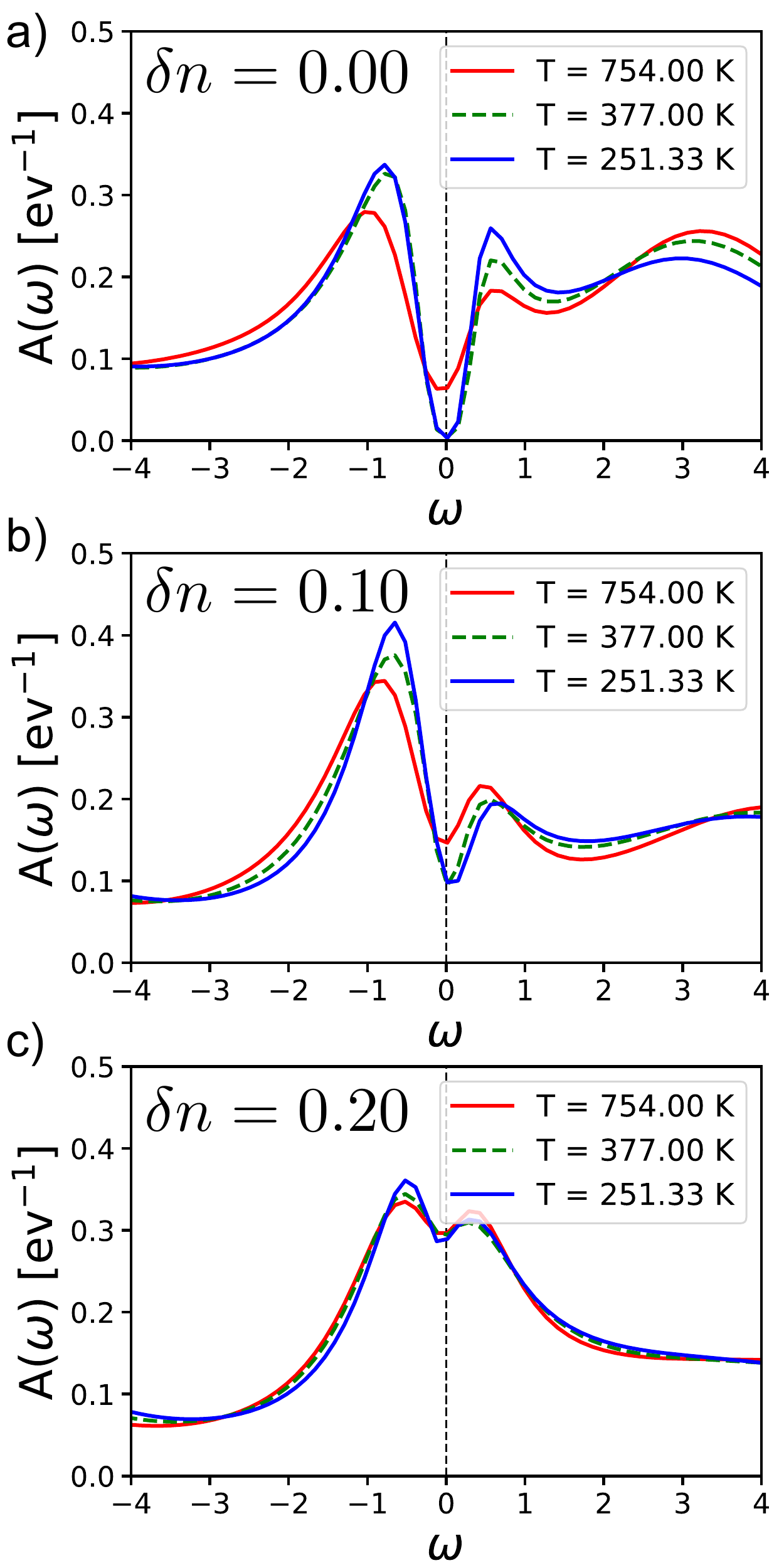}
\caption{Temperature evolution of the local spectral function in the a) insulating ($\delta n =0.00$), b) pseudogap ($\delta n =0.10$), and c) insulating ($\delta n =0.20$) regime. The local spectral functions were obtained by MaxEnt analytical continuation \cite{Reymbaut2017}.}  
\label{fig:locSpecFunc_dop}
\end{figure}

The temperature dependence of the local spectral function provides information about the metallicity of the system. In Fig.~\ref{fig:locSpecFunc_dop}, we show the temperature evolution of the low-frequency (total) local spectral function obtained by MaxEnt analytical continuation \cite{Reymbaut2017} in the (a) insulating, (b) pseudogap, and (c) metallic regime. 

Fig.~\ref{fig:locSpecFunc_dop}(a) shows that upon cooling, the local spectral function at the Fermi level decreases towards zero, leading to the emergence of the insulating gap. For the range of temperatures we studied in this work, the insulating gap remains constant upon further cooling. In the pseudogap regime, Fig.~\ref{fig:locSpecFunc_dop}(b), like in the insulating regime, cooling the system from $T = 754.00$ K to $T = 377.00$ K leads to a decrease of
 the local spectral function at the Fermi level; however, unlike the insulating regime, upon further cooling to the lowest temperature we considered in this work ($T = 251.33$ K), the local spectral function saturates and does not display a real gap. In the metallic regime,  Fig.~\ref{fig:locSpecFunc_dop}(c), lowering the temperature of the system from $T = 754.00$ K to $T = 251.33$ K, the local spectral function always displays a finite value at the Fermi level and does not considerably change.

\subsection{Orbital character}
\begin{figure}[b!]
\centering
\includegraphics[width=.9\columnwidth]{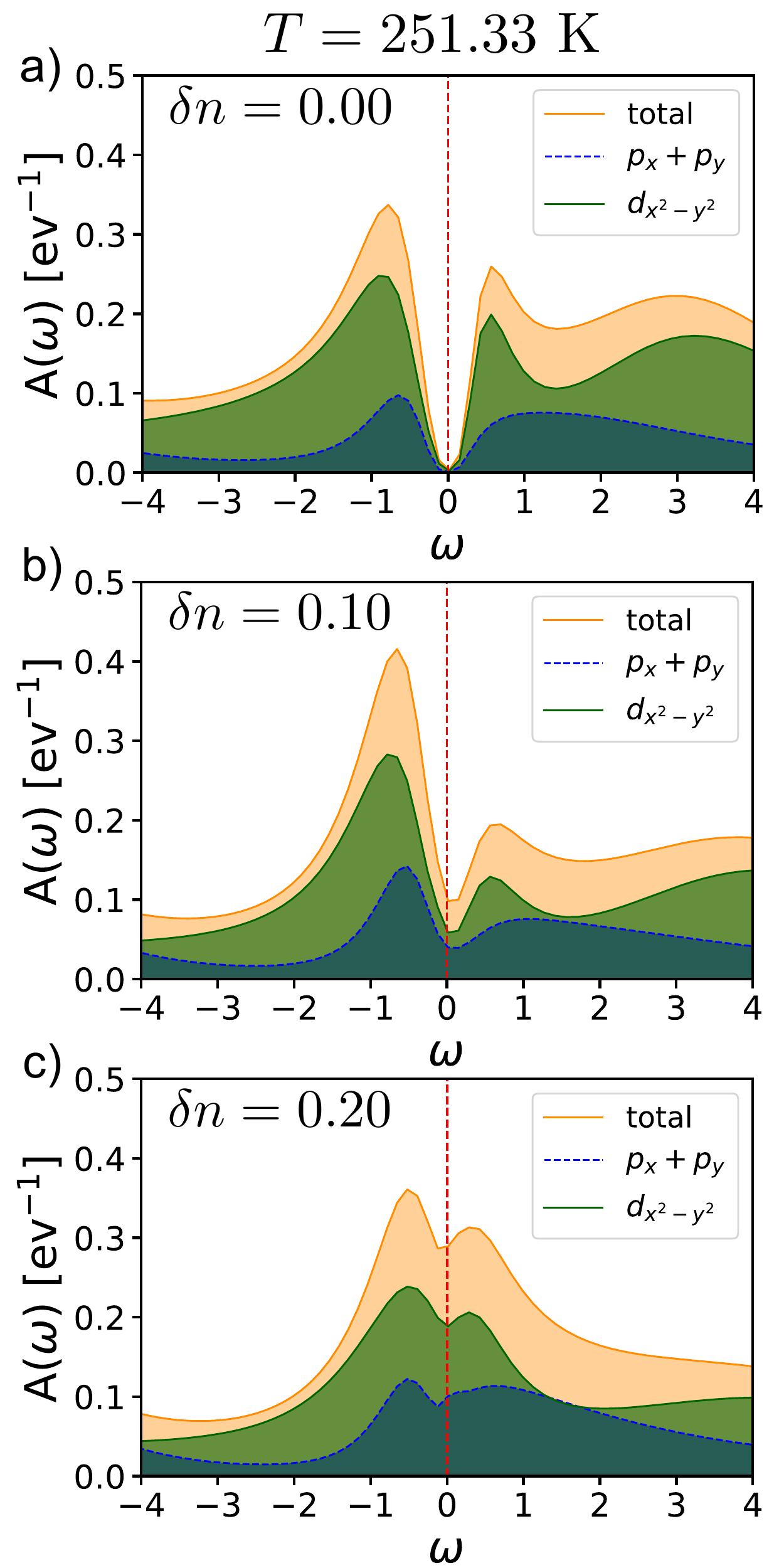}
\caption{Hole-doping evolution of the local spectral function at the lowest temperature, T = 251.33 K ($\beta t_{\text{pp}}$ = 30), in the a) insulating ($\delta n =0.00$), b) pseudogap ($\delta n =0.10$), and c) insulating ($\delta n =0.20$) regime. The local spectral functions were obtained by MaxEnt analytical continuation \cite{Reymbaut2017}.}  
\label{fig:loc_spec_func}
\end{figure}

One of the main features of cuprate superconductors is the charge-transfer insulator character of the undoped (parent) compound. Such a feature requires more realistic models that, such as the Emery model, include the copper-oxygen hybridization of the single band that crosses the Fermi level. In order to investigate the orbital differentiation of our model, in Fig.~\ref{fig:loc_spec_func}, we show the orbital resolved local spectral function around the Fermi level obtained by MaxEnt analytical continuation \cite{Reymbaut2017} at $T=251.33$ K for the undoped system $\left(\delta n=0.00\right)$ as well as deep inside the pseudogap ($\delta n = 0.10$) and the metallic $\left(\delta n=0.20\right)$ phase. 

From Fig.~\ref{fig:loc_spec_func}(a) we note that, for the lowest temperature we consider in this work ($T = 251.33$ K), the local spectral function of the undoped system ($\delta n = 0.00$) displays a gap at the Fermi level. This gap arises from the feedback of the long-ranged spin fluctuations (see Fig.~\ref{fig:real_space_stat_suscept}) included via the ladder D$\Gamma$A self-energy [see Eq.~(\ref{eq:sig_ladder_final})]. From the orbital resolved local spectral function, we observe that right below the Fermi level, the orbital character is dominated by both $d_{x^{2}-y^{2}}$- and $p_{x}+p_{y}$-states. In this insulating state, the low-lying electronic excitations should present a strongly hybridized character. 

Deep within the pseudogap regime ($\delta n = 0.10$), Fig.~\ref{fig:loc_spec_func}(b), the local spectral function displays a dip at the Fermi level, and its orbital contribution, at the Fermi level, is slightly dominated by $d_{x^{2}-y^{2}}$-states. In addition, in this regime, right below the Fermi level ($\sim 0.5$ eV), we also observe a mixed $d_{x^{2}-y^{2}}$- and $p_{x}+p_{y}$-orbital character but with a much higher contribution of $d_{x^{2}-y^{2}}$-states.

In Fig.~\ref{fig:loc_spec_func}(c), the metallic regime ($\delta n = 0.20$) is marked by a finite local spectral function at the Fermi level. Unlike in the pseudogap regime [see Fig.~\ref{fig:loc_spec_func}(b)], in the metallic regime, besides that both orbitals contribute to the local spectral function at the Fermi level, the $d_{x^{2}-y^{2}}$-states are dominant. Furthermore, for this doping and temperature, no  coherent quasi-particle peak is observed.

\section{Doping and temperature evolution of the static spin susceptibility}
\label{sec:susc}

\begin{figure}[b!]
\centering
\includegraphics[width=.8\columnwidth]{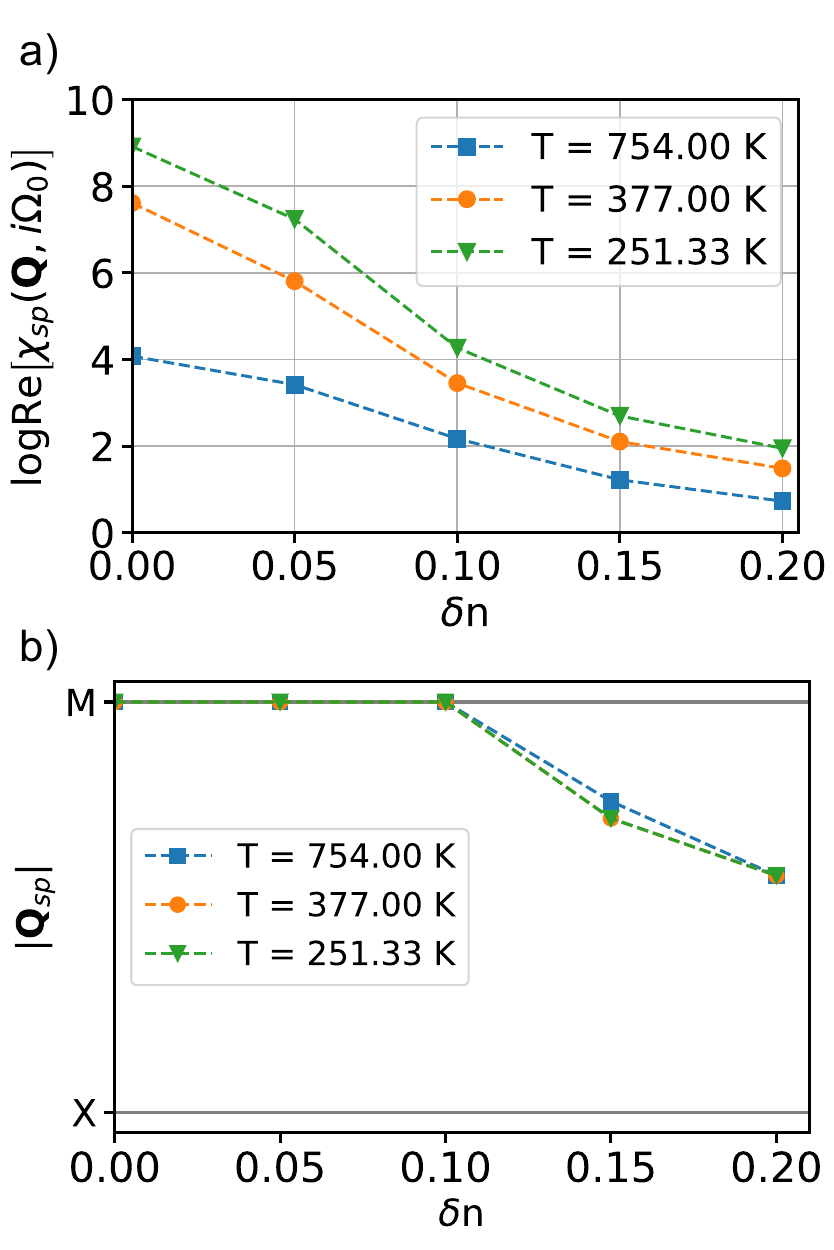}
\caption{Doping evolution of the a) maximum  and b) position of the maximum of the static spin susceptibility for different temperatures.}
\label{fig:max_stat_suscept}
\end{figure}

\begin{figure}[t!]
\centering
\includegraphics[width=1.\columnwidth]{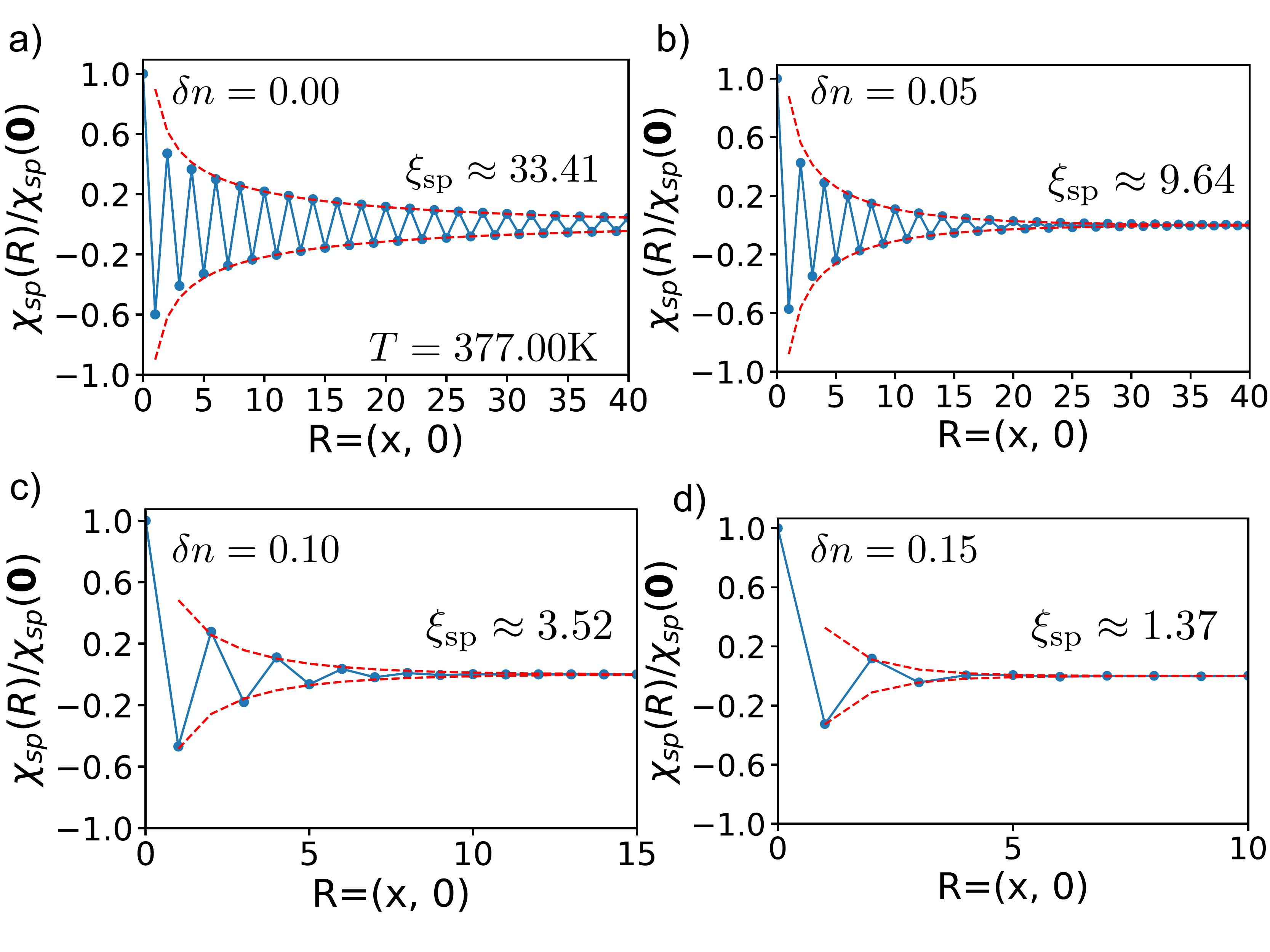}
\caption{Doping evolution of the real space static spin susceptibility for  a) $\delta n$ = $0.0$, b) $\delta n$ = $0.05$, c) $\delta n$ = $0.1$ and d) $\delta n$ = $0.15$. The red dashed line represents the fit of the asymptotic behavior of the real-space magnetic susceptibility, $\left|\chi_{\text{m}}\left(\mathbf{r} \rightarrow \infty,i\Omega_{0}\right)\right|\propto\sqrt{\frac{\xi}{r}}e^{-\frac{r}{\xi}}.$} \label{fig:real_space_stat_suscept}
\end{figure}

Fig.~\ref{fig:max_stat_suscept}(a) shows the hole-doping dependence of the maximum of the static lattice magnetic susceptibility $\chi_{\text{m}}\left(\mathbf{Q},i\Omega_{0}\right)$ [see Eq.~(\ref{eq:BSE})] for different temperatures. For all the temperatures shown in Fig.~\ref{fig:max_stat_suscept}(a), the static lattice magnetic susceptibility reaches its maximum value in the undoped compound ($\delta n = 0.0$) and monotonously decreases upon hole-doping. Turning to the temperature dependence, we note that for all the hole-doping levels displayed in Fig.~\ref{fig:max_stat_suscept}(a), the static lattice magnetic susceptibility increases upon cooling the system.

In order to investigate the nature of the magnetic correlations, Fig.~\ref{fig:max_stat_suscept}(b)
displays, for different temperatures, the hole-doping dependence of the amplitude of the 2D momentum vector $\left(\left|\mathbf{Q}\right|=\sqrt{Q_{x}^{2}+Q_{y}^{2}}\right)$ where the static lattice magnetic susceptibility reaches its maximum value in the Brillouin zone (BZ). For all temperatures displayed in Fig.~\ref{fig:max_stat_suscept}(b) we observe that either in the insulating or pseudogap regime, $0.0\leq\delta n\leq0.1$, antiferromagnetic correlations peaked at $\mathbf{q}=\mathbf{M}=(\pi,\pi)$ dominate the magnetic susceptibility, while inside the metallic regime ($\delta n\geq0.15 $) they become incommensurate and peaked at $\mathbf{q} =(\pi,\pi\pm\delta),$ $(\pi\pm\delta,\pi)$.

\begin{figure*}[t!]
\includegraphics[width=.9\textwidth]{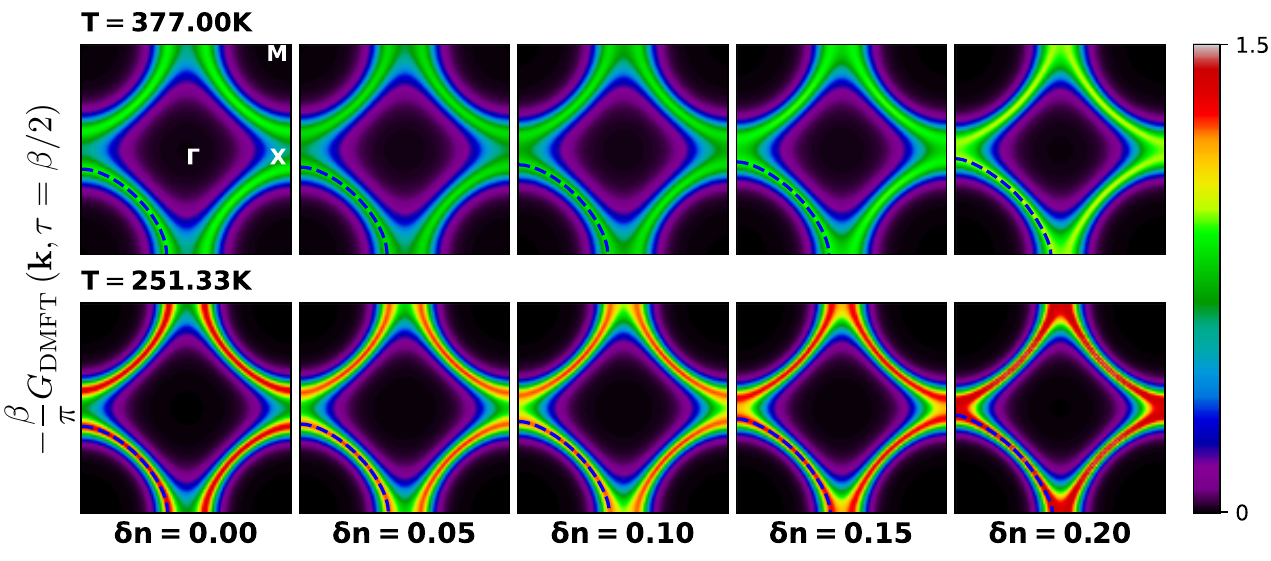}
\caption{Hole-doping evolution of DMFT $\mathbf{k}$-resolved spectral intensities at the Fermi energy for $T=t_{pp}/20=377$ K (upper panels) and $T=t_{pp}/30=251.33 $ K (lower panels) with displayed interacting Fermi plus Luttinger surface (dashed white line).} 
\label{fig:spectral_functions_dmft}
\end{figure*}

To extend our analysis, in Fig.~\ref{fig:real_space_stat_suscept} we show, along the $x$-direction, the spatial decaying the real space magnetic correlation function, $\chi_{\text{m}}\left(\mathbf{r},i\Omega_{0}\right)=\int_{0}^{\beta}d\tau \left<S_{z}\left(\mathbf{r},\tau\right)S_{z}\left(0,0\right)\right>$, at $T=377.00$ K and representative hole-doping levels. In fact, when the system is either in the insulating or pseudogap regime, $0.0\leq\delta n\leq0.1$, the real-space magnetic susceptibility displays an alternating sign, which is a hallmark feature of predominant antiferromagnetic fluctuations. The spatial extension of these magnetic fluctuations is quantified by the magnetic correlation length. Indeed, the correlation length can be extracted from the long-distance behavior of the real-space magnetic susceptibility which can be approximated by its asymptotic form $\left|\chi_{\text{m}}\left(\mathbf{r} \rightarrow \infty,i\Omega_{0}\right)\right|\propto\sqrt{\frac{\xi}{r}}e^{-\frac{r}{\xi}}$ \cite{Altland2008}. From Fig.~\ref{fig:real_space_stat_suscept}, we note that the insulating nature of the parent compound ($\delta n$ = 0.0) is driven by the long-ranged antiferromagnetic fluctuations ($\xi \approx 34 $ lattice sites) while deep inside the pseudogap phase ($\delta n$ = 0.10) these fluctuations become rather short-ranged ($\xi \approx 2.5 $ lattice sites). Such behavior of the correlation length hints that, in ladder D$\Gamma$A, the short-ranged dynamical magnetic correlations drives the pseudogap regime in the Emery model, and, hence, its pseudogap can be classified as a result of strong-coupling physics.

\section{Doping and temperature dependence of DMFT spectral function}
\label{sec:dmft_spec}

The absence of non-local fluctuations in DMFT leads to an approximation to the self-energy that includes all temporal (quantum) fluctuations but that it is local in real space (momentum independent). Non-local fluctuations arising from the calculation of susceptibilities via Bethe-Salpeter ladders also do not feed back to the single-particle level (self-energy) in DMFT. In this section we display the DMFT spectral functions so that they can be contrasted to our D$\Gamma$A results shown in the main text.

Fig.~\ref{fig:spectral_functions_dmft} shows the momentum dependence of the orbital-traced DMFT lattice Green function [see Eq.~(\ref{eq:DMFT_latt_GF})] in imaginary time $\tau\!=\!\beta/2$ $\left(\beta\!=\!1/T\right)$ for different temperatures: $T=377$ K, $251.33$ K and various values of hole-doping: $\delta n=0.00, 0.05, 0.10, 0.15, 0.20$ with $\delta n=5-n$. This quantity is directly related to the spectrum in an energy interval $\sim\!T$ around the Fermi energy, $-\frac{\beta}{\pi}G\left(\mathbf{k},\tau=\beta/2\right)=\frac{\beta}{2\pi}\int\text{d}\omega\frac{A\left(\mathbf{k},\omega\right)}{\cosh\left(\beta\omega/2\right)}$, and avoids the perils of the ill-conditioned analytical continuation. Fig.~\ref{fig:spectral_functions_dmft} also displays the correlated Fermi plus Luttinger surface, indicated as dashed blue lines and determined as $\text{Re }G_{\text{DMFT}}\left(\mathbf{k}_{\text{FS}},i\omega_{n}\rightarrow0\right)=0$, where $i\omega_{n}\rightarrow0$ is obtained as a linear extrapolation of the two first fermionic Matsubara frequencies. For both temperatures,  $T=377$ K and $251.33$ K, we observe that the parent compound ($\delta n = 0.0$) is metallic, and upon doping, it remains metallic, not giving rise to a pseudogap phase. This observation can be directly traced back to the lack of feedback of non-local spin fluctuations to the single-particle level.

Please also note that, upon cooling, for all hole-doping levels, the spectrum intensity over the Fermi surface enhances, which is consistent with a metallic state. In addition, for both temperatures and hole-doping levels the Fermi surface has a hole-like shape.

\FloatBarrier
\bibliography{supplemental_material.bib}